\documentclass[%
 aip, 
 amsmath,amssymb,
eqsecnum, 
reprint,%
onecolumn, 
11pt, 
author-year,%
tightenlines
]{revtex4-2}

\usepackage{graphicx}
\usepackage{dcolumn}

\usepackage{epstopdf,epsfig}

\allowdisplaybreaks

\usepackage{esint}  
\usepackage{stmaryrd} 
\usepackage[nointegrals]{wasysym} 
\usepackage{color} 

\usepackage{array}
\newcolumntype{C}[1]{>{\centering\let\newline\\\arraybackslash\hspace{0pt}}m{#1}}
\newcolumntype{L}[1]{>{\raggedright\let\newline\\\arraybackslash\hspace{0pt}}m{#1}}

\makeatletter
\renewcommand{\p@subsection}{}
\renewcommand{\p@subsubsection}{}
\makeatother

\renewcommand{\vec}[1]{\boldsymbol {#1}}
\newcommand{\vect}[1]{\boldsymbol {#1}}
\newcommand{\tens}[1]{\boldsymbol{#1}}
\newcommand{\vnabla}{\nabla}

\newcommand{\F}{\mathcal{F}}
\renewcommand{\H}{\mathcal{H}}

\renewcommand{\P}{\mathbb{P}}
\newcommand{\R}{\mathbb{R}}

\newcommand{\D}{\mathcal{D}}
\newcommand{\Dc}{\mathcal{D}_c}
\newcommand{\Dct}{\mathcal{D}_{c,t}}
\newcommand{\Db}{\mathcal{D}_b}
\newcommand{\Dbt}{\mathcal{D}_{b,t}}
\newcommand{\Dk}{\text{$D_k$}}  
\newcommand{\Dkstar}{\text{$D_{k}^*$}}  
\newcommand{\Dkt}{\text{$D_{k,t}$}}  
\newcommand{\Dl}{\text{$D_{\ell}$}}  
\newcommand{\Dlt}{\text{$D_{\ell,t}$}}  
\newcommand{\Dst}{\mathcal{D}_{s,t}}

\newcommand{\dsigma}{\dot{\sigma}}

\renewcommand{\i}{\vec{i}}
\renewcommand{\j}{\vec{j}}
\newcommand{\J}{\vec{J}}
\renewcommand{\k}{\vec{k}}
\newcommand{\m}{\vec{m}}
\newcommand{\n}{\vec{n}}
\renewcommand{\r}{\vec{r}}
\renewcommand{\u}{\vec{u}}
\renewcommand{\v}{\vec{v}}
\newcommand{\x}{\vec{x}}
\newcommand{\C}{\vec{C}}
\newcommand{\T}{\vec{T}}
\newcommand{\U}{\vec{U}}
\newcommand{\V}{\vec{V}}
\newcommand{\Z}{Z}

\newcommand{\freq}{\tilde{\nu}}
\newcommand{\Stef}{\tilde{\sigma}}
\newcommand{\abs}{\tilde{\alpha}}
\newcommand{\emiss}{\tilde{\epsilon}}
\newcommand{\refl}{\tilde{\rho}}
\newcommand{\transm}{\tilde{\tau}}

\newcommand{\e}{\tens{e}}
\newcommand{\w}{\tens{w}}

\newcommand{\Cu}{\mathsf{C}}

\newcommand{\slope}{\mathsf{S}}

\newcommand{\hatPi}{\hat{\Pi}}

\newcommand{\tension}{\varsigma}

\newcommand{\FF}{\digamma}
\newcommand{\G}{\widetilde{G}}

\newcommand{\tr}{\text{tr}}
\newcommand{\sign}{\text{sign}}

\newcommand{\tto}{\rightsquigarrow}

\definecolor{orange}{rgb}{1.0, 0.5, 0.0}  

\definecolor{Green}{cmyk}{0.70, 0.05, 0.99, 0.18}  

\newcommand{\Y}{}  
\newcommand{\YY}{}  

\DeclareMathOperator{\ad}{ad}
\DeclareMathOperator{\compr}{compr}
\DeclareMathOperator{\cond}{cond}
\DeclareMathOperator{\conv}{conv}
\DeclareMathOperator{\disp}{disp}
\DeclareMathOperator{\dyn}{dynam}
\DeclareMathOperator{\entrop}{entrop}
\DeclareMathOperator{\explos}{explos}
\DeclareMathOperator{\final}{final}
\DeclareMathOperator{\flow}{flow}
\DeclareMathOperator{\geom}{geom}

\DeclareMathOperator{\group}{group}
\DeclareMathOperator{\info}{info}
\DeclareMathOperator{\init}{init}
\DeclareMathOperator{\jump}{jump}
\DeclareMathOperator{\kin}{kinem}
\DeclareMathOperator{\lat}{lat}
\DeclareMathOperator{\net}{net}
\DeclareMathOperator{\open}{open}
\DeclareMathOperator{\rad}{rad}
\DeclareMathOperator{\rot}{rot}
\DeclareMathOperator{\ROU}{ROU}
\DeclareMathOperator{\ship}{ship}
\DeclareMathOperator{\Sh}{Sh}
\DeclareMathOperator{\shock}{shock}
\DeclareMathOperator{\signal}{signal}
\DeclareMathOperator{\steady}{steady}
\DeclareMathOperator{\stor}{stor}
\DeclareMathOperator{\strat}{strat}
\DeclareMathOperator{\sys}{sys}
\DeclareMathOperator{\tot}{tot}
\DeclareMathOperator{\univ}{univ}

\begin{document}


\title[Dimensionless Groups by Entropic Similarity]{Dimensionless Groups by Entropic Similarity}

\author{Robert K. Niven}
 \email{r.niven@adfa.edu.au}
\affiliation{ 
School of Engineering and Information Technology, The University of New South Wales, Canberra, ACT, 2600, Australia.
}%

\date{26 January 2023}

\begin{abstract}
We propose an additional category of dimensionless groups based on the principle of {\it entropic similarity}, defined by ratios of (i) entropy production terms; (ii) entropy flow rates or fluxes; or (iii) information flow rates or fluxes. 
Since all processes involving work against friction, dissipation, diffusion, dispersion, mixing, separation, chemical reaction, gain of information or other irreversible changes are driven by (or must overcome) the second law of thermodynamics, it is appropriate to analyse these processes directly in terms of competing entropy-producing and transporting phenomena and the dominant entropic regime, rather than indirectly in terms of forces.  
In this study, we derive entropic groups for a wide variety of phenomena relevant to fluid mechanics, classified into diffusion and chemical reaction processes, dispersion mechanisms and wave phenomena. 
It is shown that many dimensionless groups traditionally derived by kinematic or dynamic similarity (including the Reynolds number) can also be recovered by entropic similarity -- albeit with a different entropic interpretation -- while a large number of new dimensionless groups are also identified.
The analyses significantly expand the scope of dimensional analysis and similarity arguments for the resolution of new and existing problems in science and engineering.
\end{abstract}

\maketitle

\section{\label{sect:intro} Introduction} 

Since the seminal work of work of \cite{Buckingham_1914}, built on the insights of many predecessors \citep[e.g.,][]{Fourier_1822, Rayleigh_1877, Rayleigh_1892, Bertrand_1878, Carvallo_1892, Vaschy_1892a, Vaschy_1892b, Federman_1911, Riabouchinsky_1911}, dimensional analysis and similarity arguments based on dimensionless groups have provided a powerful tool -- and in many cases the most important tool -- for the analysis of physical, chemical, biological, geological, environmental, astronomical, mechanical and thermodynamic systems, especially those involving fluid mechanics. The dimensionless groups obtained are usually classified into those arising from {\it geometric similarity}, based on ratios of length scales (or areas or volumes); {\it kinematic similarity}, based on ratios of velocities or accelerations; and {\it dynamic similarity}, based on ratios of forces \citep[e.g.,][]{Pao_1961, Street_etal_1996, Furbish_1997, Streeter_etal_1998, White_2006, Munson_etal_2010, Douglas_etal_2011}. Thus for example, the Reynolds number \citep{Reynolds_1883} is generally interpreted as:
\begin{align}
Re = \frac{\text{inertial force} }{\text{viscous force}} 
=\frac{F_I}{F_\nu}
\sim \frac{\rho d^3 (U^2/d)}{\mu (U/d) d^2} 
= \frac{\rho U d}{\mu}
= \frac{U d}{\nu}
\label{eq:Re}
\end{align}
where $\sim$ indicates ``of the order of'' (discarding numerical constants), 
$\rho$ is the fluid density [SI units: kg m$^{-3}$]\footnote{In the author's experience as a university educator, it is generally more instructive -- and enables higher fidelity -- to write the dimensions of each quantity in SI units rather than in ``dimensions'' notation.}, 
$\mu$ is the dynamic viscosity [Pa s], $\nu$ is the kinematic viscosity [m$^{2}$ s$^{-1}$], $U$ is a velocity scale [m s$^{-1}$] and $d$ is an applicable length scale [m]. 
In common with many dimensionless groups, \eqref{eq:Re} provides an identifier of the flow regime, reflecting the fact that viscous forces -- causing laminar flow -- are dominant at Reynolds numbers below some critical value $Re_c$, while inertial forces -- leading to flows of increasingly turbulent character -- will be dominant above this critical value. 

The aim of this work is to provide a new interpretation for a large class of dimensionless groups based on the principle of {\it entropic similarity}, involving ratios of entropic terms. Since {all} processes involving work against friction, dissipation, diffusion, dispersion, mixing, separation, chemical reaction, gain of information 
or other irreversible changes are driven by (or must overcome) the second law of thermodynamics, it is appropriate to analyse these processes directly in terms of competing entropy-producing and transporting phenomena and the dominant entropic regime, rather than indirectly in terms of forces.  As will be shown, an entropic perspective enables the reinterpretation of many known dimensionless groups --  including the Reynolds number \eqref{eq:Re} -- as well as the formulation of a large assortment of new groups. These significantly expand the scope of dimensional analysis and similarity arguments for the resolution of new and existing problems.

We further note that while the energetic formulation of thermodynamics developed over the last 150 years -- especially by \cite{Gibbs_1875} -- has been of tremendous utility in all branches of science and engineering, its underlying basis is entropic, as was recognised by Gibbs and many other prominent researchers  \citep[e.g.,][]{Massieu_1869, Planck_1922, Jaynes_1957, Tribus_1961a, Tribus_1961b, Guggenheim_1967, Prigogine_1967, Callen_1985, Kondepudi_P_2015}. Such researchers understood that ``dissipation'', the irreversible loss of organised energy, matter or information, 
is not a driving force in its own right, but is a consequence of the second law of thermodynamics. It is therefore appropriate that dissipative phenomena be expressed in terms of entropic rather than energetic quantities. In this manner, the {\it quality} or {\it temper} of the degradation -- expressed by a thermodynamic integration factor such as $1/T$, where $T$ is the absolute temperature [K] -- is recognised explicitly \citep{Tribus_1961b}. 

This work is set out as follows. Firstly, in \S\ref{sect:theory} the theoretical foundations of the entropic perspective are examined in detail. This includes the generic definitions of entropy given by \cite{Boltzmann_1877} and \cite{Shannon_1948}, the maximum entropy method of \cite{Jaynes_1957}, the information-theoretic interpretation of the second law given by \cite{Szilard_1929}, and the thermodynamic entropy balance equation and the definitions of the entropy production in global and local form. 
The importance of dimensional arguments and the known definitions of similarity (geometric, kinematic and dynamic) are examined in \S\ref{sect:entrop_sim}, following which the principle of entropic similarity is established, with three interpretations.
In \S\ref{sect:EP}, a number of entropy-producing and entropy-transporting phenomena are then examined in detail, including diffusion and chemical reaction processes (\S\ref{sect:diffusion}), an assortment of dispersion mechanisms (\S\ref{sect:dispersion}) and a variety of wave phenomena and their associated information-theoretic flow regimes and frictional losses (\S\ref{sect:wave}). A number of dimensionless groups are derived for these phenomena on the basis of entropic similarity, and compared to the traditional groups obtained by other methods, providing interesting contrasts between the two approaches and several important new insights. The conclusions are given in \S\ref{sect:concl}. 

We note that the analyses in this study span many branches of science and engineering, and some clashes of standard symbols cannot be avoided; where present, these are mentioned explicitly. 

\section{\label{sect:theory} Theoretical Foundations} 
\subsection{\label{sect:entropy_infm} Dimensionless Entropy and Information} 

We first review the entropy concept and its connections to combinatorics and information theory. 
The (dimensionless) discrete entropy function was given by \cite{Shannon_1948} as:
\begin{equation}
\H_{\Sh} = - \sum\limits_{i = 1}^n {p_i \ln p_i } 
\label{eq:Shannon}
\end{equation}
where $p_i$ is the probability of the $i$th outcome or category, from $n$ such categories.  
For a system with unequal prior probabilities $q_i$ for each category, it is necessary to adopt the relative entropy function or negative Kullback-Leibler divergence \citep{Kullback_L_1951, Kapur_K_1992, Jaynes_2003}:
\begin{equation}
\H = -\sum\limits_{i = 1}^n {p_i \ln \frac{{p_i }}{{q_i }}} 
\label{eq:rel_entropy}
\end{equation}
Eqs.\ \eqref{eq:Shannon}-\eqref{eq:rel_entropy} can be extended to multivariate discrete variables by summation over all indices. 
For a continuous variable $\x \in \R^m$ with $m \in \mathbb{N}$, both \eqref{eq:Shannon}-\eqref{eq:rel_entropy} converge to the continuous relative entropy \citep{Sanov_1957, Jaynes_2003}:
\begin{equation}
\H= -\int\limits_{\Omega_{\x}} d\x  \; {p(\x) \ln \frac{{p(\x) }}{{q(\x) }}} 
\label{eq:rel_entropy_int}
\end{equation}
where $\Omega_{\x}$ is the domain of $\x$, $d\x=dx_1 ... dx_m$, and $p(\x)$ and $q(\x)$ are respectively the posterior and prior probability density functions (pdfs). 
We note in passing that the ``continuous entropy'' $\H_{\Sh}= -\int\nolimits_{\Omega_{\x}} d\x \, p(\x) \ln p(\x)$ of \cite{Shannon_1948} contains inconsistent units and is not scale invariant, and is therefore il-founded \citep[see][]{Jaynes_2003}.  

The discrete entropy functions \eqref{eq:Shannon}-\eqref{eq:rel_entropy} can be derived from the axiomatic properties of a measure of uncertainty \citep{Shannon_1948, Shore_J_1980, Kapur_K_1992}. They can also be derived from the \cite{Boltzmann_1877} combinatorial definition of entropy:
\begin{equation}
\H = \frac{1}{N} \ln \P
\label{eq:Boltzmann}
\end{equation}
where $\P$ is the governing probability distribution of the system, representing an allocation scheme for $N$ entities. For a system composed of distinguishable entities allocated to distinguishable states, $\P$ is given by the multinomial distribution $\mathbb{P} = N! \prod\nolimits_{i=1}^s {q_i ^ {n_i}}/{n_i!}$, where $n_i$ entities are allocated to the $i$th state and $\sum\nolimits_{i=1}^s n_i=N$. In the asymptotic limits $N \to \infty$ and $n_i/N \to p_i$, \eqref{eq:Boltzmann} converges to the discrete relative entropy \eqref{eq:rel_entropy} \citep{Boltzmann_1877, Planck_1901, Ellis_1985}.  Alternative entropy functions can also be derived from \eqref{eq:Boltzmann} for non-multinomial governing distributions \citep[e.g.,][]{Bose_1924, Einstein_1924, Fermi_1926, Dirac_1926, Brillouin_1930, Niven_2009_EPJB, Niven_Grendar_2009}.

In the maximum entropy (MaxEnt) method, the user maximises the appropriate entropy function \eqref{eq:Shannon}-\eqref{eq:rel_entropy_int} subject to its $(R+1)$ moment constraints, to give the inferred distribution of the system and its maximum entropy, respectively \citep{Jaynes_1957, Jaynes_1963, Jaynes_2003}:
\begin{gather}
p_i^* = \frac{q_i}{Z} \exp \biggl(- \sum_{r=0}^R \lambda_r f_{ri} \biggr)
\hspace{10pt} \text{or} \hspace{10pt}
p^*(\x) = \frac{q(\x)}{Z} \exp \biggl(- \sum_{r=0}^R \lambda_r f_r(\x) \biggr)
\label{eq:pstar}
\\
\H^* =  \ln \Z + \sum_{r=1}^R \lambda_r \langle f_r \rangle
\label{eq:MaxEnt}
\end{gather}
where $^{*}$ denotes the inferred state, 
$\Z$ 
is the partition function, and, for the $r$th category, 
$f_{ri}$ or $f_r(\x)$ is the local category value, 
$\langle f_r \rangle$ is the expected category value and
$\lambda_r$ is the Lagrangian multiplier.
In the Boltzmann interpretation, \eqref{eq:pstar} gives the most probable distribution of the system. 
The entire body of thermodynamics can readily be derived by this method, based on the thermodynamic entropy $S=k_B \H$, where $k_B$ is the Boltzmann constant [J K$^{-1}$] \citep{Jaynes_1957, Jaynes_2003, Tribus_1961a, Tribus_1961b, Davidson_1962}. The MaxEnt method has also been used to determine the stationary state of many systems beyond thermodynamics, including a large variety of hydraulic, hydrological, geological, biological, ecological and financial systems, and the analysis of water distribution, electrical and transport networks \citep[e.g.,][]{Wilson_1970, Levine_Tribus_1978, Kapur_K_1992,  Golan_etal_1996, Jaynes_2003, Harte_2011, Singh_2014, Singh_2015, Waldrip_etal_2016}. 

In information theory, it is usual to consider the binary entropy and relative entropy, expressed in binary digits or ``bits'' \citep{Cover_T_2006}:
\begin{align}
B_{\Sh} = - \sum\limits_{i = 1}^n p_i \log_2 p_i 
\hspace{20pt}  \text{or}  \hspace{20pt}
B = - \sum\limits_{i = 1}^n p_i \log_2 \frac{p_i}{q_i} 
\label{eq:rel_entropyB}
\end{align}
Comparing \eqref{eq:Shannon}-\eqref{eq:rel_entropy}, we obtain $\H_{\Sh}= B_{\Sh} \ln 2$ and $\H= B \ln 2$. 
The change in information about a system can then be defined as the negative change in its binary entropy \citep{Wiener_1948, Brillouin_1949, Brillouin_1950, Brillouin_1951a, Brillouin_1953, Schrodinger_1952}:
\begin{equation}
\Delta I  = - \Delta B_{\Sh} 
\hspace{20pt}  \text{or}  \hspace{20pt}
\Delta I  = - \Delta B
\label{eq:Weiner_rel}
\end{equation}
For example, consider a coin toss with equiprobable outcomes H and T. Before the toss, an observer will assign $p_H=p_T=\frac{1}{2}$, hence  $B_{\Sh}^{\init}=1$ bit. Once informed of the outcome, the observer must assign one probability to zero and the other to unity, hence $B_{\Sh}^{\final}=0$ bits and $\Delta B_{\Sh} =0-1=-1$ bit. From \eqref{eq:Weiner_rel}, the information gained from the binary decision is $\Delta I =  -\Delta B_{\Sh} = 1$ bit \citep{Szilard_1929, Shannon_1948, Cover_T_2006}. If instead we use the relative entropy \eqref{eq:rel_entropyB} with equal priors  $q_H=q_T=\frac{1}{2}$, we obtain $B^{\init}=0$ bits, $B^{\final}=-1$ bits and $\Delta B =-1-0=-1$ bits,  so again $\Delta I=1$ bit.

This idea was taken further by \cite{Szilard_1929} and later authors based on analyses of Maxwell's demon  \citep[e.g.,][]{Brillouin_1949, Brillouin_1950, Brillouin_1951a, Brillouin_1953, Landauer_1961, Bennett_1973}, to establish a fundamental relationship between changes in information, thermodynamic entropy and energy. From the second law of thermodynamics:
\begin{align}
\Delta S_{\univ} = \Delta S_{\sys} +  \Delta S_{\ROU} \ge 0
\label{eq:2ndlaw}
\end{align}
where $\Delta S_{\univ}$ is the change in thermodynamic entropy of the universe [J K$^{-1}$], which can be partitioned into the changes $\Delta S_{\sys}$ within a system and $\Delta S_{\ROU}$ in the rest of the universe.
Substituting for $\Delta S_{\sys}$ using \eqref{eq:rel_entropyB}-\eqref{eq:Weiner_rel} gives:
\begin{align}
k_B  \ln 2 \, \Delta B +  \Delta S_{\ROU} 
= - k_B  \ln 2 \, \Delta I +  \Delta S_{\ROU}
\ge 0
\label{eq:2ndlaw2}
\end{align}
Now in general, a system can only influence the entropy of the rest of the universe by a transfer of disordered energy $\Delta E  = T \Delta S_{\ROU}$ [J], such as that carried by heat or chemical species.
Manipulating \eqref{eq:2ndlaw2} and substituting for $\Delta E$ gives:
\begin{align}
k_B  \ln 2 \, \Delta I \le  \Delta S_{\ROU}
\hspace{10pt} \text{or} \hspace{10pt}
k_B T \ln 2 \, \Delta I \le  \Delta E 
\label{eq:Szilard3}
\end{align}
Eq.\ \eqref{eq:Szilard3} gives an information-theoretic formulation of the second law of thermodynamics, in which each bit of information gained by an observer about a system must be paid for by an energy cost of at least $k_B T \ln 2$.  This principle places an important constraint on natural and computational processes involving the transmission of information. 


\subsection{\label{sect:backgr} Thermodynamic Entropy Balance and the Entropy Production} 

We now consider flows of the thermodynamic entropy $S$ [J K$^{-1}$]. For an open system, the balance equation for thermodynamic entropy in a macroscopic control volume is:
\begin{align}
\frac{DS_{FV(t)}}{Dt} 
= \frac{\partial S_{CV}}{\partial t} + \F_{S,f}^{out} - \F_{S,f}^{in}  
\label{eq:Sbalance} 
\end{align}
where $t$ is time, $FV$ is the fluid volume, $CV$ is the control volume, $D/Dt$ is the substantial derivative, and $\F_{S,f}^{out}$ and $\F_{S,f}^{in}$ are respectively the flow rates of $S$ out of and into the control volume by fluid flow [J K$^{-1}$ s$^{-1}$]. By the de Donder method, the substantial derivative in \eqref{eq:Sbalance} can be separated into internally- and externally-driven rates of change, respectively:
\begin{align}
\frac{DS_{FV(t)}}{Dt} 
=    \frac{D_{i}S_{FV(t)}}{Dt} + \frac{D_{e}S_{FV(t)}}{Dt}
\label{eq:deDonder}
\end{align}
The first term on the right of \eqref{eq:deDonder} is the rate of entropy production, commonly denoted $\dot{\sigma}$ [J K$^{-1}$ s$^{-1}$]. 
The second term on the right of \eqref{eq:deDonder} is the total rate of change of entropy due to non-fluid flows, such as heat or chemical species:
\begin{align}
\frac{D_{e}S_{FV(t)}}{Dt}  = \F_{S,nf}^{in} - \F_{S,nf}^{out}
\label{eq:DeSDt} 
\end{align}
where $\F_{S,nf}^{out}$ and $\F_{S,nf}^{in}$ are respectively the outward and inward non-fluid flow rates of $S$ [J K$^{-1}$ s$^{-1}$].
Combining \eqref{eq:Sbalance}-\eqref{eq:DeSDt} and rearrangement gives:
\begin{align}
\dot{\sigma} 
= \frac{\partial S_{CV}}{\partial t} + \F_{S,f}^{out}  - \F_{S,f}^{in}  + \F_{S,nf}^{out} - \F_{S,nf}^{in} 
= \frac{\partial S_{CV}}{\partial t} + \F_{S,tot}^{net}  
\ge 0 
\label{eq:EPdef} 
\end{align}
where $\F_{S,tot}^{net}$ is the net total outwards entropy flow rate [J K$^{-1}$ s$^{-1}$].

From the second law of thermodynamics, the entropy production \eqref{eq:EPdef} must be nonnegative. In contrast, the rate of change of entropy within the system ${\partial S_{CV}}/{\partial t}$ can take any sign, but will vanish at the steady state. The entropy production therefore represents the irreversible rate of increase in entropy of the universe due to the system; at steady state, this is equal to the rate at which the system exports entropy to the rest of the universe. Unfortunately, there is still considerable confusion in the literature between the rate of entropy production of a system $\dot{\sigma}$ and its rate of change of entropy ${\partial S_{CV}}/{\partial t}$.

For an integral control volume commonly examined in fluid mechanics, the entropy balance equation is given by Reynolds' transport theorem \citep[e.g.,][]{Prager_1961, Aris_1962, Tai_1992}:
\begin{align}
\frac{DS_{FV(t)}}{Dt} 
= \frac{\partial }{\partial t} \iiint\limits_{CV} \rho s \, dV  + \oiint\limits_{CS} \rho s \u \cdot \vec{n} dA 
= \iiint\limits_{CV} \biggl[  \frac{\partial }{\partial t} \rho s  +   \vnabla \cdot \rho s \u \biggr] dV
\label{eq:Sbalance_int}
\end{align}
where $CS$ is the control surface, $s$ is the specific entropy [J K$^{-1}$ kg$^{-1}$], $\u$ is the fluid velocity [m s$^{-1}$], $dV$ is a volume element [m$^{3}$], $dA$ is an area element [m$^{2}$], $\vec{n}$ is an outwardly directed unit normal [-] and $\vnabla$ is the Cartesian nabla operator [m$^{-1}$].  
From \eqref{eq:deDonder}, we can write:
\begin{align}
\frac{DS_{FV(t)}}{Dt} 
= \dot{\sigma} - \oiint\limits_{FS(t)} \j_S \cdot  \vec{n}  dA
\label{eq:deDonder_int} 
\end{align}
where $\j_{S}$ is the outward non-fluid entropy flux [J K$^{-1}$ m$^{-2}$ s$^{-1}$], and $FS(t)$ is the fluid surface coincident with the control surface at time $t$. Equating \eqref{eq:Sbalance_int}-\eqref{eq:deDonder_int} and rearrangement gives \citep[e.g.,][]{Jaumann_1911, deGroot_M_1962, Prigogine_1967, Kreuzer_1981, Bird_etal_2006, Kondepudi_P_2015, Niven_Noack_2014}:
\begin{align}
\dot{\sigma} 
= \iiint\limits_{CV}\frac{\partial \rho s}{\partial t} dV  + \oiint\limits_{CS} \J_S \cdot \vec{n} dA 
= \iiint\limits_{CV} \biggl[  \frac{\partial }{\partial t} \rho s  + \vnabla \cdot \J_{S} \biggr ] dV   
\ge 0
\label{eq:EPdef_int} 
\end{align}
where $\J_{S} = \j_{S} + \rho s \u$ is the total outward entropy flux [J K$^{-1}$ m$^{-2}$ s$^{-1}$]. 


To establish a local entropy balance equation, we first define the {rate of entropy production per unit volume} 
$\hat{\dot{\sigma}}$ [J K$^{-1}$ m$^{-3}$ s$^{-1}$] by the integral:
\begin{align}
{\dsigma} = \iiint\limits_{CV} {\hat{\dot{\sigma}}} dV
\label{eq:global_local_int}
\end{align}
Applying the fundamental lemma of the calculus of variations to \eqref{eq:EPdef_int}, 
we can then extract the differential equation for the local entropy production \citep[e.g.][]{Jaumann_1911, deGroot_M_1962, Kreuzer_1981, Kondepudi_P_2015, Bird_etal_2006}:
\begin{align}
\hat{\dot{\sigma}} 
=   \frac{\partial }{\partial t} \rho s + \vnabla \cdot \J_{S} 
\ge 0
\label{eq:EPdef_local}
\end{align}
From the second law, $\hat{\dot{\sigma}}$ is non-negative, whereas the local rate $\partial s/\partial t$ can take any sign. At local steady state ${\partial (\rho s)}/{\partial t} =0$ and $\hat{\dot{\sigma}} = \vnabla \cdot \J_{S} \ge 0$.

\section{\label{sect:entrop_sim} Principle of Entropic Similarity} 

A {\it dimensionless group} is a unitless (dimensionless) parameter used to represent an attribute of a physical system, independent of the system of units used.  
By the late 19th century, researchers had established the concept of {\it similarity} or {\it similitude} between a system (prototype) and its model based on matching dimensionless groups, so that their mechanical or physical properties would be equivalent \citep{Carvallo_1892, Vaschy_1892b}. 
Such dimensional {scaling} offers the advantage of smaller-scale models, greatly simplifying the experimental requirements.
The formal method of {\it dimensional analysis} was then developed to extract the functional dependencies of a physical system from its list of parameters \citep{Buckingham_1914}. 
It also enables order reduction, reducing the number of parameters by the number of dimensions.
For over a century, these dimensional methods have been recognised as powerful tools -- and in many cases the primary tools -- for the analysis of a wide range of systems across all branches of science and engineering \citep[e.g.][]{Langhaar_1951, Zeldovich_1956, Sedov_1959, Birkhoff_1960, Gratton_1991, Pope_2000, Barenblatt_2003}.

More recently, dimensional analysis has been found to have strong connections to group theory, in particular to continuous (Lie) groups arising from symmetries in the governing equations of a system \citep{Birkhoff_1960, Ovsainnikov_1982, Blumen_Kumei_1989, Olver_1993, Burde_2002, Oliveri_2010, Niven_etal_Reynolds}. This includes a deep connection to the one-parameter Lie group of point scaling transformations \citep[e.g.,][]{Oliveri_2010, Ercan_Kavvas_2015, Polsinelli_Kavvas_2016, Ercan_Kavvas_2017, Niven_2021}.

Dimensionless groups -- commonly labelled $\Pi$ -- are usually classified as follows \citep[e.g.,][]{Pao_1961, Street_etal_1996, Furbish_1997, Streeter_etal_1998, White_2006, Douglas_etal_2011}: 
\begin{itemize}
\item Those arising from {\it geometric similarity}, based on ratios of length scales $\ell_i$ [m] or associated areas or volumes:
\begin{align}
\Pi_{\geom} = \frac{\ell_1}{\ell_2}
\hspace{15pt} \text{ or } \hspace{15pt}
\Pi_{\geom} = \frac{\ell_1^2}{\ell_2^2}
\hspace{15pt} \text{ or } \hspace{15pt}
\Pi_{\geom} = \frac{\ell_1^3}{\ell_2^3}
\end{align}

\item Those arising from {\it kinematic similarity}, based on ratios of magnitudes of velocities $U_i$ [m s$^{-1}$] or accelerations $a_i$ [m s$^{-2}$]:
\begin{align}
\Pi_{\kin} = \frac{{U}_1}{{U}_2}
\hspace{15pt} \text{ or } \hspace{15pt}
\Pi_{\kin} = \frac{{a}_1}{{a}_2}
\end{align}

\item Those arising from {\it dynamic similarity}, based on ratios of magnitudes of forces $F_i$ [N]:
\begin{align}
\Pi_{\dyn} = \frac{{F}_1}{{F}_2}
\label{eq:dyn_sim}
\end{align}

\end{itemize}
Such dimensionless groups are generally obtained in three ways: (i) assembled {\it a priori}, by an assessment of the dominant phenomena of a system; (ii) extracted from a list of parameters by dimensional analysis; or (iii) obtained by non-dimensionalisation of the governing differential equation(s). 
It is generally accepted that for a model and a system (prototype) to exhibit the same physics -- expressed in terms of identical values of dimensionless groups -- they must satisfy the conditions of geometric, kinematic and dynamic similarity. 

The aim of this work is to propose an additional category of dimensionless groups based on {\it entropic similarity}, involving ratios of entropic terms. This enables the direct analysis of phenomena involving friction, dissipation, diffusion, dispersion, mixing, separation, chemical reaction, gain of information 
or other irreversible changes governed by the second law of thermodynamics, as well as entropy transport processes. At present, these processes are usually examined indirectly by dynamic similarity \eqref{eq:dyn_sim}, requiring their conversion into forces, which for many phenomena are rather contrived.
We here distinguish three variants of entropic groups:
\begin{itemize}
\item Those defined by ratios of global \eqref{eq:EPdef_int} or local \eqref{eq:EPdef_local} entropy production terms:
\begin{align}
\Pi_{\entrop} = \frac{{\dot{\sigma}}_1}{{\dot{\sigma}}_2} 
\hspace{15pt} \text{ or } \hspace{15pt}
\hatPi_{\entrop} = \frac{\hat{\dot{\sigma}}_1}{\hat{\dot{\sigma}}_2} 
\label{eq:entropic_sim_EP}
\end{align}
where $\Pi$ represents a global or summary dimensionless group and $\hatPi$ a local group.

\item Those defined by ratios of global flow rates of thermodynamic entropy, such as in \eqref{eq:EPdef}, or by components or magnitudes of their local fluxes, such as in \eqref{eq:EPdef_local}:
\begin{align}
\Pi_{\entrop} = \frac{\F_{S,1}}{\F_{S,2}}
\hspace{15pt} \text{ or } \hspace{15pt}
\hatPi_{\entrop} (\n) = \frac{\j_{S_1} \cdot \n}{\j_{S_2} \cdot \n}
\hspace{15pt} \text{ or } \hspace{15pt}
\hatPi_{\entrop} = \frac{|| \j_{S_1} ||}{|| \j_{S_2} ||}
\label{eq:entropic_sim_flux}
\end{align}
where $\n$ is a unit normal and $||\vec{a}||=\sqrt{\vec{a}^\top \vec{a}}$ is the Euclidean norm for vector $\vec{a}$.

\item Those defined by an information-theoretic threshold (see \S\ref{sect:entropy_infm}). This is given locally by the ratio of the information flux carried by the flow $\j_{I,\flow}$ [bits m$^{-2}$ s$^{-1}$] to that transmitted by a carrier of information $\j_{I,\signal}$ [bits m$^{-2}$ s$^{-1}$], in the direction of a given unit normal $\n$:
\begin{align}
\hat{\Pi}_{\info} (\n)
= \frac{\j_{I,\flow} \cdot \n}{\j_{I,\signal} \cdot \n}
= \frac{\rho_{I,\flow} \, \u_{\flow} \cdot \n}{\rho_{I,\signal} \, \u_{\signal} \cdot \n}
\label{eq:Pi_infm_gen}
\end{align}
In this perspective, flows in which the information flux of the fluid exceeds that of a signal ($\hat{\Pi}_{\info}>1$) will experience a different information-theoretic flow regime to those in which the signal flux dominates ($\hat{\Pi}_{\info}<1$). In \eqref{eq:Pi_infm_gen}, each information flux is further reduced to the product of an information density $\rho_I$ [bits m$^{-3}$] and the corresponding fluid velocity $\u_{\flow}$ or signal velocity $\u_{\signal}$  [m s$^{-1}$]. Making the strong assumption that the two information densities are comparable, \eqref{eq:Pi_infm_gen} simplifies to give the local or summary kinematic definitions:
\begin{align}
\hat{\Pi}_{\info} (\n) \to \frac{\u_{\flow} \cdot \n}{\u_{\signal} \cdot \n}, 
\hspace{20pt}
\Pi_{\info} \to \frac{U_{\flow}}{U_{\signal}}
\label{eq:Pi_infm}
\end{align}
where $U_{\flow}$ and $U_{\signal}$ are representative flow and signal velocities [m s$^{-1}$].

\end{itemize}


\section{\label{sect:EP} Entropic Phenomena and Dimensionless Groups}

We now examine a succession of natural processes, involving competition between various entropy-producing and/or entropy-transporting phenomena. Within each class of processes, the principle of entropic similarity is invoked to construct families of dimensionless groups. These are compared to the well-known groups formed by dynamic similarity, providing some interesting contrasts between the two approaches and several important new insights.

\subsection{\label{sect:diffusion} Diffusion and chemical reaction processes}
\subsubsection{\label{sect:diffusion_phenom} Independent diffusion processes}
\newcounter{Lcount1}
\begin{list}{(\alph{Lcount1})}{\usecounter{Lcount1} \topsep 0pt \itemsep 0pt \parsep 3pt \leftmargin 0pt \rightmargin 0pt \listparindent 10pt \itemindent 20pt}
\item {\label{sect:diffusion_practical} \it \hspace{17pt} Practical diffusion relations}

We first consider diffusion processes (often termed {\it transport phenomena}) acting independently, in which a gradient in a physical field induces a flux of the corresponding (conjugate) physical quantity. Usually, the term {\it diffusion} is applied to processes acting at molecular scales, while {\it dispersion} is applied to processes acting at microscopic to macroscopic scales.
All diffusion and dispersion processes are irreversible, driven by or causing an increase in thermodynamic entropy due to mixing. 

The diffusion of heat, momentum, chemical species or charge due to the random motions of molecules, for binary species and in the absence of electromagnetic effects (thus for conservative fields), anisotropy or cross-phenomenological processes, are commonly analysed by the following practical or empirical relations \citep[e.g.,][]{Bosworth_1956, Guggenheim_1967, Levine_1978,  Newman_1991, Fogler_1992, Bejan_1993, Beek_etal_1999, Bird_etal_2006, White_2006, Kondepudi_P_2015}\footnote{Heat diffusion is commonly termed {\it conduction}. In electrical engineering, the diffusion of charge by Ohm's law is termed {\it drift}, to distinguish it from the diffusion of charge carriers under Fick's law \eqref{eq:Fick1}.}: 
\begin{align}
\j_Q  &= -k \vnabla T 		&\text{Fourier's law}  \YY
\label{eq:Fourier}
\\
\tens{\tau} &=   - \mu (\vnabla \u  + (\vnabla \u) ^\top) 	
=   - \rho \nu (\vnabla \u  + (\vnabla \u) ^\top)
&\text{Newton's law}  \YY
\label{eq:Newton}
\\
\j_c  &= -\Dc \vnabla C_c = -\Dc \vnabla (\rho m_c)	&\text{Fick's first law (binary)} \YY
\label{eq:Fick1}
\\
\i_k  & = - \kappa_k  \vnabla \Phi  =  \kappa_k \,  \vec{E}   &\text{Ohm's law} \YY
\label{eq:Ohm}
\end{align}
These are based on the following quantities\footnote{Many authors refer to fluxes as {\it flux densities}, and diffusion coefficients as {\it diffusivities}.}.
{\it Fluxes}:\ 
$\j_Q $ is the heat flux [J m$^{-2}$ s$^{-1}$], 
$\tens{\tau}$ is the momentum flux or viscous stress tensor [Pa] (defined positive in compression)\footnote{There are two sign conventions for the shear stress tensor $\tens{\tau}$. This study adopts the positive-in-compression form for consistency with other diffusion equations \citep[e.g.,][]{Bird_etal_2006}.},
$\j_c$ is the molar flux of the $c$th chemical species [(mol species) m$^{-2}$ s$^{-1}$] (relative to the molar average velocity 
$\u^* = \sum\nolimits_c C_c \u_c / \sum\nolimits_c C_c$) 
and
$\i_k$ is the charge flux or current density of the $k$th charged species (defined in the direction of positive ion flow) [C m$^{-2}$ s$^{-1}$ = A m$^{-2}$], from which $\i=\sum_k \i_k$ is the total charge flux [A m$^{-2}$]. 
{\it Fields}:\ 
$T$ is the absolute temperature [K], 
$\u$ is the (mass-average) fluid velocity [m s$^{-1}$],
$\u_c$ is the (mass-average) velocity of species $c$  [m s$^{-1}$],
$C_c$ is the molar concentration of chemical species $c$ per unit volume [(mol species) m$^{-3}$], 
$m_c= C_c/\rho$ is the molality (specific molar concentration) of chemical species $c$ [(mol species) kg$^{-1}$], 
$\rho$ is the fluid density [kg m$^{-3}$],
$\Phi$ is the electrical (electrostatic) potential [V = J C$^{-1}$] 
and
$\vec{E} = - \vnabla \Phi$ is the electric field vector [N C$^{-1}$ = V m$^{-1}$]. 
%
{\it Diffusion parameters}:\ 
$k$ is the thermal conductivity [J K$^{-1}$ m$^{-1}$ s$^{-1}$], 
$\mu$ is the dynamic viscosity [Pa s], 
$\nu$ is the kinematic viscosity or momentum diffusion coefficient [m$^{2}$ s$^{-1}$], 
$\Dc$ is the diffusion coefficient for the $c$th chemical species [m$^{2}$ s$^{-1}$] 
and
$\kappa_k$ is the electrical conductivity or specific conductance for the $k$th charged species [$\Omega^{-1}$ m$^{-1}$ = A V$^{-1}$ m$^{-1}$].
{\it Vector quantities}:\ 
$\x = [x,y,z]^\top$ is the Cartesian position vector [m], 
$\vnabla = [\partial/\partial x, \partial/\partial y, \partial/\partial z]^\top$ is the Cartesian gradient operator [m$^{-1}$] (using the $\partial (\downarrow)/\partial (\to)$ convention for vector gradients such as $\vnabla \u$),
and 
$^\top$ is the vector or matrix transpose. 

It must be emphasised that \eqref{eq:Fourier}-\eqref{eq:Ohm} are first-order correlations, and more complicated forms are required in many circumstances, e.g., for non-Newtonian fluids, inhomogeneous materials, or multicomponent or unsteady processes \citep[e.g.,][]{Hirschfelder_etal_1954, Bejan_1993}. 
In anisotropic media, it may also be necessary to redefine $k, \nu, \Dc$ or $\kappa_k$ as a second-order tensor \citep{deGroot_M_1962, Guggenheim_1967, Bird_etal_2006}. 

Fick's first law \eqref{eq:Fick1} is expressed in terms of the density of the $c$th species.
It is defined here using molar fluxes and concentrations, but can also be written in terms of mass quantities \citep{Bird_etal_2006}. 
Fourier's and Newton's laws can be rewritten respectively in terms of the energy or momentum density \citep{Beek_etal_1999, Bird_etal_2006}:
\begin{align}
\j_Q  &= -  \alpha \vnabla (\rho c_P T) \YY		&\text{Fourier's law (energy)} 
\label{eq:Fourier_v2}
\\
\tens{\tau} &=   - \nu (\vnabla  (\rho \u)  + [\vnabla  (\rho \u)] ^\top) \YY  &\text{Newton's law (momentum)} 
\label{eq:Newton_v2}
\end{align} 
where $\alpha$ is the thermal diffusion coefficient [m$^{2}$ s$^{-1}$] and
$c_P$ is the specific heat capacity of the fluid at constant pressure [J K$^{-1}$ kg$^{-1}$]. 
Ohm's law can also be written in terms of the charge density of the $k$th charged species \citep[c.f.,][]{Halliday_etal_2007, Miomandre_etal_2011}:
\begin{align}
\i_k  & 
= - \Dk  \vnabla \biggl ( \frac {z_k^2 F^2 C_k \Phi}{R T} \biggr)   \YY 
= - \Dk  \vnabla \biggl ( \frac {q_k^2 n_k \Phi}{k_B T} \biggr)  \YY
&\text{Ohm's law (charge)}
\label{eq:Ohm_v2}
\end{align}
where 
$F$ is the Faraday constant [C (mol charge)$^{-1}$], 
$R$ is the ideal gas constant [J K$^{-1}$ (mol species)$^{-1}$] 
and, for the $k$th charged species,
$C_k$ is the molar concentration [(mol species) m$^{-3}$],
$\Dk$ is the diffusion coefficient [m$^{2}$ s$^{-1}$],
$n_k$ is the number density [m$^{-3}$], 
$q_k$ is the charge [C], and
$z_k$ is the charge number (valency) [(mol charge) (mol species)$^{-1}$].
This formulation builds upon a number of electrochemical relations (see appendix \ref{sect:apx_elec}).
The first form of Ohm's law in \eqref{eq:Ohm_v2} is written on a molar basis, appropriate (with $C_k = \rho m_k$) for charged species in aqueous solution, while the second form accords with the individual carrier formulation used in electrical engineering, to analyse the drift of electrons or holes in a conductor.

As evident, the density formulations of the diffusion eqs.\ \eqref{eq:Fick1} and \eqref{eq:Fourier_v2}-\eqref{eq:Ohm_v2} contain diffusion coefficients ($\alpha, \nu, \D_c$ and $D_k$) with SI units of m$^2$ s$^{-1}$.
Comparing \eqref{eq:Fourier} to \eqref{eq:Fourier_v2} and \eqref{eq:Ohm} to \eqref{eq:Ohm_v2}, we obtain 
$\alpha \simeq {k}/{\rho c_P} $ $\YY$ 
and 
$\Dk \simeq  R T \kappa_k /z_k^2 F^2 C_k \YY = k_B T \kappa_k /q_k^2 n_k \YY$, with strict equality for spatially homogeneous properties. 
For typical natural waters $[\alpha, \nu,\Dc, \Dk] \sim [10^{-7}, 10^{-6}, 10^{-9}, 10^{-9}]$~m$^2$~s$^{-1}$.
Each diffusion coefficient can be further reduced to a function of molecular properties for different states of matter \citep{Hirschfelder_etal_1954, Hines_Maddox_1985, Bird_etal_2006}. 
Eqs.\ \eqref{eq:Fourier}-\eqref{eq:Ohm_v2} can be solved in isolation or incorporated into energy, momentum, species and charge balance calculations, e.g., the heat transfer, Navier-Stokes, advection-diffusion-reaction and drift-diffusion-convection equations.


\item {\label{sect:diffusion_thermo} \it \hspace{17pt} Thermodynamic diffusion relations}

Although of tremendous utility in practical applications, the above diffusion laws 
can be reformulated in terms of thermodynamic gradients. A thermodynamic framework has the advantage of providing a sound theoretical foundation, and can incorporate more phenomena (including cross-phenomonological effects), but it does come at the expense of  dimensionally more complicated parameters. As will be shown, it can also be connected directly to the entropy production.  The four laws become 
 \citep[e.g.,][]{Bosworth_1956, Frederick_Chang_1965, Guggenheim_1967, Bird_etal_2006, White_2006, Kondepudi_P_2015}: 
\begin{alignat}{10}
\j_Q  &=  \alpha' \vnabla \frac{1}{T} \YY		&\text{Fourier's law (thermodynamic)}
\label{eq:Fourier_th}
\\
\begin{split}
\tens{\tau} &=  - \mu (\vnabla \u  + (\vnabla \u) ^\top)
- \lambda  \, \vec{\delta} \,  \vnabla \cdot \u	 \YY
\\&={- 2 \mu \e -  \lambda  \, \vec{\delta} \, \tr({\e})} \YY
\end{split}
&\text{Newton's law (thermodynamic)}  
\label{eq:Newton_th}
\\
\j_c  &= -\D'_c \vnabla  \frac{\mu_c}{T} \Y	&\text{Fick's first law (binary, thermodynamic)} 
\label{eq:Fick1_th}
\\
\i_k  & = - \Dk'   \frac{ \vnabla \Phi}{T}  \YY &\text{Ohm's law (thermodynamic)} 
\label{eq:Ohm_th}
\end{alignat}
based on modified parameters $\alpha'$ [J K m$^{-1}$ s$^{-1}$] $\YY$, $\D'_c$ [(mol species)$^{2}$ K J$^{-1}$ m$^{-1}$ s$^{-1}$] $\YY$ and $\Dk'$ [C$^2$ K J$^{-1}$ m$^{-1}$ s$^{-1}$] $\YY$, in which $\lambda=\kappa -\frac{2}{3} \mu \YY$ 
is the second viscosity or first Lam\'e coefficient [Pa s], $\kappa$ is the dilatational or volume viscosity [Pa s], 
$\vec{\delta}$ is the Kronecker delta tensor, and $\mu_c$ is the chemical potential of species $c$ [J (mol species)$^{-1}$].  In the second form of Newton's law \eqref{eq:Newton_th}, $\nabla \u = \e + \w \YY$ is decomposed into symmetric and antisymmetric components (strain rate and spin tensors), respectively $\e=\frac{1}{2}( \nabla \u + (\nabla \u)^\top) \YY$ and $\w=\frac{1}{2}( \nabla \u  -(\nabla \u)^\top) \YY$ \citep{Spurk_1997, Pope_2000}. 
Eq.\ \eqref{eq:Newton_th} thus imposes symmetry of  $\tens{\tau}$, but not necessarily of $\vnabla \u$. The second viscosity is not needed for incompressible flows ($\vnabla \cdot \u =0 \YY$) or monatomic ideal gases ($\kappa=0 \YY$) \citep{Frederick_Chang_1965, Bird_etal_2006, White_2006}.  

We require simplified expressions for the above thermodynamic parameters. Firstly, comparison of the empirical and thermodynamic laws for the diffusion of heat \eqref{eq:Fourier}, \eqref{eq:Fourier_v2} and \eqref{eq:Fourier_th} gives 
$\alpha' = k T^2 \YY \simeq \alpha \rho c_p T^2 \YY$.   
For chemical diffusion, a thermodynamic analysis (see appendix \ref{sect:apx_thermod_diffusion}) yields $\D'_c \simeq \rho m_c \Dc /R \YY=  p_c \Dc /R^2 T \YY$ respectively for solutes or gaseous species, with strict equality for spatially-invariant activity or fugacity coefficients, fluid density and temperature. Finally, comparing \eqref{eq:Ohm}, \eqref{eq:Ohm_v2} and \eqref{eq:Ohm_th} for Ohm's law gives
$\Dk' \simeq  \kappa_k T \YY  
\simeq z_k^2 F^2 C_k  \Dk/R \YY
= q_k^2 n_k  \Dk/k_B \YY 
$, for vanishing concentration and inverse temperature gradients.

The action of Fick's and Ohm's laws on the $c$th charged species can also be written in terms of the electrochemical potential $\widetilde{\mu}_c=\mu_c + z_c F \Phi \YY$ [J (mol species)$^{-1}$] \citep{Bosworth_1956, deGroot_M_1962, Guggenheim_1967, Atkins_1982, Kondepudi_P_2015}:
\begin{align}
\j_c  &= -\D''_c \, \vnabla  \frac{\widetilde{\mu}_c}{T} \Y	&\text{Fick's + Ohm's laws (thermodynamic)} 
\label{eq:Fick_Ohm_th}
\end{align}
with the combined diffusion parameter $\D''_c$ [(mol species)$^{2}$ K J$^{-1}$ m$^{-1}$ s$^{-1}$] $\YY$.
Comparing \eqref{eq:Fick1_th}-\eqref{eq:Ohm_th} gives $\Dc''= \Dc' \simeq D_c'/z_c^2 F^2 \YY$.

\item {\label{sect:diffusion_EP} \it \hspace{17pt} Entropy production by diffusion}

For each of the above phenomena in isolation, the local unsteady entropy production \eqref{eq:EPdef_local} can be shown to reduce to \citep{Hirschfelder_etal_1954, deGroot_M_1962, Prigogine_1967, Kreuzer_1981, Callen_1985, Hines_Maddox_1985, Bird_etal_2006, Kondepudi_P_2015}:
\begin{align}
\hat{\dot{\sigma}}_{\alpha}  
&= \j_Q \cdot {\vnabla}  {\dfrac{1}{T}}  \YY
\label{eq:hatsigma_Q}
\\
\hat{\dot{\sigma}}_{\nu}  
&=  - \tens{\tau}: \dfrac{ \vnabla  \u}{T} \YY		
\label{eq:hatsigma_tau}
 \\
\hat{\dot{\sigma}}_{\Dc}  
&= - {\j_c} \cdot  {\vnabla}  \dfrac{\mu_c}{ T} \YY
\label{eq:hatsigma_c}
\\
\hat{\dot{\sigma}}_{\Dk}  
&= - \i_k  \cdot   \dfrac{ \vnabla \Phi}{T}   \YY
\label{eq:hatsigma_k}
\end{align}
where $\vec{a} \cdot \vec{b} = \vec{a}^\top \vec{b} = \sum\nolimits_{i} a_i b_i$ 
is the vector scalar product, $\tens{A} : \tens{B}= \tr (\tens{A}^\top \tens{B}) = \sum\nolimits_{i}  \sum\nolimits_{j} A_{ij} B_{ij}$ is the Frobenius form of the tensor scalar product \citep{Zwillinger_2003}\footnote{The Frobenius inner product, instead of the common convention $\sum\nolimits_{i}  \sum\nolimits_{j} A_{ji} B_{ij}$ \citep[][eq.\ A.3-14]{Bird_etal_2006}, is chosen for consistency with its norm. Note that \cite{Bird_etal_2006} also uses a transposed velocity gradient, hence \eqref{eq:hatsigma_tau} has the same form.}, and $\tr(\tens{A})=\sum\nolimits_{i}   A_{ii}$ is the trace. 
Substituting \eqref{eq:Fourier_th}-\eqref{eq:Ohm_th} and the above relations into \eqref{eq:hatsigma_Q}-\eqref{eq:hatsigma_k} gives:
\begin{align}
\hat{\dot{\sigma}}_{\alpha}  
&\simeq   {\alpha \rho c_p T^2} \, \biggl | \biggl | \vnabla \dfrac{1}{T} \biggr| \biggr|^2 \YY
=   \frac{1}{\alpha \rho c_p T^2} \, || \j_Q ||^2 \YY
\label{eq:hatsigma_Q2}
\\
\begin{split}
\hat{\dot{\sigma}}_{\nu}  
&=    \dfrac{\rho \nu}{T} \Bigl[ || \vnabla \u||^2 + \tr((\vnabla \u) ^2) \Bigr]
+ \dfrac{\lambda}{T}  \, (\vnabla \cdot \u)^2  \YY
{= \frac{1}{T}{\bigl[2 \mu \e+\lambda \,\vec{\delta}\, \tr(\e) \bigr] : \vnabla \u} }\YY
\\&
= \dfrac{1}{2 \rho \nu T}  \bigl [ || \tens{\tau}||^2 + \lambda \, \tr( \tens{\tau})  \, \tr (\e)  \bigr]
- \dfrac{ \tens{\tau} :  \w }{T}  \YY
%
\end{split}
\label{eq:hatsigma_tau2}
 \\
\begin{split}
\hat{\dot{\sigma}}_{\Dc}  
& \simeq \dfrac{ \rho m_c \Dc}{R} \, \biggl | \biggl| \vnabla  \dfrac{\mu_c}{T} \biggr| \biggr| ^2 \YY
= \dfrac{ p_c \Dc}{R^2 T} \, \biggl | \biggl| \vnabla  \dfrac{\mu_c}{T} \biggr| \biggr| ^2 \YY
=  \dfrac{R}{ \rho m_c \Dc} \,  || {\j_c} ||^2 \YY
=  \dfrac{R^2 T}{ p_c \Dc} \,  || {\j_c} ||^2 \YY
\end{split}
\label{eq:hatsigma_c2}
\\
\begin{split}
\hat{\dot{\sigma}}_{\Dk}  
&\simeq \frac{ z_k^2 F^2 C_k  \Dk}{RT^2}  \, ||   \vnabla \Phi || ^2  \YY
= \frac{ q_k^2 n_k  \Dk}{k_B T^2}  \, ||   \vnabla \Phi ||^2 \YY
= \frac{R}{ z_k^2 F^2 C_k  \Dk} \, || \i_k ||^2  \YY
= \frac{k_B}{ q_k^2 n_k  \Dk} \, || \i_k ||^2  \YY
\end{split}
\label{eq:hatsigma_k2}
\end{align} 
where $||\vec{a}||=\sqrt{\vec{a}^\top \vec{a}}$ is the Euclidean norm for vector $\vec{a}$, and 
$||\tens{A}||= \sqrt{\tr(\tens{A}^\top \tens{A})}$ is the Frobenius norm for tensor $\tens{A}$.
In \eqref{eq:hatsigma_tau2}, if $\vnabla \u$ is symmetric, $ \tr((\vnabla \u) ^2) =|| \vnabla \u||^2 $ and $\w=\vec{0}$. 
The first result or pair of results for each phenomenon in \eqref{eq:hatsigma_Q2}-\eqref{eq:hatsigma_k2} apply to a fixed local gradient, thus to a {\it gradient-controlled system}, while the second result(s) apply to a fixed local flux, thus to a {\it flux-controlled system} \citep{Niven_2009}. 

\item {\label{sect:diffusion_Pi_EP} \it \hspace{17pt} Entropic similarity in diffusion (based on entropy production terms)}

We can now apply the principle of entropic similarity to construct dimensionless groups between individual diffusion processes, to assess their relative importance. For this we consider the following ratios of local entropy production terms:
\begin{equation}
\begin{aligned}
\hatPi_{\nu/\alpha} &= \frac{\hat{\dot{\sigma}}_{\nu} }{{\hat{\dot{\sigma}}_{\alpha}}}, \hspace{10pt} &
\hatPi_{\nu/\Dc}&= \frac{\hat{\dot{\sigma}}_{\nu} }{{\hat{\dot{\sigma}}_{\Dc}}}, \hspace{10pt} &
\hatPi_{\nu/\Dk} &= \frac{\hat{\dot{\sigma}}_{\nu} }{{\hat{\dot{\sigma}}_{\Dk}}}, \hspace{10pt} &
\hatPi_{\alpha/\Dc} &= \frac{{\hat{\dot{\sigma}}_{\alpha}}}{{\hat{\dot{\sigma}}_{\Dc}}}, \hspace{10pt} &
\\
\hatPi_{\alpha/\Dk} &= \frac {{\hat{\dot{\sigma}}_{\alpha}}}{{\hat{\dot{\sigma}}_{\Dk}}}, \hspace{10pt} &
\hatPi_{\Dc/\Dk} &= \frac {{\hat{\dot{\sigma}}_{\Dc}}}{\hat{\dot{\sigma}}_{\Dk} }, \hspace{10pt} &
\hatPi_{\Dc/\Db} &= \frac{{\hat{\dot{\sigma}}_{\Dc}}}{{\hat{\dot{\sigma}}_{\Db}}}, \hspace{10pt} &
\hatPi_{\Dk/\Dl} &= \frac {{\hat{\dot{\sigma}}_{\Dk}}}{{\hat{\dot{\sigma}}_{\Dl}}} \hspace{10pt} &
\end{aligned}
\label{eq:Pi_diffusion}
\end{equation}
where the second-last and final groups represent the relative effects of diffusion of chemical species $c$ and $b$, and diffusion of charged species $k$ and $\ell$. 
Applying \eqref{eq:hatsigma_Q2}-\eqref{eq:hatsigma_k2} under the assumption of constant gradients and various other properties, \eqref{eq:Pi_diffusion} reduce to ratios of the corresponding diffusion coefficients:
\begin{equation}
\begin{aligned}
\hatPi_{\nu/\alpha} &\to Pr = \dfrac{\nu}{\alpha}, \YY & 
\hatPi_{\nu/\Dc} &\to Sc_c = \dfrac{\nu}{\Dc}, \YY & 
\hatPi_{\nu/\Dk} &\to Sc_k = \dfrac{\nu}{\Dk}, \YY&
\\
\hatPi_{\alpha/\Dc} &\to Le_c = \dfrac{\alpha}{\Dc}, \YY& 
\hatPi_{\alpha/\Dk} &\to Le_k = \dfrac{\alpha}{\Dk}, \YY&
\hatPi_{\Dc/\Dk} &\to \dfrac{\Dc}{\Dk} ,\YY &
\\
\hatPi_{\Dc/\Db} &\to  \dfrac{\Dc}{\Db}, \YY&
\hatPi_{\Dk/\Dl} 
&\simeq\dfrac{{ z_k^2  C_k  \Dk}   }{{ z_\ell^2 C_\ell  \Dl}  } \YY
\to \dfrac{\Dk}{\Dl} \YY
\end{aligned}
\label{eq:Pi_diffusion2}
\raisetag{30pt}
\end{equation}
These respectively give the Prandtl, Schmidt (species), Schmidt (charge), Lewis (species) and Lewis (charge) numbers\footnote{The Lewis number can also be defined as the reciprocal \citep[e.g.,][]{Eckert_1963, White_2006}.}, and ratios of diffusion coefficients of chemical and/or charged species
\citep[e.g.,][]{Bosworth_1956, Eckert_1963, Schlichting_1968, Newman_1991, Fogler_1992, Streeter_etal_1998, Incropera_DeWitt_1990, Incropera_DeWitt_2002, Bird_etal_2006, White_2006}: 
Alternatively, applying \eqref{eq:hatsigma_Q2}-\eqref{eq:hatsigma_k2} for fixed fluxes rather than gradients, with other constant properties, \eqref{eq:Pi_diffusion} reduce to reciprocals of the groups in \eqref{eq:Pi_diffusion2}, and so -- by convention -- can also be represented by these groups$\YY$. Additional groups can be defined for the second viscosity, e.g.,
from \eqref{eq:hatsigma_tau2} for a flow-controlled system:
\begin{equation}
\hatPi_{\lambda/\mu} 
= \frac{  \lambda  (\vnabla \cdot \u)^2}
 {\mu \bigl[ || \vnabla \u||^2 + \tr((\vnabla \u) ^2) \bigr] }
 \to  \dfrac{\lambda}{\mu} \YY
\label{eq:Pi_diffusion_2ndvisc2}
\end{equation}

It must be emphasised that the entropic groups based on \eqref{eq:hatsigma_Q2}-\eqref{eq:hatsigma_k2} contain additional functional dependencies, for example on $T$, $\rho$, $c_p$, $\vnabla \cdot \u$, $\tens{w}$, $\lambda$, $m_c$, $z_k$ and $C_k$, as well as the gradients and/or fluxes. Depending on the system, it may be necessary to preserve these parameters, using the primary definitions of the entropic dimensionless groups given in \eqref{eq:Pi_diffusion}.  Furthermore, if one phenomenon is controlled by a flux and the other by a gradient, we obtain a hybrid dimensionless group, for example:
\begin{align}
\hatPi_{\alpha/\Dc} 
= \dfrac{ \alpha T^2 \rho^2 c_P m_c D_c}{R} 
\frac{ \, \bigl | \bigl | \vnabla {T}^{-1} \bigr| \bigr|^2}{   || {\j_c} ||^2 } \YY
\label{eq:Pi_diffusion_mixed}
\end{align}
Such groups are not readily reducible to the ratios of diffusion coefficients \eqref{eq:Pi_diffusion2} obtained by dynamic similarity.


\item {\label{sect:diffusion_Pi_flux} \it \hspace{17pt} Entropic similarity in diffusion (based on entropy fluxes)}

Also of interest for diffusion are the local non-fluid entropy fluxes, given respectively for independent flows of heat, chemical species and charged particles by the product of the flux and its corresponding intensive variable or field
 \citep[e.g.,][]{deGroot_M_1962, Prigogine_1967, Kreuzer_1981, Bird_etal_2006, Kondepudi_P_2015}:
\begin{align}
\j_{S,\alpha}  
= \j_Q  {\dfrac{1}{T}}, \YY
\hspace{10pt} 
\j_{S,\Dc}  
= - {\j_c} \dfrac{\mu_c}{ T}, \YY
\hspace{10pt} 
\j_{S,\Dk}  
= - {\i_k} \dfrac{\Phi}{ T} \YY
\label{eq:fluxS}
\end{align}
From the fluxes in \eqref{eq:Fourier_th}, \eqref{eq:Fick1_th} and \eqref{eq:Ohm_th} these give, as functions of the gradients:
\begin{align}
\j_{S,\alpha}  
\simeq \alpha \rho c_p T \vnabla \frac{1}{T}, \YY
\hspace{10pt}
\j_{S,\Dc}  
\simeq \frac{ \rho m_c \Dc}{R}  \dfrac{\mu_c}{ T} \vnabla  \frac{\mu_c}{T}, \YY
\hspace{10pt} 
\j_{S,\Dk}  
\simeq \frac{ z_k^2 F^2 C_k  \Dk}{R}   \dfrac{\Phi}{ T} \frac{\vnabla \Phi}{T}  \YY
\label{eq:fluxS2}
\end{align}
While not dissipative, the fluid-borne entropy flux $\j_{S,f} = \rho s \u \YY$ in \eqref{eq:Sbalance_int} is also important for entropy transport. 
The principle of entropic similarity can now be applied to assess the interplay between these entropy fluxes. 
Firstly considering the flux magnitudes, for fixed fluxes, gradients and other parameters these give the following dimensionless groups: 
\begin{equation*}
\begin{aligned}
\hatPi_{\j_{S,\alpha}/\j_{S,\Dc}} 
&= \dfrac{ ||\j_{S,\alpha} || }{|| \j_{S,\Dc}|| } 
\to Le_c = \dfrac{\alpha}{\Dc},  \YY \hspace{5pt} &
\hatPi_{\j_{S,\alpha}/\j_{S,\Dk}} 
&= \dfrac{ ||\j_{S,\alpha} || }{|| \j_{S,\Dk}|| } 
\to Le_k = \dfrac{\alpha}{\Dk}, \YY \hspace{5pt} &
\\
\hatPi_{\j_{S,\Dc}/\j_{S,\Dk}} 
&= \dfrac{ ||\j_{S,\Dc}  || }{ || \j_{S,\Dk} ||}
\to \dfrac{\Dc}{\Dk}, \YY \hspace{5pt} &
\hatPi_{\j_{S,\Dc}/\j_{S,\Db}} 
&= \dfrac{ ||\j_{S,\Dc}  || }{ || \j_{S,\Db} ||}
\to \dfrac{\Dc}{\Db}, \YY  \hspace{5pt} &
\\
\hatPi_{\j_{S,\Dk}/\j_{S,\Dl}} 
&= \dfrac{ ||\j_{S,\Dk}  || }{ || \j_{S,\Dl} ||}
\simeq\dfrac{{ z_k^2  C_k  \Dk}   }{{ z_\ell^2 C_\ell  \Dl}  }
\to \dfrac{\Dk}{\Dl} \YY \hspace{5pt} &
\end{aligned} 
\end{equation*}
\begin{equation}
\begin{aligned}
\hatPi_{\j_{S,\alpha}/\j_{S,f}} 
&= \dfrac{ ||\j_{S,\alpha}  ||}{|| \j_{S,f} ||} 
\simeq \dfrac{|| \j_Q ||}{\rho s T \, ||\u||} \Y
\simeq \dfrac{\alpha c_p T }{s} \dfrac{|| \vnabla {T^{-1}}||}{||\u||}, \Y
\\
\hatPi_{\j_{S,\Dc}/\j_{S,f}} 
&= \dfrac{|| \j_{S,\Dc}  || }{|| \j_{S,f} ||}
\simeq   \dfrac{\mu_c \, || \j_c || }{\rho s T \, ||\u||} \Y
\simeq \dfrac{{ \Dc m_c  \mu_c} }{ s R T }   \dfrac{\Bigl | \Bigl |\vnabla  \dfrac{\mu_c}{T} \Bigr | \Bigr |}{||\u||}, \Y 
\\
\hatPi_{\j_{S,\Dk}/\j_{S,f}} 
&= \dfrac{|| \j_{S,\Dk}  || }{|| \j_{S,f} ||}
\simeq \dfrac{\Phi \, || \i_k || }{\rho s T \, ||\u||} \Y
\simeq \dfrac{ \Dk z_k^2 F^2 C_k  \Phi}{\rho s R T^2} \dfrac{ ||\vnabla   \Phi|| }{ ||\u||} \Y
\end{aligned} 
\label{eq:Pi_diffusion_fluxes}
\raisetag{40pt}
\end{equation}
The first five groups in \eqref{eq:Pi_diffusion_fluxes} reduce to the Lewis numbers and ratios of diffusion coefficients listed in \eqref{eq:Pi_diffusion2}. In contrast, the last three groups in \eqref{eq:Pi_diffusion_fluxes} contain fluxes or gradients, the fluid velocity and other parameters, and are less easily interpreted by kinematic or dynamic similarity. The relative importance of the fluid and non-fluid entropy fluxes can also be expressed by the composite group:
\begin{align}
\hatPi_{\j_{S,f}/\j_{S}} 
= \dfrac{|| \j_{S,f}  || }{|| \j_{S} ||}
= \dfrac{|| \rho s \u || }{|| \j_{S,\alpha}+ \sum\nolimits_c \j_{S,\Dc}   + \sum\nolimits_k \j_{S,\Dk}  ||   } \YY
\label{eq:Pi_diffusion_fluxes3}
\end{align}
As previously noted, if the fluxes, gradients, fluid velocity or other parameters are not constant, it may be necessary to retain the unsimplified groups defined in \eqref{eq:Pi_diffusion_fluxes}-\eqref{eq:Pi_diffusion_fluxes3}.

The foregoing definitions in \eqref{eq:Pi_diffusion_fluxes}-\eqref{eq:Pi_diffusion_fluxes3} discard the flux directions. A broader set of directional dimensionless groups can be constructed by reference to a unit normal $\n$, e.g.:
\begin{align}
\hatPi_{\j_{S,\alpha}/\j_{S,\Dc}}(\n)
&= \dfrac{ \j_{S,\alpha} \cdot \n} {\j_{S,\Dc} \cdot \n},
\label{eq:Pi_diffusion_fluxes_dotprod}
\end{align}
or alternatively by the use of normed dot products, e.g.:
\begin{align}
\widehat{\Pi}_{\j_{S,\alpha}/\j_{S,\Dc}} 
&= \dfrac{ \j_{S,\alpha} \cdot \j_{S,\Dc}} {||\j_{S,\Dc}||^2},
\label{eq:Pi_diffusion_fluxes_dotprod}
\end{align}
The former captures the directional dependence, but will exhibit singularities associated with the direction (relative to $\n$) of the denominator flux. The latter definition provides a more robust dimensionless group, attaining a maximum when the component fluxes are equidirectional, decreasing to zero as the fluxes become orthogonal, and decreasing further to a minimum (negative) value for antiparallel fluxes. 
\end{list}

\subsubsection{\label{sect:chem_rn} Chemical reactions}
\newcounter{Lcount2}
\begin{list}{(\alph{Lcount2})}{\usecounter{Lcount2} \topsep 0pt \itemsep 0pt \parsep 3pt \leftmargin 0pt \rightmargin 0pt \listparindent 10pt \itemindent 20pt}
\item {\label{sect:reaction_fund} \it \hspace{17pt} Thermodynamic fundamentals}

Although of different character, chemical reactions commonly occur in conjunction with one or more diffusion processes examined in \S\ref{sect:diffusion_phenom}. Continuous spontaneous chemical reactions also have a non-zero entropy production, which can be expressed as the product of conjugate variables in an manner analogous to diffusion processes. For these reasons, we here examine the interplay between chemical reaction and diffusion processes from the perspective of entropic similarity. 

The driving force for a chemical reaction is commonly represented by a {\it free energy}. A generalised dimensionless free energy concept, known as the generalised work or negative Massieu function $\FF$, is obtained for a given set of constraints (in thermodynamics, a specific ``ensemble'') by rearrangement of the maximum entropy relation \eqref{eq:MaxEnt} \citep[e.g.,][]{Jaynes_1957, Jaynes_1963, Tribus_1961a, Tribus_1961b}:
\begin{equation}
\FF = - \ln Z = -\lambda_0 = - \H^* + \sum_{r=1}^R \lambda_r \langle f_r \rangle
\label{eq:Massieu}
\end{equation}
Two philosophical interpretations of \eqref{eq:Massieu}, based the concept of generalised work or the change in entropy of the universe, are examined in detail elsewhere \citep[e.g.,][]{Jaynes_1957, Tribus_1961a, Tribus_1961b, Jaynes_2003, Niven_2009}; in each case, these show that spontaneous change occurs in the direction $d\FF<0$. For an ensemble defined by a constant temperature and volume, $k_B \FF /T$ reduces to the Helmholtz free energy $F$, while for constant temperature and pressure, it gives the Gibbs free energy $G$. The latter (or the Planck potential $G/T$ or $-G/T$, or a positive or negative affinity $A_d$) is most commonly used to analyse chemical reactions. Other free energy functions can be derived for different sets of constraints \citep[e.g.,][]{Gibbs_1875, Hill_1956, Davidson_1962, Callen_1985, Niven_2009}. 

\item {\label{sect:reaction_EP} \it \hspace{17pt} Entropy production by chemical reactions}

For a continuous process, the local entropy production of the $d$th chemical reaction is given by \citep{deGroot_M_1962, Prigogine_1967, Kreuzer_1981, Bird_etal_2006, Lebon_etal_2010, Kondepudi_P_2015}\footnote{Strictly, from the definition \eqref{eq:Massieu} of the Massieu function, \eqref{eq:hatsigma_chem} should be written as $\Delta (\G_d/T)$ rather than the common notation $\Delta \G_d/T$; for constant temperature these become equal.}:
\begin{align}
\hat{\dot{\sigma}}_{d}  
&= - {\hat{\dot{\xi}}_d} \frac{\Delta \G_d}{T} \Y
\label{eq:hatsigma_chem}
\end{align}
where $\Delta \G_d$ is the change in molar Gibbs free energy for the $d$th reaction [J (mol reaction)$^{-1}$], 
and $\hat{\dot{\xi}}_d$ is its rate of reaction per unit volume [(mol reaction) m$^{-3}$ s$^{-1}$]. 
To reduce \eqref{eq:hatsigma_chem}, we express the change in Gibbs free energy as a function of chemical potentials in the reaction, then from \eqref{eq:activity} in terms of chemical activities:
\begin{align}
\Delta \G_d = \sum\limits_c \nu_{cd} \mu_c
 = \Delta \G_d^\minuso + RT \ln \prod\limits_c  \alpha_c^{\nu_{cd}} \YY
\label{eq:DeltaG}
\end{align}
where $\nu_{cd}$ is the stoichiometric coefficient of species $c$ in the $d$th reaction [(mol species) (mol reaction)$^{-1}$], with $\nu_{cd}>0$ for a product and $\nu_{cd}<0$ for a reactant, $\Delta \G_d^\minuso = \sum\nolimits_c \nu_{cd} \mu_c^\minuso$ is the change in molar Gibbs free energy for the reaction under standard conditions, and $\prod\nolimits_c  \alpha_c^{\nu_{cd}}$ is termed the reaction quotient.  
The sum or product in \eqref{eq:DeltaG} is taken over all species $c$ participating in the reaction. 

Some authors have postulated a linear relation between the chemical reaction rate and driving force, analogous to Fick's first law \citep[e.g.,][]{Lebon_etal_2010}:
\begin{align}
{\hat{\dot{\xi}}}_d = - \ell_d \frac{\Delta \G_d}{T} 
\label{eq:rate_chem_linear}	 
\end{align}
where $\ell_d$ is a linear rate coefficient [(mol reaction)$^2$ K J$^{-1}$ m$^{-3}$ s$^{-1}$].  
However, in contrast to diffusion processes, in general this assumption is not reasonable. Instead, the rate is generally expressed by a kinetic equation of the form \citep[e.g.,][]{Levine_1978, Atkins_1982, Fogler_1992}:
\begin{align}
{\hat{\dot{\xi}}}_d = k_d \prod_c {C_c}^{\beta_{cd}} \YY
\label{eq:rate_chem}
\end{align}
where $k_d$ is the rate constant, $C_c$ is the molar concentration of chemical species $c$ [(mol species) m$^{-3}$] and $\beta_{cd} \in \R$ is a power exponent [--] which must be found by experiment. 
A variety of concentration variables have been used in \eqref{eq:rate_chem}, including molar or mass concentrations, mole fractions, partial pressures or (rarely) activities or fugacities \citep[e.g.,][]{Atkins_1982, Fogler_1992}. 
The units of $k_d$ depend on the concentration units and power exponents in \eqref{eq:rate_chem}, also with an implicit correction between (mol reaction) and (mol species) units. 
In general, the exponents $\beta_{cd}$ are unequal to the stoichiometric coefficients $\nu_{cd}$, but become equal for an elementary chemical reaction.
For complicated reactions, the total rate is the sum of rates for all individual steps or mechanisms with index $n$, such as forward and backward processes:
\begin{align}
{\hat{\dot{\xi}}}_d = \sum\limits_n {\hat{\dot{\xi}}}_{nd} = \sum\limits_n k_{nd} \prod_c {C_c}^{\beta_{ncd}}
\label{eq:rate_chem2}
\end{align}
Inserting \eqref{eq:DeltaG} and \eqref{eq:rate_chem2} into \eqref{eq:hatsigma_chem} gives:
\begin{align}
\hat{\dot{\sigma}}_{d}  
&= - \biggl( \sum\limits_n k_{nd} \prod_c {C_c}^{\beta_{ncd}} \biggr)  \frac{\Delta \G_d}{T}
= - \biggl( \sum\limits_n k_{nd} \prod_c {C_c}^{\beta_{ncd}} \biggr) \biggl( \frac{\Delta \G_d^\minuso}{T} + R \ln \prod\limits_c  \alpha_c^{\nu_{cd}} \biggr) \YY
\label{eq:hatsigma_chem2}
\end{align}
The activities can be converted to molalities or partial pressures using \eqref{eq:activity}-\eqref{eq:fugacity}, while the rate constants are often expressed in terms of activation energies by the Arrhenius equation \citep[e.g.,][]{Levine_1978, Atkins_1982, Fogler_1992}.

\item {\label{sect:reaction_Pi_EP} \it \hspace{17pt} Entropic similarity in chemical reactions and diffusion (based on entropy production terms)}

We now apply the principle of entropic similarity to construct entropic dimensionless groups between chemical reactions and transport phenomena:
\begin{equation}
\begin{aligned}
\hatPi_{d/e} &= \frac{\hat{\dot{\sigma}}_{d} }{{\hat{\dot{\sigma}}_{e}}}, \hspace{10pt} &
\hatPi_{nd/md} &= \frac{\hat{\dot{\sigma}}_{nd} }{{\hat{\dot{\sigma}}_{md}}}, \hspace{10pt} &
\hatPi_{d/\alpha}&= \frac{\hat{\dot{\sigma}}_{d} }{{\hat{\dot{\sigma}}_{\alpha}}}, \hspace{10pt} 
\\
\hatPi_{d/\nu} &= \frac{\hat{\dot{\sigma}}_{d} }{{\hat{\dot{\sigma}}_{\nu}}}, \hspace{10pt} &
\hatPi_{d/\Dc} &= \frac{{\hat{\dot{\sigma}}_{d}}}{{\hat{\dot{\sigma}}_{\Dc}}}, \hspace{10pt} &
\hatPi_{d/\Dk} &= \frac {{\hat{\dot{\sigma}}_{d}}}{\hat{\dot{\sigma}}_{\Dk} } \hspace{10pt} 
\end{aligned}
\label{eq:Pi_reaction}
\end{equation}
The first two groups represent the competition between two single-mechanism chemical reactions $d$ and $e$, or two mechanisms $n$ and $m$ for the same chemical reaction $d$.  The remaining groups represent the entropic competition between the $d$th reaction and a diffusion process.  
Inserting \eqref{eq:hatsigma_chem2} and the gradient forms of \eqref{eq:hatsigma_Q2}-\eqref{eq:hatsigma_k2} into \eqref{eq:Pi_reaction} gives:
\begin{align}
\begin{split}
%
\hatPi_{d/e} 
&= 
\dfrac{{\hat{\dot{\xi}}_d} {\Delta \G_d}}{{\hat{\dot{\xi}}_e} {\Delta \G_e}} 
=
\dfrac
{\bigl( k_{d} \prod_c {C_c}^{\beta_{cd}} \bigr) \Delta \G_d }
{\bigl( k_{e} \prod_c {C_c}^{\beta_{ce}} \bigr) \Delta \G_e}, \YY \hspace{10pt}
\hatPi_{nd/md} 
= \dfrac{{\hat{\dot{\xi}}_{nd}} }{{\hat{\dot{\xi}}_{md}} } 
=
\dfrac
{ k_{nd} \prod_c {C_c}^{\beta_{ncd}}}
{ k_{md} \prod_c {C_c}^{\beta_{mcd}}} \YY
\end{split}
\label{eq:Pi_reaction_EP}
\end{align}
\begin{align*}
\hatPi_{d/\alpha} 
&=
\dfrac{{\hat{\dot{\xi}}_d} {\Delta \G_d}}{{\hat{\dot{\sigma}}_\alpha} } 
=\dfrac
{- \bigl( \sum\limits_n k_{nd} \prod_c {C_c}^{\beta_{ncd}} \bigr)\Delta \G_d }
{\alpha \rho c_p T^3 \, || \vnabla {T^{-1}} ||^2}
&\tto
\dfrac{ k_{d} \,C_c^{\beta_{cd}} \,\Gamma_{\alpha d}}{ \alpha},  \YY 
\\
\hatPi_{d/\nu} 
&=
\dfrac{{\hat{\dot{\xi}}_d} {\Delta \G_d}}{{\hat{\dot{\sigma}}_\nu} } 
=\dfrac
{- \bigl( \sum\limits_n k_{nd} \prod_c {C_c}^{\beta_{ncd}} \bigr)\Delta \G_d }
{\rho \nu [ || \vnabla \u||^2 + \tr((\vnabla \u) ^2) ] +\lambda \, (\vnabla \cdot \u)^2 }
&\tto
\dfrac{ k_{d} \,C_c^{\beta_{cd}} \,\Gamma_{\nu d}}{ \nu}, \YY 
\\
\hatPi_{d/\Dc} 
&=\dfrac{{\hat{\dot{\xi}}_d} {\Delta \G_d}}{{\hat{\dot{\sigma}}_{\Dc}} } 
=\dfrac
{- R \bigl( \sum\limits_n k_{nd} \prod_c {C_c}^{\beta_{ncd}} \bigr)\Delta \G_d }
{C_c \Dc T \Bigl | \Bigl | \vnabla \dfrac{\mu_c}{T} \Bigr| \Bigr|^2}
&\tto
\dfrac{ k_{d} \,C_c^{\beta_{cd}-1} \,\Gamma_{cd}}{ \Dc}, \YY 
\\
\hatPi_{d/\Dk} 
&=\dfrac{{\hat{\dot{\xi}}_d} {\Delta \G_d}}{{\hat{\dot{\sigma}}_{\Dk}} } 
=\dfrac
{- R T \bigl( \sum\limits_n k_{nd} \prod_c {C_c}^{\beta_{ncd}} \bigr)\Delta \G_d }
{z_k^2 F^2 C_k \Dk || \vnabla \Phi ||^2}
&\tto
\dfrac{ k_{d} \,C_c^{\beta_{cd}} \,\Gamma_{kd}}{ C_k \Dk}  \YY 
\end{align*}
where $\tto$ indicates evaluation for a single mechanism with single-species kinetics, 
using the scaled free energy terms 
$\Gamma_{\alpha d} = - \Delta \G_d / (\rho c_p T^{3}  || \vnabla T^{-1} ||^{2} )\YY$ [m$^{5}$ mol$^{-1}$],
$\Gamma_{\nu d} \simeq - \Delta \G_d /(\rho [ || \vnabla \u||^2 + \tr((\vnabla \u) ^2) ]) \YY$ [m$^{5}$ mol$^{-1}$],   
$\Gamma_{cd}=-{R \Delta \G_d} / ( T || \vnabla ({\mu_c}/{T}) || ^{2}) \YY$ [m$^2$]
and
$\Gamma_{kd}=-{R T \Delta \G_d} /( z_k^{2} F^{2}$ $|| \vnabla \Phi || ^{2}) \YY$ [m$^2$].
The last four groups 
can be interpreted as local modified Damk\"ohler numbers, which compare the rate of entropy production by chemical reaction to that respectively from the diffusion of heat, momentum, chemical species or charge \citep[c.f.,][]{Fogler_1992, Bird_etal_2006}. 


For fixed fluxes rather than gradients, the flux terms in  \eqref{eq:hatsigma_Q2}-\eqref{eq:hatsigma_k2} yield a different set of groups, e.g.:
\begin{equation}
\hatPi_{d/\Dc} 
=\dfrac
{- \bigl( \sum\limits_n k_{nd} \prod_c {C_c}^{\beta_{ncd}} \bigr)\Delta \G_d C_c \Dc}
{ RT || \j_c ||^2}
\tto
\dfrac
{- \bigl(  k_{d} {C_c}^{\beta_{cd}+1} \bigr)\Delta \G_d \Dc}
{ RT || \j_c ||^2} \YY
\label{eq:Pi_reaction_EP_fixedflux}
\end{equation}
Hybrid groups based on other assumptions \citep[e.g.,][]{Fogler_1992} are also possible. Similarly to \eqref{eq:Pi_diffusion2}, for multispecies kinetics, variable composition or in other situations, it may be necessary to adopt the unreduced forms of the above groups.

\item {\label{sect:reaction_Pi_flux} \it \hspace{17pt} Entropic similarity in chemical reactions and diffusion (based on entropy production and flux terms)}

Entropic dimensionless groups can also be constructed by comparing the entropy production by chemical reaction \eqref{eq:hatsigma_chem2} to the entropy flux of a diffusion or fluid transport process \eqref{eq:fluxS2}, by inclusion of a length scale $\ell$ [m]. This gives the following groups:
\begin{align}
\begin{split}
\hatPi_{\hat{\dot{\sigma}}_{d}/\j_{S,\alpha}} 
&= \dfrac{ \hat{\dot{\sigma}}_{d} \ell}{|| \j_{S,\alpha}|| } 
=\dfrac
{- \bigl( \sum\limits_n k_{nd} \prod_c {C_c}^{\beta_{ncd}} \bigr)\Delta \G_d \ell}
{\alpha \rho c_p T^2 \, || \vnabla {T^{-1}} ||} \YY
\\
\hatPi_{\hat{\dot{\sigma}}_{d}/\j_{S,\Dc}} 
&= \dfrac{ \hat{\dot{\sigma}}_{d} \ell}{|| \j_{S,\Dc}|| } 
=\dfrac
{- R \bigl( \sum\limits_n k_{nd} \prod_c {C_c}^{\beta_{ncd}} \bigr)\Delta \G_d \ell }
{C_c \Dc \mu_c \Bigl | \Bigl | \vnabla \dfrac{\mu_c}{T} \Bigr| \Bigr|} \YY
\\
\hatPi_{\hat{\dot{\sigma}}_{d}/\j_{S,\Dk}} 
&= \dfrac{ \hat{\dot{\sigma}}_{d} \ell}{|| \j_{S,\Dk}|| } 
=\dfrac
{- R T \bigl( \sum\limits_n k_{nd} \prod_c {C_c}^{\beta_{ncd}} \bigr)\Delta \G_d \ell }
{z_k^2 F^2 C_k \Dk \Phi || \vnabla \Phi ||} \YY
\\
\hatPi_{\hat{\dot{\sigma}}_{d}/\j_{S,f}} 
&= \dfrac{ \hat{\dot{\sigma}}_{d} \ell}{|| \j_{S,f}|| } 
=\dfrac
{- \bigl( \sum\limits_n k_{nd} \prod_c {C_c}^{\beta_{ncd}} \bigr)\Delta \G_d \ell}
{\rho s T || \u ||} \YY
\end{split}
\label{eq:Pi_reaction_fluxes}
\end{align}
Comparing the first three groups to their analogues in \eqref{eq:Pi_reaction_EP}, we see that the length scales in \eqref{eq:Pi_reaction_fluxes} are provided by the normalised gradient $\ell^{-1} \sim ||\nabla X||/X$, where $X$ is the intensive variable or field. The last group in \eqref{eq:Pi_reaction_fluxes} has no equivalent in \eqref{eq:Pi_reaction_EP}.

\item {\label{sect:reaction_transp} \it \hspace{17pt} Entropic similarity in chemical reactions and flow processes (based on reaction rates)}

For chemical reactions in which the free energy $\Delta \G_d$ can be considered constant -- such as for steady-state conditions -- the free energy component of the entropy production \eqref{eq:hatsigma_chem} can be disregarded, enabling the formation of entropic groups based solely on the reaction rate ${\hat{\dot{\xi}}_d}$. The comparison of chemical reactions $d$ and $e$, or reaction mechanisms $m$ and $n$, then leads by entropic similarity to the following groups (compare \eqref{eq:Pi_reaction_EP}):
\begin{align}
\begin{split}
\hatPi_{d/e} 
&= 
\dfrac{{\hat{\dot{\xi}}_d} }{{\hat{\dot{\xi}}_e} } 
=
\dfrac
{k_{d} \prod_c {C_c}^{\beta_{cd}} }
{k_{e} \prod_c {C_c}^{\beta_{ce}}}, \YY \hspace{10pt}
\hatPi_{nd/md} 
= \dfrac{{\hat{\dot{\xi}}_{nd}} }{{\hat{\dot{\xi}}_{md}} } 
=
\dfrac
{ k_{nd} \prod_c {C_c}^{\beta_{ncd}}}
{ k_{md} \prod_c {C_c}^{\beta_{mcd}}} \YY
\end{split}
\label{eq:Pi_reactions_ratio2}
\end{align}
For process engineering applications, the reaction rate can also be compared to fluid transport processes by an extended Damk\"ohler number \citep[c.f.,][]{Fogler_1992, Bird_etal_2006}:
\begin{align}
\hatPi_{d/\theta} 
= \dfrac{{\hat{\dot{\xi}}_d} \theta}{\prod_c {C_c}} 
= \Bigl(\sum\limits_n k_{nd} \prod_c {C_c}^{\beta_{ncd}-1} \Bigr) \theta
&\tto
Da_d =
 k_{d} \,C_c^{\beta_{cd}-1} \theta  \Y
\end{align}
where $\theta$ is a transport or residence time scale [s].  For a single mechanism and single-species kinetics this reduces to the standard Damk\"ohler number. Examples of the residence time include $\theta=V/Q$ for a continuously mixed flow reactor, where $V$ is its volume [m$^3$] and $Q$ is the volumetric flow rate [m$^3$ s$^{-1}$], or $\theta=L/U$ for a plug-flow reactor, where $L$ is its length [m] and $U$ is the superficial velocity [m s$^{-1}$] \citep{Fogler_1992, Bird_etal_2006}.  

\end{list}

\subsubsection{\label{sect:cross-diffusion} Diffusion and chemical reaction cross-phenomena}


\newcounter{Lcount3}
\begin{list}{(\alph{Lcount3})}{\usecounter{Lcount3} \topsep 0pt \itemsep 0pt \parsep 3pt \leftmargin 0pt \rightmargin 0pt \listparindent 10pt \itemindent 20pt}
\item {\label{sect:Onsager_rels} \it \hspace{17pt} Description and Onsager relations}

In thermodynamic systems, it is necessary to consider the occurrence of diffusion and chemical reaction cross-phenomena, in which a thermodynamic force conjugate to one quantity induces the flow of a different quantity, or vice versa. A number of pairwise interactions -- based on the phenomena examined previously and also pressure gradients and magnetic fields -- are listed in table \ref{tab:cross-phenom}. Some ternary interactions 
(e.g., thermomagnetic convection; Nernst effect; Ettinghausen effect) 
are also known \citep{Bosworth_1956}.


\begin{turnpage}
\renewcommand{\arraystretch}{0.75}
\begin{table}
  \begin{tabular*}{610pt}{ L{2.5cm} | C{2.5cm} C{2.5cm} C{2.5cm} C{2.5cm} C{2.5cm} C{2.5cm} C{2.5cm} }  
       {\bf Phenomenon}  & Fluid flux & Heat  flux & Momentum flux (viscous stress tensor) & Chemical flux & Charge flux & Magnetic flux & Chemical reaction rate \\ 
       \hline
       Pressure gradient 
       & Fluid flow (Poiseuille) $\Y$ & Fluid flow (dissipative); Forced convection & Turbulent flow $\Y$ 
       & Reverse osmosis & Streaming current $\Y$ &Magnetic pressure &Chemical-mechanical coupling \\  
       Temperature gradient				& Thermal transpiration; Thermal osmosis; {Free convection} $\Y$ & Thermal conduction $\Y$ && Thermo-diffusion /  Soret effect $\Y$ & Thermo-electric (Seebeck) effect $\Y$ & Thermo-magnetism & Reaction-induced temperature gradient $\Y$ \\ 
       Velocity gradient 				& Fluid flow (Couette) $\Y$ && Momentum diffusion $\Y$ & Tollert effect $\Y$ \\ 
       Chemical potential gradient 		&Osmosis $\Y$ & Dufour effect; $\Y$ Latent heat transfer $\Y$ 	&& Chemical diffusion; Co-diffusion $\Y$ & Electro-chemical transference; Donnan effect $\Y$ &&Reaction-induced chemical gradient $\Y$\\ 
       Electric field  					&Electro-osmosis $\Y$ &Thermo-electric (Peltier) effect $\Y$	& Electro-viscous coupling $\Y$ &Hittorf  transference $\Y$ & Charge diffusion (drift) $\Y$ & Electro-magnetism & Electrolytic cell \\ 
       Magnetic field					&Magneto-osmosis; Magneto-hydrodynamic effect &Thermo-magnetism & Magneto-viscous coupling $\Y$ &&Electro-magnetic induction $\Y$ & Magnetism $\Y$	\\ 
       Free energy difference			&Reaction osmosis & Reaction-induced heat flux $\Y$ && Reaction-induced chemical flux $\Y$ & Electro-chemical (galvanic) cell && Chemical reaction; $\Y$ Coupled reactions $\Y$ \\ 
	\hline
  \end{tabular*}
  \caption{Reported examples of thermodynamic cross-phenomena, showing pairwise couplings between thermodynamic forces and fluxes \citep[e.g.,][]{Onsager_1931a, Onsager_1931b, Bosworth_1956, Miller_1959, Guggenheim_1967, Pommaret_2001, Lebon_etal_2010, Kondepudi_P_2015}.}
  \label{tab:cross-phenom}
\end{table}
\end{turnpage}

For small gradients (or small ``distance from equilibrium''), the pairwise interactions are commonly represented by the \cite{Onsager_1931a, Onsager_1931b} relations, derived from the assumption of microscopic reversibility 
\citep{Bosworth_1956, Miller_1959, deGroot_M_1962, Demirel_2002, Lebon_etal_2010, Bikken_Lyapilin_2014, Kondepudi_P_2015}:
\begin{align}
j_{r{\imath \jmath}} = \sum\limits_{m{\kappa \ell}} L_{r{\imath \jmath}, m{\kappa \ell}} \, f_{m{\kappa \ell}} \YY
\label{eq:Onsager}
\end{align}
where 
$j_{r{\imath \jmath}}$ $\in \{ \j_Q, \tens{\tau}, \{ \j_c \}, \{ \i_k\}, \{ \hat{\dot{\xi}}_d \} \}$
is the $\imath \jmath$th component of the $r$th flux or reaction rate, 
$f_{m{\kappa \ell}}$ $\in \bigl\{  {\vnabla}  {{T^{-1}}} , -{\vnabla \u}/{T}, \{ -{\vnabla} ({\mu_c}/{T}) \}, -\vnabla {\Phi}/{T}, \{ - \Delta \G_d/T \} \bigr \}$ 
is the $\kappa \ell$th component of the $m$th thermodynamic force, 
and $L_{r{\imath \jmath},m{\kappa \ell}}$ is the phenomenological coefficient or generalised conductance, where $\imath,\jmath, \kappa, \ell \in \{x,y,z\}$ are Cartesian coordinate components with $\jmath, \ell$ redundant for vectors and all indices redundant for scalars. The sum is calculated over all terms.

Eqs.\ \eqref{eq:Onsager} provide linear approximations to a more general result, in which each partial derivative $\partial \langle j_{r{\imath \jmath}} \rangle/\partial f_{s{\kappa \ell}}$ based on the expected flux $\langle j_{r{\imath \jmath}} \rangle$ is proportional to the second derivative of the Massieu function $\lambda_0=\ln Z$, and thence to the covariance of the corresponding flux pair \citep{Jaynes_1957, Jaynes_1963, Tribus_1961b, Niven_2009}. By the equivalence of cross-derivatives, these reduce to the \cite{Onsager_1931a, Onsager_1931b} reciprocal relations $L_{r{\imath \jmath},m{\kappa \ell}}=L_{m{\kappa \ell},r{\imath \jmath}}$ for time-symmetric systems, or the Casimir relations $L_{r{\imath \jmath},m{\kappa \ell}}=-L_{m{\kappa \ell},r{\imath \jmath}}$ for time-antisymmetric systems \citep{Miller_1959, deGroot_M_1962}.

By reindexing of the components $j_{r{\imath \jmath}} \mapsto j_a$, $f_{m{\kappa \ell}} \mapsto f_b$ and $L_{r{\imath \jmath}, m{\kappa \ell}} \mapsto L_{a,b}$, \eqref{eq:Onsager} can be assembled into the vector-tensor form: 
\begin{align}
\tens{j} = \tens{L} \, \tens{f}
\label{eq:Onsager_mat}
\end{align}





\noindent where $\tens{j}$ is the flux vector, $\tens{f}$ is the thermodynamic force vector and $\tens{L}$ is a two-dimensional matrix of phenomenological coefficients. 
To ensure nonnegativity of \eqref{eq:hatsigma_Onsager}, the matrix $\tens{L}$ (or equivalently its inverse $\tens{L}^{-1}$) must be positive semidefinite.
This requires $\tens{L}$ to satisfy several conditions, including nonnegative eigenvalues, nonnegative trace, $L_{a,a} \ge 0, \forall a$, and $L_{a,a} L_{b,b} \ge \frac{1}{4}(L_{a,b}+L_{b,a})^2, \forall a, b$ with $a \ne b$
\citep{Onsager_1931a, Bosworth_1956, deGroot_M_1962, Lebon_etal_2010, Bikken_Lyapilin_2014, Kondepudi_P_2015}.

Many authors restrict \eqref{eq:Onsager}-\eqref{eq:Onsager_mat} using the ``Curie principle'' \citep{Curie_1908}, commonly interpreted to allow coupling only between phenomena of the same tensorial order \citep{Bosworth_1956, deGroot_M_1962, Lebon_etal_2010}. Under this principle, the velocity gradient $\vnabla \u$, a second order tensor, is unable to influence the fluxes of heat $\j_Q$ or chemical species $\j_c$, these being first order tensors (vectors). In turn, the fluxes are unable to influence the scalar chemical reaction rates $\hat{\dot{\xi}}_d$. However, the velocity divergence $\vnabla \cdot \u$, a scalar, is able to influence the scalar reaction rates $\hat{\dot{\xi}}_d$. 
The Curie principle can be derived under the linear assumption \eqref{eq:Onsager} and the invariance of an isotropic tensor to coordinate inversion \citep{deGroot_M_1962, Lebon_etal_2010}.
However, due to continued controversy over its definition and validity \citep[e.g.][]{Moszynski_etal_1963, Jardetzky_1964, Acland_1966, Vaidhyanathan_Sitaramam_1992, Martyushev_Gorbich_2003, Klika_2010}, and deeper connections between tensorial order and group theory \citep[e.g.,][]{Brandmuller_1966, Wadhawan_1987, Simon_1997, Pommaret_2001}, this study adopts the most general formulation.
We also recall the concerns expressed in \S\ref{sect:chem_rn} on the validity of the linear assumption \eqref{eq:rate_chem_linear}  for chemical reactions, hence the reaction components of \eqref{eq:Onsager}-\eqref{eq:Onsager_mat} may have limited applicability.

The Onsager relations \eqref{eq:Onsager}-\eqref{eq:Onsager_mat} can also be inverted to give linear relations for the thermodynamic forces \citep{Bosworth_1956}:
\begin{align}
f_{m{\kappa \ell}} 
= \sum\limits_{r \imath \jmath} R_{m{\kappa \ell}, r{\imath \jmath}} \, j_{r{\imath \jmath}} 
\hspace{15pt} \text{ or } \hspace{15pt}
\tens{f} = \tens{R} \, \tens{j}
\label{eq:Onsager_inverse}
\end{align}
where $R_{m{\kappa \ell}, r{\imath \jmath}}$ is the generalised resistance defined by elements of $\tens{R}=\tens{L}^{-1}$. 
For oscillatory phenomena including heat, fluid, chemical and electrical flows, each resistance can be extended to define a generalised (complex) impedance, the sum of generalised resistance and reactance terms \citep{Bosworth_1956}. The latter provide additional oscillatory contributions to \eqref{eq:Onsager_inverse}, and have been shown to enable mutual inductance, the coupling of two transport cycles by a second transport process, for several phenomena \citep{Bosworth_1956}. 

\item {\label{sect:cross-diffusion_EP} \it \hspace{17pt} Entropy production and entropic similarity}

Incorporating diffusion and chemical reaction cross-phenomena, the total local entropy production is: 
\begin{align}
\hat{\dot{\sigma}}_{\tot}
= \j \cdot \vect{f}
= ( \tens{L} \, \tens{f})^\top \vect{f}
= \tens{f}^\top \tens{L}^\top \vect{f}
= \j^\top \tens{L}^{-1} \j
\ge 0
\label{eq:hatsigma_Onsager}
\end{align}
This subsumes diffusion and reaction processes acting in isolation \eqref{eq:hatsigma_Q}-\eqref{eq:hatsigma_k} and \eqref{eq:hatsigma_chem}. However, \eqref{eq:hatsigma_Onsager} also allows some phenomena within a coupled group to decrease the entropy production, provided that the total is nonnegative.

Applying the principle of entropic similarity, the total local entropy production \eqref{eq:hatsigma_Onsager} can be scaled by any reference entropy production term, giving the dimensionless group:
\begin{align}
{\hatPi}_{\tot} 
=  \frac{\hat{\dot{\sigma}}_{\tot}}{\hat{\dot{\sigma}}_{\text{ref}}} 
= \frac{\tens{f}^\top \tens{L}^\top \vect{f}}{\j_{\text{ref}} \cdot \vect{f}_{\text{ref}}}
\label{eq:EP_phenom}
\end{align}
Generally it will be necessary to carefully handle a mix of different units in $\j$, $\vec{f}$ and $\tens{L}$. 
In some cases $\j$ and $\vec{f}$ can be chosen so that all phenomenological coefficients $L_{a,b} \in \tens{L}$ have consistent units such as [m$^2$ s$^{-1}$].
By entropic similarity, an extended family of groups can also be derived for competing cross-diffusion processes, analogous to  \eqref{eq:Pi_diffusion}-\eqref{eq:Pi_diffusion2}:
\begin{align}
\begin{split}
\hatPi_{{L_{c,d}}/{L_{a,b}}} 
= \frac{{{{\hat{\dot{\sigma}}_{L_{c,d}}}}} }{{\hat{\dot{\sigma}}_{L_{a,b}}}} 
\to \dfrac{L_{c,d}}{L_{a,b}}
\end{split}
\end{align}
assuming constant gradients and other properties, and common units for $L_{\imath,\jmath}$. 



\end{list}

\subsection{\label{sect:dispersion} Dispersion processes}

We now consider {\it dispersion} processes, which encompass mixing or spreading phenomena which mimic the effect of diffusion, but arise from processes acting at microscopic to macroscopic rather than molecular scales. Commonly, these are represented by dispersion coefficients of the same dimensions [m$^{2}$ s$^{-1}$] as practical diffusion coefficients. Several dispersion processes are discussed in turn. 
(The phenomenon of wave dispersion is quite different, and is deferred to \S\ref{sect:grav}.)

\subsubsection{\label{sect:inertial} \it Inertial dispersion} 

\begin{figure}
\begin{center}
\includegraphics[width=50mm]{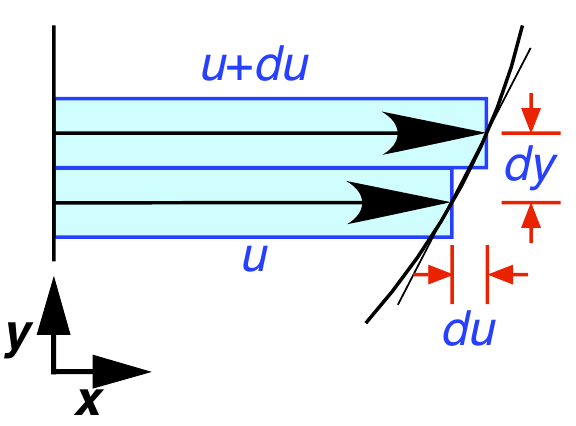}
\end{center}
\caption{Differential motion of parallel fluid elements and the resulting orthogonal velocity gradient ${du}/{dy}$.}
\label{fig:fluid_els}
\end{figure}

In fluid flow, the foremost dispersion process -- here termed {\it inertial dispersion} -- involves the spreading of momentum and any fluid-borne properties (e.g., heat, chemical species, charge) due to inertial flow. This can be demonstrated by the classical diagram of inertial effects in a fluid flow, given in figure \ref{fig:fluid_els} \citep[e.g.,][]{Street_etal_1996, Demirel_2002}.
As shown, the velocity difference between two adjacent fluid elements moving in (say) the $x$-direction will create a velocity gradient normal to the flow, here drawn in the $y$ direction. Above a threshold, the resulting shear stresses will produce inertial motions of fluid normal and opposite to the velocity gradient, causing the lateral transfer of momentum from regions of high to low momentum. Similarly, a velocity gradient in the flow direction, say $x$, will tend to be counterbalanced by opposing inertial flows. Taken together, these effects cause the transfer of momentum in opposition to the velocity gradient tensor, producing a flow field which is more spatially uniform in the mean, but with a tendency towards turbulent flow.

\newcounter{Lcount4}
\begin{list}{(\alph{Lcount4})}{\usecounter{Lcount4} \topsep 4pt \itemsep 0pt \parsep 3pt \leftmargin 0pt \rightmargin 0pt \listparindent 10pt \itemindent 20pt}

\item {\label{sect:internal} \it \hspace{17pt} Entropic similarity for inertial dispersion in internal flows}

We first consider {\it internal flows}, involving flow in a conduit with solid walls under a pressure gradient (Poiseuille flow). 
For steady irrotational incompressible flow in a cylindrical pipe, the total entropy production is \citep{Bejan_1982, Bejan_1996, Bejan_2006, Niven_2010_JNET, Niven_2021}:
\begin{align}
\dot{\sigma}_{\text{int}} 
= \frac {p_L Q }{T} 
= \frac {\rho g H_L Q }{T} 
= \frac {\pi \rho g H_L U d^2 }{4T}  \YY
\label{eq:EP_int}
\end{align}
where $p_L$ is the pressure loss [Pa], $H_L$ is the head loss [m], $Q$ is volumetric flow rate [m$^{3}$ s$^{-1}$], $U$ is the mean velocity [m s$^{-1}$] and $d$ is the pipe diameter [m].
The head loss by inertial flow is given by the Darcy-Weisbach equation \citep[e.g.,][]{Pao_1961, Street_etal_1996, Spurk_1997, Streeter_etal_1998, White_2006, Munson_etal_2010, Douglas_etal_2011}:
\begin{align}
H_{L,I} = \frac{f L}{2gd} {U^2} \YY
\label{eq:DW}
\end{align}
where $f$ is the Darcy friction factor [--] for the inertial regime, itself a function of the flow and pipe properties \citep{Colebrook_1939}, and $L$ is the pipe length [m]. 
Substituting in \eqref{eq:EP_int} gives the entropy production by inertial dispersion: 
\begin{align}
\dot{\sigma}_{\text{int},I} 
= \frac {\pi \rho d f L}{8T}  {U^3}  \YY
\label{eq:EP_int_inertial}
\end{align}
Eqs.\ \eqref{eq:EP_int}-\eqref{eq:EP_int_inertial} allow for flow reversal, with $p_L<0$, $H_L<0$ and $f<0$ corresponding to $U<0$ and $Q<0$, giving $\dot{\sigma}_{\text{int},I} \ge 0$ in all cases. 
For laminar flow involving purely viscous diffusion, the head loss and entropy production are, analytically \citep{Pao_1961, Schlichting_1968, Street_etal_1996}:  
\begin{align}
H_{L, \nu} 
= \frac{32 \nu L}{g d^2} U ,\YY 
\hspace{15pt}
\dot{\sigma}_{\text{int},\nu} 
= \frac{16 \pi \rho \nu L}{ T} U^2 \YY
\label{eq:EP_int_lam}
\end{align}
The relative importance of inertial dispersion and viscous diffusion can thus be examined by the entropic group:
\begin{align}
\Pi_{\text{int},I/\nu} 
=  \frac{ \dot{\sigma}_{\text{int},I} }{ \dot{\sigma}_{\text{int},\nu}}
= \frac{  f \,  U  d    }{  64 \nu  } 
\sim f Re, 
\hspace{10pt}
\text{with } 
Re = \frac{   U  d    }{   \nu  }  \YY
\label{eq:Pi_I_nu}
\end{align}
where $Re$ is the Reynolds number \eqref{eq:Re}. 
Groups proportional to $f Re$ have been termed the {Poiseuille number} \citep{Churchill_1988, White_2006}. 
For purely viscous diffusion (laminar flow), $f=64/Re$ \citep{Pao_1961, Street_etal_1996} and $\Pi_{\text{int},I/\nu}=1$. 
For increasingly inertial flows $\Pi_{\text{int},I/\nu}>1$, $f$ is commonly correlated as a function of $Re$ and surface roughness \citep[e.g.,][]{Colebrook_1939}. For non-circular conduits, $d$ is replaced by the hydraulic diameter $d_H=4A/P_w$, where $A$ is the area [m$^2$] and $P_w$ is the wetted perimeter [m] \citep{Schlichting_1968, Spurk_1997, White_2006}. 
For wall shear flows, \eqref{eq:Pi_I_nu} is written using the friction velocity $u^*=\sqrt{\tau_0/\rho} = U \sqrt{f/8}$, where $\tau_0$ is the wall shear stress [Pa], giving $\Pi_{\text{int},I/\nu} \sim \sqrt{f} \, Re^* \Y$ with $Re^* = {u^* d}/{\nu} \Y$ \citep[e.g.,][]{Street_etal_1996}. For flow in porous media, the pressure loss is given by the \cite{Ergun_1952} equation, and replacing $d$ by a void length scale and $U$ by an interstitial velocity gives a more natural representation \citep{Churchill_1988, Niven_2002}.  

Interpreting \eqref{eq:Pi_I_nu} from the perspective of entropic similarity, within an internal flow the velocity will decrease from the order of its mean velocity $U$ to zero, over distances of the order of its length scale $d$ (see figure \ref{fig:fluid_els}). Above a critical threshold, these will produce inertial flows with a macroscopic dispersion coefficient of order $Ud$. In consequence, the Reynolds number can be interpreted from an entropic perspective simply as the ratio of the inertial dispersion and viscous diffusion coefficients. Multiplication by $f$ to give the entropic group $\Pi_{\text{int},I/\nu}$ incorporates the resistance of the conduit to the imposed flow, including the effects of flow regime, pipe geometry and surface roughness. 

\item {\label{sect:external} \it \hspace{17pt} Entropic similarity for inertial dispersion in external flows}

We can also consider {\it external flows}, involving flow around a solid object \citep{Pao_1961, Street_etal_1996, White_2006, Munson_etal_2010, Douglas_etal_2011}. For a simplified steady one-dimensional flow, the entropy production is \citep{Bejan_1982, Bejan_1996, Bejan_2006, Niven_2021}:
\begin{align}
\dot{\sigma}_{\text{ext}} = \frac{{F}_D U}{T}  \YY
\label{eq:EP_ext}
\end{align}
where $F_D$ is the magnitude of the drag force [N] and $U$ is the mean velocity of the ambient fluid [m s$^{-1}$].  The drag force is commonly scaled by the dynamic pressure to give a dimensionless drag coefficient \citep[e.g.][]{Pao_1961, Schlichting_1968, Street_etal_1996}: 
\begin{align}
C_D = \frac{{F}_D}{\tfrac{1}{2} \rho A_s U^2} \YY
\label{eq:C_D}
\end{align}
where $A_s$ is the cross-sectional area of the solid normal to the flow [m$^2$]. Taking $A_s = \pi d^2/4$ for a sphere, where $d$ is the diameter [m], the inertial entropy production is:
\begin{align}
\dot{\sigma}_{\text{ext},I} 
= \frac{ \rho A_s C_D U^3}{2T} 
= \frac {\pi \rho d^2 C_D U^3}{8T} \YY
\label{eq:EP_ext_inertial}
\end{align}
Eqs.\ \eqref{eq:EP_ext}-\eqref{eq:EP_ext_inertial} allow for flow reversal, with $F_D<0$ and $C_D<0$ corresponding to $U<0$, whence $\dot{\sigma}_{\text{ext},I} \ge 0$ in all cases. For steady laminar (Stokes) flow around a sphere, the drag force and entropy production are, analytically \citep[e.g.,][]{Pao_1961, Schlichting_1968, Street_etal_1996, White_2006}:
\begin{align}
F_D =3 \pi \rho \nu d U, \YY
\hspace{15pt}
\dot{\sigma}_{\text{ext},\nu} = \frac{3 \pi \rho \nu d U^2}{T} \YY
\label{eq:EP_ext_lam}
\end{align}
The relative importance of inertial dispersion and viscous diffusion can thus be assessed by the entropic group \citep{Duan_etal_2015}:
\begin{align}
\Pi_{\text{ext},I/\nu} 
= \frac{ \dot{\sigma}_{\text{ext},I} }{ \dot{\sigma}_{\text{ext},\nu}}
= \frac{  C_D \,  U  d    }{  24 \nu  } 
\sim C_D Re,
\hspace{10pt}
\text{with } 
Re = \frac{   U  d    }{   \nu  }  \YY
\label{eq:Pi_I_nu2} 
\end{align}
This is analogous to \eqref{eq:Pi_I_nu}, with an equivalent entropic interpretation. For non-spherical solids, $A_s$ can be used directly, or $d$ assigned to a representative length scale for the solid.
For purely viscous diffusion, $C_D=24/Re$ \citep{Pao_1961, Schlichting_1968, Street_etal_1996} and $\Pi_{\text{ext},I/\nu}=1$. For increasingly inertial flows $\Pi_{\text{ext},I/\nu}>1$, $C_D$ is commonly correlated as a function of $Re$ for various solid shapes \citep[e.g.,][]{Birkhoff_1960, Clift_etal_1978}.

For both internal and external flows, the characteristic plot of $f$ against $Re$ (the \cite{Moody_1944} diagram) or $C_D$ against $Re$ can be interpreted as an {\it entropic similarity diagram}. Some authors have also presented plots of $f Re$ or $C_D Re$ against $Re$  \citep[e.g.,][]{Duan_etal_2015}, or other variants \citep{PaulusJr_G_2004, Niven_2010_JNET}, to directly examine the entropy production. Such diagrams are useful in assessing the interplay between entropic phenomena for different geometries and flow conditions.

The above treatment can be extended for more complicated flows.  For two- or three-dimensional steady irrotational external flows, it is necessary to consider the drag force and lift force(s) respectively aligned with and normal to a reference direction, each with a corresponding drag or lift coefficient. A vector formulation is warranted. Consider a solid object moving at velocity $\V$ [m s$^{-1}$] in a uniform flow field of ambient velocity $\u$ \protect{[m s$^{-1}$]}, producing the vector drag-lift force $\vec{F}_D$ [N] on the object. The vector drag-lift coefficient   $\vec{C}_D = [C_D, C_L]^\top$ or $[C_D, C_{L1}, C_{L2}]^\top$ and inertial entropy production are \citep{Niven_2021}:
\begin{gather}
\vec{C}_{D} = \dfrac{\vec{F}_D}{\frac{1}{2} \rho A_s ||\u-\V||^2}
\label{eq:CD_vect}
\\
\dot{\sigma}_{\text{ext},I} 
= \frac{\vec{F}_D \cdot (\u - \V)}{T} 
= \frac{ \frac{1}{2} \rho A_s ||\u-\V||^2 \, \vec{C}_D \cdot (\u - \V)}{T} 
\label{eq:EP_ext_I_vect}
\end{gather}
while the Stokes viscous force and viscous entropy production for a sphere are:
\begin{gather}
\vec{F}_D =3 \pi \rho \nu d (\u-\V)
\hspace{10pt} \text{and} \hspace{10pt}
\dot{\sigma}_{\text{ext},\nu} 
= \frac{ 3 \pi \rho \nu d \, || \u-\V ||^2}{T} 
\label{eq:EP_ext_nu_vect}
\end{gather}
The entropic dimensionless group \eqref{eq:Pi_I_nu2}, taking $A_s = \pi d^2/4$, becomes:
\begin{align}
\Pi_{\text{ext},I/\nu} 
=  \frac{ \dot{\sigma}_{\text{ext},I} }{ \dot{\sigma}_{\text{ext},\nu}}
= \frac{ \vec{C}_D \cdot (\u - \V) d} {24  \nu  } 
\sim \vec{C}_D \cdot \vec{Re}
\hspace{10pt}
\text{with } 
\vec{Re} = \frac{ (\u - \V)  d} { \nu  } 
\label{eq:Pi_I_nu_vect}
\end{align}
where $\vec{Re}$ is a vector Reynolds number, to account for the ratio of the inertial dispersion and viscous diffusion coefficients in each direction. 
The drag and lift directions are quite distinct, since lift forces are unrelated to viscous friction \citep{Rubinow_Keller_1961} but are governed by the fluid circulation ${\Gamma=-\oint_{\mathcal{C}} \u \cdot d\vec{s}}$ on any closed path $\mathcal{C}$ around the solid, where $\vec{s}$ is the path coordinate~\citep{Pao_1961, Street_etal_1996, Douglas_etal_2011, Spurk_1997}. 


Now consider purely rotational motion of a rigid sphere of radius vector $\r$ [m] about its centroid $\x_G$, at the angular velocity $\vec{\omega}(t)$ [s$^{-1}$]. 
The entropy production is: 
\begin{align}
\dot{\sigma}_{\text{ext}}^{\text{rot}} 
= \frac{ \T \cdot \vec{\omega} }{T} 
\label{eq:EP_ext_rot}
\end{align}
where $\T$ is the torque on the solid [N m]. 
The viscous torque on the sphere is $\T = 8 \pi \mu ||\r||^3 \vec{\omega}$ \citep{Kirchhoff_1876, Rubinow_Keller_1961, Sawatzki_1970}. The ratio of rotational inertial to viscous effects can therefore be represented by the entropic group:
\begin{align}
\begin{split}
\Pi_{\text{ext},I/\nu}^{\text{rot}}  
=  \frac{ \dot{\sigma}^{\text{rot}}_{\text{ext},I} }{ \dot{\sigma}^{\text{rot}}_{\text{ext},\nu}}
= \frac{\C_T \cdot \vec{Re}_T} {16 \pi} 
\sim \C_T \cdot \vec{Re}_T 
\\
\text{with}  \hspace{15pt}
\vec{Re}_T  = \frac{  ||\r||^2 \vec{\omega} }{\nu}, \hspace{15pt}
\C_T = \frac{ \T} { \tfrac{1}{2} \rho ||\r||^5 \, || \vec{\omega} ||^2}
\end{split}
\label{eq:Pi_rot}
\end{align}
where $\vec{Re}_T$ is a rotational Reynolds (or Taylor) number and $\C_T$ is a torque coefficient, such that $\vec{Re}_T$ and $\C_T$ are pseudovectors \citep[c.f.,][]{Schlichting_1968, Sawatzki_1970, Lukerchenko_etal_2012}.  Clearly, $\vec{Re}_T$ can be interpreted as the ratio of the rotational inertial dispersion coefficient $||\r||^2 \vec{\omega}$ to the viscous diffusion coefficient $\nu$, while its multiplication by $\C_T$ incorporates the resistance to rotation. 
For purely viscous diffusion, $\C_T=16\pi \vec{Re}_T /||\vec{Re}_T||^2$ \citep{Sawatzki_1970} and $\Pi_{\text{ext},I/\nu}^{\text{rot}} =1$. 
Eq.\ \eqref{eq:Pi_rot} can be extended to different solid shapes and centres of rotation based on moments of inertia.

For unsteady irrotational external flows, it is necessary to consider the inertial drag due to the ``added mass'' of fluid, a history-dependent force and the acceleration of the local fluid \citep[e.g.,][]{Boussinesq_1885, Basset_1888, Corrsin_Lumley_1956,  Odar_Hamilton_1964, Maxey_Riley_1983, Mei_1994}. For combined translational and rotational flows, it is necessary to consider the Magnus force due to rotation-induced lift \citep{Rubinow_Keller_1961, Lukerchenko_etal_2012}, with additional contributions from flow-induced or imposed vibrations \citep{Naudasher_Rockwell_2005} and deformable solids such as flapping wings. 

For boundary-layer flows such along a flat plate, it is usual to consider a local drag coefficient $C_{D}(x)$ and boundary layer thickness $\delta(x)$ as functions of $x$ \citep[e.g.,][]{Schlichting_1968, Street_etal_1996}.  These give the local dimensionless groups $\Pi_{I,x/\nu}(x) = C_{D}(x) Re_{x}(x)$ and $\Pi_{I,\delta/\nu}(x) = C_{D}(x) Re_\delta(x)$, where $Re_{x}(x) = U_{\infty} x/\nu$, $Re_\delta(x) = U_{\infty} \delta (x)/\nu$ and $U_{\infty}$ is the free-stream velocity.  The Reynolds numbers assess the importance of local inertial dispersion, measured by $U_{\infty} x$ or $U_{\infty} \delta (x)$, relative to viscous diffusion.

\item  {\label{sect:inertial_and_other} \it \hspace{17pt} Entropic similarity for inertial dispersion with other diffusion processes}

Since inertial dispersion dramatically enhances the spreading of other fluid properties, it is necessary to consider its influence relative to diffusion processes. 
Making an analogy between macroscopic heat, chemical species or ionic diffusion processes and laminar viscous diffusion in an internal flow \eqref{eq:EP_int_lam}, this can be represented by the entropic groups \citep[c.f.,][]{Bosworth_1956, Schlichting_1968, Clift_etal_1978, Fogler_1992, Furbish_1997, Bird_etal_2006}:
\begin{align}
\begin{array}{llll}
\Pi_{\text{int},I/\alpha} =\dfrac{ \dot{\sigma}_{\text{int},I} }{ \dot{\sigma}_{\text{int},\alpha}} \sim f Pe_\alpha, \hspace{5pt} &&\text{with } Pe_\alpha = \dfrac{U d}{\alpha} = Re Pr,\YY
\\
\Pi_{\text{int},I/c} =\dfrac{ \dot{\sigma}_{\text{int},I} }{ \dot{\sigma}_{\text{int},\Dc}} \sim f Pe_c, \hspace{5pt} &&\text{with } Pe_c = \dfrac{U d}{\Dc} = Re Sc_c,\YY
\\
\Pi_{\text{int},I/k} =\dfrac{ \dot{\sigma}_{\text{int},I} }{ \dot{\sigma}_{\text{int},D_k}} \sim f Pe_k, \hspace{5pt} &&\text{with } Pe_k = \dfrac{U d}{\Dk} = Re Sc_k \Y
\end{array}
\label{eq:Pe} 
\end{align}
where $d$ is a length scale [m], and $Pe_{\alpha}, Pe_c$ and $Pe_k$ are P\'eclet numbers respectively for heat, chemical species $c$ or charged species $k$.  
Each P\'eclet number is the ratio of the inertial dispersion coefficient $Ud$ to the heat, chemical species or charge diffusion coefficient, equivalent to the product of the Reynolds number \eqref{eq:Pi_I_nu} and the corresponding Prandtl or Schmidt number \eqref{eq:Pi_diffusion2}. 
In process engineering, the length scales in \eqref{eq:Pe} are substituted by a length scale $L$ of the reactor \citep{Fogler_1992}. 


\end{list}

\subsubsection{\it Turbulent dispersion} \label{sect:turb}


An important mixing process in fluid flow is {\it turbulent dispersion}, often termed {\it eddy dispersion} or {\it eddy diffusion}, caused by turbulent motions of the fluid.  This provides the dominant mixing mechanism for many natural flow systems, including flows in streams, lakes, oceans and atmosphere \citep{Streeter_etal_1998}. 
Turbulent dispersion is closely related to the inertial dispersion but is analysed at local scales, using the Reynolds decomposition of each physical quantity $a = \overline{a} + a'$, where $\overline{a}$ is a Reynolds average (such as the time or ensemble average) and $a'$ is the fluctuating component \citep{Schlichting_1968, Furbish_1997, Pope_2000}. As shown by \cite{Reynolds_1895}, averaging of the incompressible Navier-Stokes equations reveals an additional contribution to the mean stress tensor $ \overline{\tens{\tau}_{t}} = \rho \overline{\vect{u}' \vect{u}'}$ [Pa], now termed the Reynolds stress. This is often correlated empirically by the \cite{Boussinesq_1877} approximation \citep{Pope_2000, Davidson_2004}\footnote{Recall the sign convention used here, in which the stress tensor is positive in compression. Eq.\ \eqref{eq:tau_turb} is symmetric, consistent with the shear stress tensor \eqref{eq:Newton}. The last term in \eqref{eq:tau_turb}, omitted from many references, gives the correct normal stresses \citep{Pope_2000, Davidson_2004}.}:
\begin{align}
    \overline{\tens{\tau}_{t}}
=   \rho \overline{\vect{u}' \vect{u}'} 
\approx
  - \rho  \nu_t \bigl(\vnabla \overline{\u}  
  + (\vnabla \overline{ \u}) ^\top \bigr) 
  + \tfrac{1}{3} \rho \vec{\delta} \, \tr (\overline{\vect{u}' \vect{u}'})  \YY
\label{eq:tau_turb}
\end{align}
where $\nu_t$ is the turbulent dispersion coefficient or eddy viscosity [m$^2$ s$^{-1}$]. 
%
Similarly, taking the Reynolds average of the conservation laws for heat, chemical species and charge reveals mean-fluctuating or Reynolds fluxes, associated with turbulent mixing. These can be correlated empirically by:
\begin{align}
\begin{aligned}
    \overline{\j_{Q,t}}  
    &=  \overline{\u' (\rho c_p T)'}
&&\approx  -  \alpha_t \vnabla (\overline{\rho c_P T}) 
\simeq -k_t \vnabla \overline{T} 	\YY
\\
\overline{\j_{c,t}}  &=\overline{\u' C_c'}
&&\approx -{\Dct} \vnabla \overline{C_c} \YY
\\
\overline{\i_{k,t}}  & = \overline{ \u' \biggl ( \frac {z_k^2 F^2 C_k \Phi}{R T} \biggr)'} \Y
&&\approx  - D_{k,t}  \vnabla \overline{ \biggl ( \frac {z_k^2 F^2 C_k \Phi}{R T} \biggr)} \Y
\simeq - \kappa_{k,t}  \vnabla \overline{\Phi}    \Y
\end{aligned}
\label{eq:fluxes_turb}
\end{align}
where 
$\alpha_t$ 
is the thermal eddy dispersion coefficient (or eddy diffusivity) [m$^{2}$ s$^{-1}$], 
$k_t$ is the thermal eddy conductivity [J K$^{-1}$ m$^{-1}$ s$^{-1}$], 
$\Dct$ is the eddy dispersion coefficient for the $c$th chemical species [m$^{2}$ s$^{-1}$], 
$\Dkt$ 
is the eddy dispersion coefficient for the $k$th ion [m$^{2}$ s$^{-1}$], and 
$\kappa_{k,t}$ is the electrical eddy conductivity for the $k$th ion [A V$^{-1}$ m$^{-1}$]
\citep[c.f.,][]{Bosworth_1956, Pope_2000, Davidson_2004}. 

Now consider the effect of turbulence on the local entropy production for an isolated diffusion process \eqref{eq:hatsigma_Q}-\eqref{eq:hatsigma_k}, generalised as $\hat{\dot{\sigma}}_{D_X} = \j_X \cdot \vnabla Y$, where $\j_X = \u C_X$ is the flux of $X$, $C_X$ is the concentration of $X$, $D_{X}$ is the diffusion coefficient for $X$ and $\vnabla Y$ is the gradient in $Y$ conjugate to $X$. Applying the Reynolds decomposition and averaging gives:
\begin{align}
\begin{split}
\overline{\hat{\dot{\sigma}}_{D_X}}  
&= - \overline{{\j_X}} \cdot  \overline{  {\vnabla} Y  }  
- \overline{ \j_X}  \cdot  \overline{ (\vnabla Y)' }
- \overline{ \j_X'}  \cdot  \overline{ \vnabla Y }
- \overline{ \j_X'  \cdot  (\vnabla Y)' }
\\
&= - \overline{{\j_X}} \cdot  \vnabla \overline{Y}
-0
- \overline{ (\u C_X)'}  \cdot   \vnabla \overline{Y}
- \overline{ (\u C_X)'  \cdot  (\vnabla Y)' }
\\
&= - \overline{{\j_X}} \cdot   \vnabla \overline{Y}
- \overline{ \u' C_X'}  \cdot   \vnabla \overline{Y}
- \overline{ \u \vphantom{C} }  \cdot \overline{ C_X'   (\vnabla Y)' }
- \overline{ C_X} \, \overline{ \u'  \cdot  (\vnabla Y)' }
- \overline{ C_X'  \u'  \cdot  (\vnabla Y)' }
\end{split}
\label{eq:EP_turb}
\end{align}
using $\overline{ \overline{a \vphantom{b}}\, \overline{b}} = \overline{a \vphantom{b}}\, \overline{b}$ for mean terms, $\overline{a'}=0$ for isolated fluctuating terms, and $\overline{\vnabla Y} =  \vnabla \overline{Y}$ for mean gradients \citep{Schlichting_1968, Pope_2000}. As evident, the mean entropy production is complicated by the presence of diadic and triadic Reynolds terms \citep{Niven_Noack_2014}. Examining the eddy coefficient correlations \eqref{eq:tau_turb}-\eqref{eq:fluxes_turb}, which generalise to $\overline{\j_{X,t}}  = \overline{\u' C_X'} \approx - D_{X,t} \vnabla \overline{C_X}$, where $D_{X,t}$ is the eddy diffusion coefficient, these only apply to the second term in \eqref{eq:EP_turb}, with the remaining Reynolds terms unresolved. 
If the fluxes or gradients are substituted by diffusion equations in the manner of \eqref{eq:hatsigma_Q2}-\eqref{eq:hatsigma_k2}, more complicated Reynolds terms are generated \citep{Adeyinka_Naterer_2004}. For multiple processes with cross-phenomena, the Onsager relations \eqref{eq:Onsager} give even more Reynolds terms.

Many authors adopt simplified closure models for \eqref{eq:EP_turb} based only on $\overline{\hat{\dot{\sigma}}_{D_{X,t}}}  = - \overline{ \u' C_X'}  \cdot  \vnabla \overline{Y}$, neglecting the other Reynolds terms. Applying the principle of entropic similarity by analogy with \eqref{eq:Pi_diffusion}-\eqref{eq:Pi_diffusion2}, and substituting \eqref{eq:tau_turb}-\eqref{eq:fluxes_turb} and constant mean gradients, these give: 
\begin{equation}
\begin{aligned}
\hatPi_{\nu_t/\alpha_t} 
&= \frac{ \overline{\hat{\dot{\sigma}}_{\nu_t}} }{\overline{\hat{\dot{\sigma}}_{\alpha_t}}} 
\to Pr_t = \dfrac{\nu_t}{\alpha_t}, \hspace{10pt} &
\hatPi_{\nu_t/\Dct}
&= \frac{\overline{\hat{\dot{\sigma}}_{\nu_t}} }{{\overline{\hat{\dot{\sigma}}_{\Dct}}}}
\to Sc_{c,t} = \dfrac{\nu_t}{\Dct}, \hspace{10pt} &
\\
\hatPi_{\nu_t/\Dkt} 
&= \frac{\overline{\hat{\dot{\sigma}}_{\nu_t}} }{{{\overline{{\hat{\dot{\sigma}}_{\Dkt}}}}}} 
\to Sc_{k,t} = \dfrac{\nu_t}{\Dkt}, \hspace{10pt} &
\hatPi_{\alpha_t/\Dct} 
&= \frac{{\overline{\hat{\dot{\sigma}}_{\alpha_t}}} }{{\overline{\hat{\dot{\sigma}}_{\Dct}}}}
\to Le_{c,t} = \dfrac{\alpha_t}{\Dct}, \hspace{10pt} &
\\
\hatPi_{\alpha_t/\Dkt} 
&= \frac {{\overline{\hat{\dot{\sigma}}_{\alpha_t}}} }{{{\overline{{\hat{\dot{\sigma}}_{\Dkt}}}}}} 
\to Le_{k,t} = \dfrac{\alpha_t}{\Dkt}, \hspace{10pt} &
\hatPi_{\Dct/\Dkt} 
&= \frac {{\overline{\hat{\dot{\sigma}}_{\Dct}}}}{{\overline{{\hat{\dot{\sigma}}_{\Dkt}}}} }
\to \dfrac{\Dct}{\Dkt}, \hspace{10pt} &
\\
\hatPi_{\Dct/\Dbt} 
&= \frac{{\overline{\hat{\dot{\sigma}}_{\Dct}}}}{{{\overline{{\hat{\dot{\sigma}}_{\Dbt}}}}}}
\to  \dfrac{\Dct}{\Dbt}, \hspace{10pt} &
\hatPi_{\Dkt/\Dlt} 
&= \frac {{{\overline{{\hat{\dot{\sigma}}_{\Dkt}}}}}}{{{\overline{{\hat{\dot{\sigma}}_{\Dlt}}}}}} 
\simeq\dfrac{{ z_k^2  C_k  \Dkt}   }{{ z_\ell^2 C_\ell  \Dlt}  } 
\to \dfrac{\Dkt}{\Dlt}
\hspace{10pt} &
\end{aligned}
\label{eq:Pi_turb_disp}
\raisetag{30pt}
\end{equation}
These respectively give the turbulent Prandtl, Schmidt (species), Schmidt (charge), Lewis (species) and Lewis (charge) numbers, and ratios of eddy dispersion coefficients for different chemical and/or charged species \citep[c.f.,][]{Bosworth_1956, Schlichting_1968, Streeter_etal_1998, Pope_2000}. 
By the ``Reynolds analogy'', 
some authors argue that these should all be constant for fixed turbulent conditions \citep{Schlichting_1968}.  
The same groups can also be obtained from ratios of the turbulent entropy fluxes, analogous to those in \eqref{eq:fluxS}-\eqref{eq:fluxS2}. 

An additional family of groups can be obtained from ratios of the turbulent and mean-product entropy production terms -- the second and first terms in \eqref{eq:EP_turb} -- reducing to ratios of the turbulent dispersion and molecular diffusion coefficients:
\begin{align}
\begin{split}
\hatPi_{\nu_t/\nu} 
&= \frac{\overline{\hat{\dot{\sigma}}_{\nu_t}} }{\overline{\hat{\dot{\sigma}}_{\nu}}} 
\to \dfrac{\nu_t}{\nu},
\hspace{10pt}
\hatPi_{\alpha_t/\alpha} 
= \frac{{\overline{\hat{\dot{\sigma}}_{\alpha_t}}}  }{\overline{\hat{\dot{\sigma}}_{\alpha}}} 
\to \dfrac{\alpha_t}{\alpha},
\hspace{10pt}
\hatPi_{\Dct/\Dc} 
= \frac{{\overline{{\hat{\dot{\sigma}}_{\Dct}}}} }{\overline{\hat{\dot{\sigma}}_{\Dc}}} 
\to \dfrac{\Dct}{\Dc},
\hspace{10pt}
\\
\hatPi_{\Dkt/\Dk} 
&= \frac{{\overline{{\hat{\dot{\sigma}}_{\Dkt}}}} }{\overline{\hat{\dot{\sigma}}_{\Dk}}} 
\to \dfrac{\Dkt}{\Dk}
\end{split}
\end{align}
The first group serves a similar purpose to the Reynolds number \eqref{eq:Re} or \eqref{eq:Pi_I_nu}, but defined locally, while the remaining three give local analogues of the P\'eclet numbers  $Pe_\alpha$, $Pe_c$ and $Pe_k$ respectively 
\eqref{eq:Pe}.  
If there are thermodynamic cross-phenomena, analogous groups can be defined by entropic similarity using the Onsager relations \eqref{eq:Onsager}-\eqref{eq:Onsager_mat}:
\begin{align}
\begin{split}
\hatPi_{{L_{c,d,t}}/{L_{a,b,t}}} 
= \frac{{\overline{{\hat{\dot{\sigma}}_{L_{c,d,t}}}}} }{\overline{\hat{\dot{\sigma}}_{L_{a,b,t}}}} 
\to \dfrac{L_{c,d,t}}{L_{a,b,t}},
\hspace{10pt}
\hatPi_{{L_{a,b,t}}/{L_{a,b}}} 
= \frac{{\overline{{\hat{\dot{\sigma}}_{L_{a,b,t}}}}} }{\overline{\hat{\dot{\sigma}}_{L_{a,b}}}} 
\to \dfrac{L_{a,b,t}}{L_{a,b}}
\end{split}
\end{align}
where $L_{\imath,\jmath,t}$ is the turbulent phenomenological coefficient for the $(\imath,\jmath)$th process [m$^2$ s$^{-1}$].


\subsubsection{\label{sect:conv} \it Convective dispersion} 


For heat or mass transfer processes involving fluid flow past a solid surface or fluid interface, it is necessary to consider the combined effect of diffusion and bulk fluid motion, referred to as {\it convection} \citep[e.g.,][]{Bosworth_1956, Eckert_1963, Schlichting_1968, Eckert_Drake_1972, Incropera_DeWitt_1990, Incropera_DeWitt_2002, Holman_1990, Fogler_1992, Bejan_1993, Bejan_1995, Streeter_etal_1998, Bird_etal_2006, White_2006, Cengel_etal_2012}.  
Convection can be further classified into {\it forced convection}, due to fluid flow under a pressure gradient, and {\it free} or {\it  natural convection}, due to fluid flow caused by heat-induced differences in temperature and density.
Examples include heat exchange, extraction, sorption, drying and membrane filtration, and latent heat exchange processes such as evaporation, distillation and condensation.
Convection also arises in charge transfer such as electrolysis \citep[e.g.,][]{Novev_Compton_2018}. 
Convection processes are commonly analysed by the linear transport relations:
\begin{equation}
\begin{aligned}
||{\widetilde{\j_Q}} ||&= h_Q \Delta T, \YY		\hspace{10pt} & 
||{\widetilde{\j_c}} || &= h_c \Delta \chi_c = \widetilde{h}_c \Delta C_c,\YY	\hspace{10pt} &
||{\widetilde{\i_k}}||  & =  h_k \Delta \Phi  
\end{aligned}
\label{eq:rels_convection}
\end{equation}
where $\widetilde{\j_X}$ is the convective flux of $X$ normal to the boundary, $h_Q$ is the heat transfer coefficient [J K$^{-1}$ m$^{-2}$ s$^{-1}$], $h_c$ is the mass transfer coefficient for the $c$th species [mol m$^{-2}$ s$^{-1}$], $\widetilde{h}_c$ is the mass transfer film coefficient for the $c$th species [m s$^{-1}$], $h_k$ is the charge transfer coefficient for the $k$th ion [C V$^{-1}$ m$^{-2}$ s$^{-1}$], $\chi_c = C_c / C$ is the mole fraction [--], $C= \sum\nolimits_c C_c$ is the total concentration [mol m$^{-3}$], and $\Delta$ represents a difference between two phases, e.g., between a solid surface and the free-stream fluid (beyond the boundary layer), or between two fluid phases. The transfer coefficients are specific to each process and flow geometry. 

We now apply the principle of entropic similarity to examine the transport regime during convection, based on ratios of entropy fluxes for convection and diffusion processes analogous to \eqref{eq:Pi_diffusion_fluxes}. Applying \eqref{eq:fluxS} for fixed intensive variables, using the convective fluxes \eqref{eq:rels_convection} and diffusive fluxes \eqref{eq:Fourier}, \eqref{eq:Fick1} and \eqref{eq:Ohm}, we obtain the entropic groups: 
\begin{align}
\begin{aligned}
\hatPi_{h_Q/\alpha}
&=\dfrac{||\j_{S,h_Q}||}{||\j_{S,\alpha}||}
=\dfrac{||{\widetilde{\j_Q}}|| \frac{1}{T}}{||\j_{Q}|| \frac{1}{T}}
= \dfrac{h_Q \Delta T}{k ||\vnabla T|| } \Y
&&\to 
Nu  = \dfrac{h_Q d_Q}{k}	\YY
\\
\hatPi_{h_c/\Dc}
&=\dfrac{||\j_{S,h_c}||}{||\j_{S,\Dc}||}
=\dfrac{||{\widetilde{\j_c}}|| \frac{\mu_c}{T} }{||\j_{c}|| \frac{\mu_c}{T} }
= \dfrac{h_c \Delta \chi_c}{\Dc ||\vnabla C_c|| } \Y
= \dfrac{\widetilde{h}_c \Delta C_c}{\Dc ||\vnabla C_c|| } \Y
&&\to
Sh_c = \dfrac{h_c d_c}{C \Dc}  = \dfrac{\widetilde{h}_c d_c}{\Dc}  \YY
\\
\hatPi_{h_k/D_k}
&=\dfrac{||\j_{S,h_k}||}{||\j_{S,D_k}||}
=\dfrac{||{\widetilde{\i_k}}|| \frac{\Phi}{T}}{||\i_{k}|| \frac{\Phi}{T}}
= \dfrac{h_k \Delta \Phi}{D_k ||\vnabla \Phi|| } \Y
&&\to
Sh_k = \dfrac{h_k d_k}{\kappa_k} 
\simeq \dfrac{RT h_k d_k}{z_k^2 F^2 C_k D_k} \Y
\end{aligned}
\label{eq:Pi_convection}
\end{align}
where $d_Q, d_c$ and $d_k$ are length scales [m] arising from each normalised gradient, $Nu$ is the Nusselt number, and $Sh_c$ and $Sh_k$ are the chemical and charge Sherwood numbers (see references at start of \S\ref{sect:conv}). 
For boundary layer flows, the three groups are functions of position.
$Nu$ can also be shown to represent the dimensionless temperature gradient from the surface or interface \citep{Schlichting_1968, Incropera_DeWitt_1990, Incropera_DeWitt_2002, Streeter_etal_1998, White_2006}, 
while $Sh_c$ is the dimensionless concentration gradient \citep{Incropera_DeWitt_2002, Streeter_etal_1998}. 
By the same reasoning, $Sh_k$ can be interpreted as the dimensionless electrical potential gradient from the surface or interface. 

In forced convection, $Nu$ is commonly expressed as a function of $Re$ and $Pr$ for a given flow geometry, while $Sh_c$ is correlated as a function of $Re$ and $Sc_c$ (see references at start of \S\ref{sect:conv}).
In free convection due to heat transfer, the inertia arises from temperature-induced differences in density, giving the velocity scale $U_{\conv} = \sqrt{g d |\Delta \rho|/ \rho} \YY$ [m s$^{-1}$], where $d$ is a length scale [m] and $|\Delta \rho|$ is the magnitude of the density difference between the wall and free-stream fluid [kg m$^{-3}$] \citep{White_2006}. 
For an internal flow, comparing \eqref{eq:Pi_I_nu} and \eqref{eq:Pe} we can define the entropic groups:
\begin{align}
\begin{split}
\Pi'_{\text{int},I/\nu} 
&=  \biggl( \frac{ \dot{\sigma}_{\text{int},I} }{ \dot{\sigma}_{\text{int},\nu}}\biggr)^2
\sim \biggl( \frac{  f \,  U_{\conv}  d    }{  \nu  } \biggr)^2
= f^2 Gr, \Y
\\
\text{with } \hspace{10pt}
&Gr = \biggl( \frac{U_{\conv} d}{\nu} \biggr)^2 
= \frac{g d^3 |\Delta \rho|}{\rho \nu^2} 
= \frac{g \beta d^3 |\Delta T|}{\nu^2} \YY
\end{split}
\label{eq:Pi_I_nu_conv}
\\
\begin{split}
\Pi'_{\text{int},I/\alpha,\nu} 
&=\dfrac{ \dot{\sigma}_{\text{int},I}^2}{ \dot{\sigma}_{\text{int}, \nu} \, \dot{\sigma}_{\text{int}, \alpha}} \sim f^2 Ra, \hspace{3pt} \\
\text{with } \hspace{10pt}
&Ra =  \frac{(U_{\conv} d)^2}{\nu \alpha}  
= \frac{g d^3 |\Delta \rho|}{\rho \nu \alpha} 
= \frac{g \beta d^3 |\Delta T|}{\nu \alpha}  = Gr Pr \Y
\end{split}
\label{eq:Pi_I_nu_alpha_conv}
\end{align}
using $\Delta \rho \simeq - \rho \beta \Delta T$, where $Gr$ is the Grashof number, $Ra$ is the Rayleigh number and $\beta$ is the thermal expansion coefficient [K$^{-1}$] (see references at start of \S\ref{sect:conv}). 
For boundary layer flows, analogues of \eqref{eq:Pi_I_nu_conv}-\eqref{eq:Pi_I_nu_alpha_conv} containing $C_D$ rather than $f$ are required, in which $Gr$ and $Ra$ are functions of position. 
As evident, $Gr$ is the square ratio of the inertial dispersion to viscous diffusion coefficients in a buoyancy-driven flow, while $Ra$ is a composite group based on the inertial dispersion, viscous and heat diffusion coefficients.
Traditionally, $Gr$ is interpreted by dynamic similarity as the ratio of buoyancy to viscous forces, while $Ra$ is a composite ratio of buoyancy, viscous and heat transport forces. 
In free convection, $Nu$ is generally expressed as a function of $Gr$ (or $Ra$), $Pr$ 
and geometry,
while $Sh_c$ is correlated as a function of $Gr$, $Sc_c$ and geometry
(see references at start of \S\ref{sect:conv}).
 \cite{Bejan_1995} uses a length-scale analysis to argue for correlations based on $Ra$ rather than $Gr$, 
 with a Boussinesq number $Bq=Ra Pr$ for low-$Pr$ fluids.
The Richardson number $Ri = Gr/Re^2$ can be used to characterise the flow regime as free ($Ri \gg 1$), forced  ($Ri \ll 1$) or mixed  ($Ri \sim1$) convection \citep{Schlichting_1968, Incropera_DeWitt_1990, Incropera_DeWitt_2002, White_2006, Cengel_etal_2012}. 

In free convection due to chemical gradients, mass-transfer analogues of $Gr$ and $Ra$ are defined by the $| \Delta \rho |$ forms in \eqref{eq:Pi_I_nu_conv}-\eqref{eq:Pi_I_nu_alpha_conv} \citep{Incropera_DeWitt_2002}. For variations in salinity $\mathsf{S}$ [-], applying $\Delta \rho \simeq \rho \beta' \Delta \mathsf{S}$ gives a salinity Rayleigh number $Rs = {g d^3 \beta' |\Delta S|}/{ \nu \alpha}$, where $\beta'$ is a haline contraction coefficient [-]
  \citep{Baines_Gill_1969, Turner_1974}. 

For many processes, more comprehensive formulations of the groups in \eqref{eq:Pi_convection}-\eqref{eq:Pi_I_nu_alpha_conv} may be necessary. For example, in heat transfer systems there may be multiple driving temperatures in $\hatPi_{h_Q/\alpha}$ \citep{Cengel_etal_2012}, while in chemical systems it may be necessary to introduce chemical activities into $\hatPi_{h_c/\Dc}$ or $Sh_c$ \citep{Bosworth_1956}. 
Unsteady convection processes require an extended analysis with different length and time scales \citep{Bosworth_1956}. 
Double diffusion -- such as of heat and salinity -- can induce entropy-producing instabilities, analysed by both $Ra$ and $Rs$ or a composite group \citep{Baines_Gill_1969, Turner_1974, Huppert_Turner_1981}.
For convection with chemical reaction, it is necessary to consider the three-way competition between diffusion, convection and chemical reaction processes \citep{Fogler_1992}; this may require the synthesis of groups from \eqref{eq:Pi_reaction_EP} or \eqref{eq:Pi_reaction_fluxes} with those in \eqref{eq:Pi_convection}. With thermodynamic cross-phenomena (\S\ref{sect:cross-diffusion}), it is necessary to include the total diffusive fluxes \eqref{eq:Onsager} in \eqref{eq:Pi_convection} rather than those based on individual mechanisms. 

The above analyses lead to a plethora of entropic groups for heat and mass transfer, several of which are examined in appendix \ref{sect:apx_plethora}. Analogous groups can also be defined for the convection of charge. 

\subsubsection{\it Hydrodynamic dispersion} \label{sect:hydrodynamic}

For flow in porous media such as groundwater flow or in a packed bed, the flow regime is usually not turbulent except under high hydraulic gradients, inhibiting the occurrence of turbulent dispersion. However, other mixing mechanisms arise due to the presence of the porous medium. These are commonly classified as follows:
\begin{enumerate}
\item {\it Molecular diffusion}, which occurs due to the random motions of molecules. This is generally represented by Fick's law \eqref{eq:Fick1}, but corrected to account for blocking by solid particles  \citep{Domenico_Schwartz_1998, Fetter_1999}:
\begin{align}
\j_c  &= -\Dc^* \vnabla C_c, \hspace{10pt} \text{with } \hspace{10pt} \Dc^*=\omega \Dc  \YY
\label{eq:molec_diff}
\end{align}
where $\Dc^*$ is the bulk molecular diffusion coefficient for the $c$th species [m$^2$ s$^{-1}$] and $\omega < 1$ is a correction factor [-]. Some authors correlate $\omega \approx \epsilon \varphi/ \vartheta$, where $\epsilon$ is the porosity, $\varphi$ is a pore constriction factor and $\vartheta$ is the tortuosity of the porous medium \citep{Fogler_1992}.  

\item {\it Mechanical dispersion}, which occurs due to the physical motion of fluid around solid obstacles, causing spreading in both the longitudinal and transverse directions (respectively, aligned with and normal to the flow direction) \citep{Domenico_Schwartz_1998, Fetter_1999}.  This can be identified with the early onset of inertial dispersion in a porous medium, as evidenced by an enhanced pressure loss and its dependence on a $U^2$ term \citep{Dullien_1975, Churchill_1988, Niven_2002}.  
Mechanical dispersion is generally represented by Fick's laws in each direction, with the dispersion coefficient further correlated with the fluid velocity \citep{Furbish_1997, Domenico_Schwartz_1998, Fetter_1999}:
\begin{align}
\begin{split}
j_{c,\imath}  &= -\mathsf{D}_{c,\imath}^m \dfrac{\partial C_c}{\partial \imath}, 
\hspace{10pt} \text{with } \hspace{10pt} \mathsf{D}_{c,\imath}^m = U \mathsf{a}_\imath  \YY
\end{split}
\label{eq:mech_disp}
\end{align}
where, for the $\imath$th direction, $\mathsf{D}_{c,\imath}^m$ is the mechanical dispersion coefficient for the $c$th species [m$^2$ s$^{-1}$], $\mathsf{a}_\imath$ is the dispersivity [m] and $U$ is the superficial fluid velocity [m s$^{-1}$].  These are commonly defined for the longitudinal and transverse directions $\imath \in \{L,T\}$ or three-dimensional space $\imath \in \{L,T_1,T_2\}$, with 
$\mathsf{a}_L > \{ \mathsf{a}_{T_1}, \mathsf{a}_{T_2} \}$. 
The dispersivities $\mathsf{a}_\imath$ are not constant, but increase with the measurement scale \citep{Domenico_Schwartz_1998, Fetter_1999}. 
\end{enumerate}

Collectively the two spreading processes are termed {\it hydrodynamic dispersion}, represented by a hydrodynamic dispersion coefficient $\mathsf{D}_{c,\imath}$ [m$^2$ s$^{-1}$] for the $c$th species in the $\imath$th direction \citep{Furbish_1997, Domenico_Schwartz_1998, Fetter_1999, Nazaroff_Alvarez-Cohen_2001}:
\begin{align}
\begin{split}
\mathsf{D}_{c,\imath} = \mathsf{D}_{c,\imath}^m + \Dc^* = U \mathsf{a}_\imath  + \Dc^* \YY
\end{split}
\label{eq:hydrodynam_disp_chem}
\end{align}
with an analogous relation for the $k$th charged species. For heat, it is necessary to account for conduction through the solid $s$ and fluid $f$ \citep{Furbish_1997, Domenico_Schwartz_1998}:
\begin{align}
\begin{aligned}
\mathfrak{a}_{\imath} &= \mathfrak{a}_{\imath}^m + \alpha^* = U \mathsf{a}_\imath  + \alpha^*, 
\hspace{10pt}  \text{with } \hspace{10pt}  
\alpha^*= \epsilon \alpha_f + (1-\epsilon) \alpha_s \YY 
\end{aligned}
\label{eq:hydrodynam_disp_heat}
\end{align}
where $\alpha^*$, $\mathfrak{a}_{\imath}^m$ and $\mathfrak{a}_{\imath}$ are the bulk diffusion, mechanical and hydrodynamic dispersion coefficients for heat  [m$^2$ s$^{-1}$]. 
Eqs.\ \eqref{eq:hydrodynam_disp_chem}-\eqref{eq:hydrodynam_disp_heat} can be incorporated into advection-dispersion equations for the modelling of contaminant, charge or heat transport in porous media, in the last case with forced or free convection \citep{Domenico_Schwartz_1998} (\S\ref{sect:conv}).

Applying entropic similarity, the relative importance of mechanical dispersion and diffusion can be represented by the  P\'eclet numbers \citep[e.g.,][]{Bear_Bachmat_1991, Furbish_1997}:
\begin{align}
\begin{split}
\hatPi_{\mathsf{D}_{c,\imath}} 
&= \frac{{\hat{\dot{\sigma}}_{\mathsf{D}_{c,\imath}^m}} }{{\hat{\dot{\sigma}}_{\Dc^*}}} 
\to Pe_{\mathsf{D}_{c,\imath}} = \dfrac{\mathsf{D}_{c,\imath}^m}{\Dc^*} = \dfrac{U \mathsf{a}_\imath}{\Dc^*},\YY
\hspace{20pt}
\hatPi_{\mathfrak{a}_{\imath}} 
= \frac{{\hat{\dot{\sigma}}_{\mathfrak{a}_{\imath}^m}} }{{\hat{\dot{\sigma}}_{\alpha^*}}} 
\to Pe_{\mathfrak{a}_{\imath}} = \dfrac{\mathfrak{a}_{\imath}^m}{\alpha^*} = \dfrac{U \mathsf{a}_\imath}{\alpha^*},
\\
\hatPi_{\mathsf{D}_{k,\imath}} 
&= \frac{{{{\hat{\dot{\sigma}}_{\mathsf{D}_{k,\imath}^m}}}} }{{\hat{\dot{\sigma}}_{\Dkstar}}} 
\to Pe_{\mathsf{D}_{k,\imath}} = \dfrac{\mathsf{D}_{k,\imath}^m}{\Dkstar} = \dfrac{U \mathsf{a}_\imath}{\Dkstar} \Y
\end{split}
\label{eq:Peclet_hydrodynam}
\end{align}
These can also be defined using total hydrodynamic dispersion coefficients such as $\mathsf{D}_{c,\imath}/\Dc^*$, 
or as ratios of inertial and diffusion terms $U \ell/\Dc^*$
where $\ell$ is a length scale [m] 
\citep{Furbish_1997, Domenico_Schwartz_1998, Fetter_1999}. 
Some authors use the interstitial velocity $U/\epsilon$ \citep{Fogler_1992}. 
For contaminant migration in clay soils, generally $Pe_{\mathsf{D}_{c,\imath}} \ll1$, dominated by diffusion, while for sands and gravels $Pe_{\mathsf{D}_{c,\imath}} \gg 1$, dominated by mechanical dispersion. 
Additional groups can be defined for competition with chemical reactions, analogous to \eqref{eq:Pi_reaction}-\eqref{eq:Pi_reaction_EP}, or entropy fluxes, analogous to those in \eqref{eq:Pi_diffusion_fluxes}.

\subsubsection{\it Shear-flow dispersion} \label{sect:shear_disp}

An additional mechanism of mixing in internal and open channel flows is {\it shear-flow dispersion}, arising from the difference in fluid velocities between the centreline and solid walls \citep{Fischer_1979, French_1985, Streeter_etal_1998, Nazaroff_Alvarez-Cohen_2001}. Employing a Reynolds decomposition $a = \langle a \rangle + a''$ based on the cross-sectional average $\langle a \rangle$ and deviation $a''$ rather than a temporal decomposition, this is correlated as:
\begin{align}
\langle j_{c,x, \text{shear}} \rangle
= \langle u'' C_c'' \rangle 
\approx - K \dfrac{\partial \langle {C_c} \rangle}{\partial x} \YY
\label{eq:shear_disp}
\end{align}
where, for the $c$th species and downstream direction $x$, 
$\langle j_{c,x, \text{shear}} \rangle$ is the mean flux relative to the flow [mol m$^{-2}$ s$^{-1}$], 
$\langle u'' C_c'' \rangle$ is the mean Reynolds flux [mol m$^{-2}$ s$^{-1}$], 
$K$ is the shear dispersion coefficient [m$^2$ s$^{-1}$] 
and $\langle C_c \rangle$ is the mean concentration [mol m$^{-3}$]. 
%
Applying entropic similarity using \eqref{eq:shear_disp} and cross-sectional averages of \eqref{eq:fluxS2}
or \eqref{eq:fluxes_turb} yields: 
\begin{align}
\hatPi_{K/\Dc} 
= \dfrac{\langle \hat{\dot{\sigma}}_{K} \rangle}{\langle \hat{\dot{\sigma}}_{\Dc} \rangle}
\to  \dfrac{K}{\Dc},\Y
\hspace{20pt}
\hatPi_{K/\Dct} 
&= \dfrac{\langle \hat{\dot{\sigma}}_{K} \rangle}{\langle {\hat{\dot{\sigma}}_{\Dct}} \rangle} 
\to  \dfrac{K}{\Dct} \Y
\label{eq:Pi_shear_disp}
\end{align}
In many natural water bodies such as rivers and estuaries $K \gg \Dct \gg \Dc$, whence $\hatPi_{K/\Dc} \gg \hatPi_{K/\Dct} \gg 1$ \citep{Fischer_1979, Streeter_etal_1998}.


\subsubsection{\label{sect:dispersed} \it Dispersion of bubbles, drops and particles}

Consider a system containing a dispersed phase composed of bubbles, drops or solid particles of density $\rho_d$ [kg m$^{-3}$] and length scale $d$ [m] within a continuous fluid of density $\rho_c$ [kg m$^{-3}$] and kinematic viscosity $\nu_c$ [m$^2$ s$^{-1}$].
From the analysis of convection \S\ref{sect:conv}, the difference in densities creates a buoyancy-driven inertia between the phases, represented by the velocity scale $U_{\disp} = \sqrt{g d |\Delta \rho|/ \rho_c}$ [m s$^{-1}$], where $|\Delta \rho| = |\rho_d-\rho_c |$. For external flow around dispersed phase particles with drag coefficient $C_D$, comparison of the entropy production by intrinsic to external inertial dispersion \eqref{eq:EP_ext_inertial}, or by viscous diffusion \eqref{eq:EP_ext_lam}, gives:
\begin{align}
\begin{aligned}
\Pi_{\text{ext},I/\disp, I} 
&=  \dfrac{ \dot{\sigma}_{\text{ext},I} }{\dot{\sigma}_{\disp,I} } 
= Fr_{\disp}^3 \Y
&\text{with } 
Fr_{\disp} 
&=  \dfrac{ U}{U_{\disp}}
=  \dfrac{ U}{ \sqrt{g d |\Delta \rho|/ \rho_c} } \Y
\\
\Pi_{\disp,I/\text{ext},\nu_c} 
&=  \biggl( \frac{ \dot{\sigma}_{\disp,I} }{ \dot{\sigma}_{\text{ext},\nu_c}}\biggr)^2
\sim C_D^2 Ar Fr_{\disp}^{-4} \Y
%
&\text{with } \hspace{12pt}
Ar 
&= \biggl( \frac{U_{\disp} d}{\nu_c} \biggr)^2 
= \frac{g d^3 |\Delta \rho|}{\rho_c \nu_c^2}  \Y
\end{aligned}
\label{eq:Pi_Frdisp_Ar}
\end{align}
where $Fr_{\disp}$ is the densimetric particle Froude number 
and $Ar$ is the Archimedes number 
\citep{Clift_etal_1978, Pavlov_etal_1979, Churchill_1988, Sturm_2001}. As evident, $Fr_{\disp}$ is the ratio of inertial dispersion by external to intrinsic sources, traditionally interpreted by dynamic similarity as the ratio of inertial to buoyancy forces. 
$Ar$ is of the same form as $Gr$ \eqref{eq:Pi_I_nu_conv}, 
traditionally interpreted as buoyancy relative to viscous forces.  
Comparing \eqref{eq:Pi_I_nu_alpha_conv}, we identify $Fr_{\disp}^2= Re^2/Ar \Y=Re Pe_{\alpha}/Ra \Y = Ri^{-1} \Y$, for $Ra$ and $Ri$ now defined using $U_{\disp}$. 
The above groups -- often written in terms of the friction velocity $u^*$ 
(\S\ref{sect:inertial}) instead of $U$ -- are widely used for the analysis of dispersed phase entrainment, transport and sediment bed forms \citep[e.g.,][]{Shields_1936, Henderson_1966, Yalin_1977, Yang_1996}. 

The vertical distribution of sediment in an internal or channel flow is often modelled by a Reynolds-averaged advection-dispersion equation \citep{Yang_1996, Raudkivi_1990, Sturm_2001}:
\begin{align}
 \dfrac{\partial \overline {\mathsf{C}_s} }{\partial t} 
 =   w_s \frac{\partial \overline{\Cu_s}}{\partial z}  - \dfrac{\partial }{\partial z} \overline{w' \Cu_s'}
 \approx w_s  \frac{\partial \overline{\Cu_s}}{\partial z} + \dfrac{\partial }{\partial z} \biggl( \Dst \dfrac{\partial \overline {\Cu_s} }{\partial z} \biggr) \Y
\label{eq:sediment_disp}
\end{align}
where $\Cu_s$ is the sediment concentration [kg m$^{-3}$], 
$w_s$ is the settling velocity [m s$^{-1}$] and
$\Dst$ is the eddy dispersion coefficient [m$^2$ s$^{-1}$].
Integrating \eqref{eq:sediment_disp} with ${\partial \overline {\Cu_s} }/{\partial t} =0$ and $\Dst \sim \nu_t$ gives the \cite{Rouse_1937} equation for the equilibrium sediment distribution \citep{Yang_1996, Raudkivi_1990, Sturm_2001}.
Now compare the inertial entropy production per particle \eqref{eq:EP_ext_inertial} for $U=w_s$, multiplied by the mean particle number density $\overline{n_s}=6 \overline{\Cu_s}/\rho_s \pi d_s^3 \Y$, where $\rho_s$ is the solid density [kg m$^{-3}$] and $d_s$ is the solid diameter [m], to the entropy production by sediment dispersion, which from \eqref{eq:hatsigma_c2_approx} can be written as $\overline{\hat{\dot{\sigma}}_{\Dst}} = R^* D_{s,t} (\partial \overline{\Cu_s}/\partial z)^2/ \overline{\Cu_s} \Y$, where $R^*$ is the specific gas constant [J K$^{-1}$ kg$^{-1}$]. This gives the hybrid entropic group:
\begin{align}
\hatPi_s 
&= \dfrac{ \overline{n_s} \, {{\dot{\sigma}}_{\text{ext},I}} }{\overline{\hat{\dot{\sigma}}_{\Dst}}} 
=\dfrac{3 C_D w_{s}^{3} d_{s} }{4   R^* T \D_{s,t}} \,
\dfrac{\rho_c}{\rho_s} \,
\dfrac{\overline{\Cu_{s}}^{2}}{ ({\partial \overline{\Cu_s}}/{\partial z})^{2} \, d_{s}^2} \Y
\label{eq:Pi_hydrid_sediment}
\end{align}
This is very different to simplified groups for the competition between settling and dispersion, such as a turbulent sediment P\'eclet number $Pe_s = {w_s d_s}/{\Dst}$.

For drops or bubbles with surface or interfacial tension $\tension$ [J m$^{-2}$], 
the external or intrinsic rates of entropy production needed to maintain the dispersed phase are:
\begin{align}
\dot{\sigma}_{\text{ext}, \tension} 
= \frac{A_d \tension}{T \theta_{\text{ext}}}  
\sim \frac{ \tension U d}{T},
\hspace{20pt}
\dot{\sigma}_{\disp, \tension} 
= \frac{A_d \tension}{T \theta_{\disp}}  
\sim \frac{  \tension U_{\disp} d }{T}
\label{eq:EP_gamma}
\end{align}
defined for two choices of time scale $\theta_{\text{ext}} \sim d/U$ or $\theta_{\disp} \sim d/U_{\disp}$ [s], where $A_d \sim d^2$ is the surface area [m${^2}$].
Applying entropic similarity to combinations of the intrinsic or external inertial dispersion \eqref{eq:EP_ext_inertial}, viscous diffusion \eqref{eq:EP_ext_lam} and tension \eqref{eq:EP_gamma} gives the groups:
\begin{align}
\Pi_{\text{ext},I/\text{ext},\tension} 
&=  \biggl( \frac{ \dot{\sigma}_{\text{ext},I} }{ \dot{\sigma}_{\text{ext},\tension}}\biggr)
\sim C_D We
\hspace{10pt} \text{with } \hspace{10pt}
We = \dfrac{  { \rho_c U^2 d} }{ { \tension  } }  = Eo Fr_{\disp}^2  \Y
\\
\Pi_{\disp,I/\disp,\tension} 
&=  \biggl( \frac{ \dot{\sigma}_{\disp,I} }{ \dot{\sigma}_{\disp,\tension}}\biggr)
\sim C_D Eo
\hspace{10pt} \text{with } \hspace{10pt}
Eo = \dfrac{  { \rho_c U_{\disp}^2 d} }{ { \tension  } }
= \dfrac{  {  g d^2  |\Delta \rho|} }{ {   \tension  } } \Y
\label{eq:Pi_disp_tension}
\\
\Pi_{\text{ext}, \nu_c/\text{ext}, \tension} 
&=  \biggl( \frac{ \dot{\sigma}_{\text{ext}, \nu_c} }{ \dot{\sigma}_{\text{ext}, \tension}} \biggr)
\sim Ca = \dfrac{  { \rho_c \nu_c U} } {{  \tension  }  } \Y
\end{align}
where $We$ is the Weber number, $Eo$ is the E\"otv\"os or Bond number and $Ca$ is the capillary number. These and the Morton number $M=g \nu_c^4 \rho_c^2 |\Delta \rho| / \tension^3 = Eo^3/Ar^2$ are widely used to characterise bubble and droplet shapes, flow regimes and their entrapment in porous media \citep{Clift_etal_1978, Fetter_1999}. 
%



\subsection{\label{sect:wave} Wave motion and information-theoretic flow regimes}

To complete this survey of entropy-producing processes, it is necessary to examine wave motion.  A {\it wave} can be defined as an oscillatory process that facilitates the transfer of energy through a medium or free space.  Generally, this is governed by the wave equation:
\begin{align}
\frac{\partial^2 \phi}{\partial t^2} &= c^2 \vnabla^2 \phi	
\label{eq:wave}	 
\end{align}
where $\phi$ is a displacement parameter and $c$ is a characteristic velocity or celerity [m s$^{-1}$]\footnote{Due to common practice, \S\ref{sect:wave} makes several excursions from the notation of previous sections.}. Wave motion, in its own right, does not produce entropy, although wave interactions with materials or boundaries can be dissipative in some situations. 
However, a wave is also a {\it carrier of information}, communicating the existence and strength of a disturbance or source of energy. 
Its celerity therefore provides an intrinsic velocity scale for the rate of transport of information $U_{\signal} = c$ through the medium. 
For a fluid flow with local velocity $\u$, the celerity provides a threshold between two different information-theoretic flow regimes, governed by processes which can ($||\u||<c$) or cannot ($||\u||>c$) be influenced by downstream disturbances. 
Adopting the information-theoretic formulation of similarity \eqref{eq:Pi_infm}, this can be represented by the local and macroscopic dimensionless groups:
\begin{align}
\hat{{\Pi}}_{\info}(\n) = \dfrac{\u \cdot \n}{c} \xrightarrow{\text{max}} \hat{{\Pi}}_{\info} = \dfrac{||\u||}{c},
\hspace{20pt} 
\Pi_{\info} = \dfrac{U}{c}
\label{eq:Pi_infm2}
\end{align}
where $\n$ is a given unit normal. The first group gives a local definition, 
while the second group gives a summary criterion in terms of a representative velocity $U$ [m s$^{-1}$]. 
In some situations, a sharp junction can be formed between the two information-theoretic flow regimes (e.g., a shock wave or hydraulic jump), with a high rate of entropy production. Wave-carrying flows are also subject to friction, with distinct differences between the two flow regimes.

We examine several types of waves from this perspective.

\subsubsection{\it Acoustic waves} \label{sect:acoustic}

An acoustic or sound wave carries energy through a material by longitudinal 
compression and decompression at the sonic velocity $a = \sqrt{{d p}/{d \rho}} = \sqrt{{K}/{\rho}}$ [m s$^{-1}$], 
where $p$ is the pressure [Pa] and $K$ is the bulk modulus of elasticity of the material [Pa].
For isentropic (adiabatic and reversible) changes in an ideal gas, this reduces to $a = \sqrt{\gamma p /\rho} = \sqrt{\gamma R^* T}$, where $\gamma$ is the adiabatic index [-] and $R^*$ is the specific gas constant [J K$^{-1}$ kg$^{-1}$].
By information-theoretic similarity \eqref{eq:Pi_infm2}, this defines the local and macroscopic Mach numbers:
\begin{align}
\hat{{\Pi}}_{a} =\hat{{M}} 
= \dfrac{|| \u ||}{a} 
\xrightarrow[\text{isentropic}]{\text{ideal gas}} \dfrac{|| \u ||}{\sqrt{\gamma R^* T}}, \YY
\hspace{20pt} 
\Pi_{a} = M_{\infty} 
= \dfrac{U_{\infty}}{a_{\infty}}  
\xrightarrow[\text{isentropic}]{\text{ideal gas}} \dfrac{U_{\infty}}{\sqrt{\gamma R^* T_{\infty}}} \YY
\label{eq:Pi_Mach}
\end{align}
where $U_{\infty}$ is the free-stream fluid velocity [m s$^{-1}$] and $T_{\infty}$ is the free-stream temperature [K] \citep{vonMises_1958, Pao_1961, White_1986, Street_etal_1996, Anderson_2001}.
These groups discriminate between two flow regimes:
\begin{enumerate}
\item {\it Subsonic flow} (locally $\hat{{M}}<1$ or summarily $M_{\infty} \lesssim 0.8$), subject to the influence of the downstream pressure, of lower $\u$ and often of higher $p$, $\rho$, $T$ and $s$; and  
\item {\it Supersonic flow} (locally $\hat{{M}}>1$ or summarily $M_{\infty} \gtrsim 1.2$), which cannot be influenced by the downstream pressure, of higher $\u$ and often of lower $p$, $\rho$, $T$ and $s$.
\end{enumerate}

Locally $\hat{{M}} = 1$ is termed {\it sonic flow}, while summarily $0.8 \lesssim M_{\infty} \lesssim 1.2$ indicates {\it transonic flow} and $M_{\infty} \gtrsim 5$ {\it hypersonic flow} \citep{Anderson_2001}.
Instead of \eqref{eq:Pi_Mach}, some authors use the Cauchy number $C_{\infty} = M_{\infty}^2 = \rho U_{\infty}^2 / K$, traditionally interpreted by dynamic similarity as the ratio of inertial to elastic forces \citep{Pao_1961, Street_etal_1996}. 

In many compressible flows it is possible to effect a smooth, isentropic transition between subsonic and supersonic flow (or vice versa) using a nozzle or diffuser, described as a {\it choke} \citep{Pao_1961, White_1986, Street_etal_1996}. However, the transition from supersonic to subsonic flow is often manifested as a {\it normal shock wave}, a sharp boundary normal to the flow with discontinuities in $\u$, $p$, $\rho$, $T$ and $s$ \citep{Anderson_2001}. From the entropy production \eqref{eq:EPdef_int} at steady state:
\begin{align}
\dot{\sigma}_{\steady}
=  \oiint\limits_{CS} \J_S \cdot \vec{n} dA  =  \oiint\limits_{CS} (\j_S + \rho s \u )\cdot \vec{n} dA \ge 0  \YY
\label{eq:EPdef_int_stst} 
\end{align}
Adopting a control volume for a normal shock wave of narrow thickness, with inflow 1, outflow 2 and no non-fluid entropy fluxes ($\j_S=0$), 
\eqref{eq:EPdef_int_stst} gives the entropy production per unit area across the shock 
[J K$^{-1}$ m$^{-2}$ s$^{-1}$] 
\citep[c.f.,][]{Niven_Noack_2014}:
\begin{align}
{\breve{\dot{\sigma}}}_{\shock} 
= \Delta (\rho s \u) \cdot \vec{n} 
= \rho_2 s_2 u_2 - \rho_1 s_1 u_1 \Y
\label{eq:EP_shock} 
\end{align}
Using relations for $u$, $p$, $\rho$ and $T$ derived from the conservation of fluid mass, momentum and energy for  inviscid adiabatic steady-state flow across the shock \citep{Rankine_1870, Hugoniot_1887, Hugoniot_1889, Shapiro_1953, vonMises_1958, Pao_1961, Zeldovich_R_1968, Daneshyar_1976, Churchill_1980, White_1986, Anderson_2001, Greitzer_etal_2004, Douglas_etal_2011}, \eqref{eq:EP_shock} can be rescaled by the internal entropy flux to give the entropic group:
\begin{align}
\begin{split}
\breve{\Pi}_{\shock}
&= \dfrac{{\breve{\dot{\sigma}}}_{\shock}}{\rho_1 c_p u_1}
= \dfrac{s_2-s_1}{c_p} \Y
=  \ln \biggl( \frac{(2+(\gamma -1) \hat{M}_{1}^{2})}{(\gamma +1) \hat{M}_{1}^{2}} \biggr)
+ \dfrac{1}{\gamma} \ln \biggl(\frac{2 \gamma  \hat{M}_{1}^{2}-\gamma +1}{\gamma +1}\biggr) \YY
\\
\text{with  } \hspace{10pt}
\hat{M}_2  &=\sqrt{\frac{2+(\gamma -1) \hat{M}_{1}^{2}}{2 \gamma  \hat{M}_{1}^{2}-\gamma +1}} \YY
\end{split}
\label{eq:Pi_shock} 
\raisetag{30pt}
\end{align}
From \eqref{eq:Pi_shock}, $\breve{\Pi}_{\shock} > 0$ and ${\breve{\dot{\sigma}}}_{\shock} > 0$ for $\hat{M}_1 > 1$ and $\hat{M}_2 < 1$, allowing the formation of an entropy-producing normal shock in the transition from supersonic to subsonic flow. However, $\breve{\Pi}_{\shock} < 0$ and ${\breve{\dot{\sigma}}}_{\shock} < 0$ for $\hat{M}_1 < 1$ and $\hat{M}_2 > 1$, so the formation of a normal shock in the transition from subsonic to supersonic flow (a rarefaction shock $p_2<p_1$) is prohibited by the second law \eqref{eq:EPdef_int} \citep[c.f.,][]{Hirschfelder_etal_1954, vonMises_1958, Pao_1961, Zeldovich_R_1968, White_1986, Anderson_2001}. 

By inwards deflection of supersonic flow (a concave corner), it is also possible to form an {\it oblique shock wave}, a sharp transition to a different supersonic or subsonic flow with increasing $p$, $\rho$ and $T$ \citep{Anderson_2001, Douglas_etal_2011}. This satisfies the same relations and entropy production \eqref{eq:EP_shock}-\eqref{eq:Pi_shock} as a normal shock, but written in terms of velocity components $u_1$ and $u_2$ normal to the shock, thus with normal Mach components $\hat{M}_{n1}  = \hat{M}_1 \sin \beta$ and $\hat{M}_{n2}  = \hat{M}_2 \sin(\beta - \theta)$, 
where
$\beta$ is the shock wave angle  
and $\theta$ is the deflection angle. 
From the second law $\breve{\Pi}_{\shock} > 0$, an oblique shock wave is permissible for $\hat{M}_1>\hat{M}_2$,  
in general with a supersonic transition Mach number, 
and either two or no $\beta$ solutions depending on $\theta$ \citep{Anderson_2001}. 
In contrast, outwards deflection of supersonic flow (a convex corner) creates an {\it expansion fan}, a continuous isentropic transition $\hat{M}_2 > \hat{M}_1$
with decreasing $p$, $\rho$ and $T$ \citep{White_1986, Anderson_2001, Douglas_etal_2011}. 

This topic also highlights the confusion in the aerodynamics literature between the specific entropy and the local entropy production. For a non-equilibrium flow system, the second law is defined exclusively by ${\dot{\sigma}} \ge 0$  \eqref{eq:EPdef_int}, reducing for a sudden transition to ${\breve{\dot{\sigma}}}_{\shock} \ge 0$ \eqref{eq:EP_shock}, while the change in specific entropy can (in principle) take any sign $\Delta s = s_2 -s_1 \lesseqqgtr 0$. From the above analysis, $\Delta s \ge 0$ implies ${\breve{\dot{\sigma}}}_{\shock} \ge 0$ only for flow transitions satisfying \eqref{eq:EP_shock} and local continuity $\rho_1 u_1 = \rho_2 u_2$; these include normal and oblique shocks. For transitions involving a change in fluid flux, or if there are non-fluid entropy fluxes in \eqref{eq:EP_shock}, $\Delta s$ and ${\breve{\dot{\sigma}}}_{\shock}$ can have different signs. Similarly, an isentropic process $\Delta s=0$ need not indicate zero entropy production ${\breve{\dot{\sigma}}}_{\shock}=0$. The above analyses are also complicated by fluid turbulence (\S\ref{sect:turb}), which generates additional Reynolds entropy flux terms in \eqref{eq:EP_shock}-\eqref{eq:Pi_shock}  \citep{Niven_Noack_2014} (see \S\ref{sect:turb}).

For frictional compressible internal flow at steady state, the local entropy production is $\hat{\dot{\sigma}} = \vnabla \cdot \J_{S}$ \eqref{eq:EPdef_local}. For one-dimensional adiabatic flow of an ideal gas without chemical or charge diffusion, the conservation of fluid mass, momentum and energy with friction \eqref{eq:DW} \citep{Shapiro_1953, White_1986, Douglas_etal_2011} with scaling gives the local entropic group:
\begin{align}
\begin{split}
\hat{\Pi}_{\compr}(x)
&= \dfrac{\hat{\dot{\sigma}}(x) \, \ell}{\rho_a c_p a} \Y
=  \dfrac{\ell}{\rho_a c_p a} \dfrac{d \bigl(\rho(x) s(x) u(x) \bigr)}{d x} \Y
=  \dfrac{\ell}{c_p} \dfrac{d s(x)}{d x} \Y
\\&= \frac{2 \ell \, (\gamma -1)  ( 1-\hat{M}(x)^{2})}{\gamma  \bigl(2+  (\gamma -1)\hat{M}(x)^{2} \bigr) \, \hat{M}(x)} \dfrac{d \hat{M}(x)}{d x} \Y
= \ell \, \Theta(x) \, \dfrac{d \hat{M}(x)}{d x} 
= \tfrac{1}{2} { (\gamma-1) f \hat{M}(x)^2} \Y
\end{split} 
\label{eq:Pi_compr}
\raisetag{45pt}
\end{align}
where $x$ is the flow coordinate [m], $\ell$ is a pipe length scale [m], $f$ is the Darcy friction factor [-] \eqref{eq:DW} and $\rho_a$ is the fluid density at $\hat{M}=1$ [kg m$^{-3}$].  Note the different roles of the specific entropy and the local entropy production. Analysis of the group $\Theta(x)$ in \eqref{eq:Pi_compr} for $\gamma>1$ reveals the following effects of friction:
\begin{enumerate}
\item The last term in \eqref{eq:Pi_compr} is positive for all $\hat{M}>0$, hence $\hat{\Pi}_{\compr}>0$ and $\hat{\dot{\sigma}}>0$; i.e., the entropy production cannot be zero for finite flow.
\item For subsonic flow $\hat{M}<1$, $\Theta>0$, hence the second law $\hat{\Pi}_{\compr} > 0$ or $\hat{\dot{\sigma}} > 0$ implies ${d \hat{M}}/{d x} > 0$, and so $\hat{M}$ will increase with $x$ towards $\hat{M} = 1$;
\item For supersonic flow $\hat{M}>1$, $\Theta<0$, hence the second law $\hat{\Pi}_{\compr} > 0$ or $\hat{\dot{\sigma}} > 0$ implies ${d \hat{M}}/{d x} < 0$, and so $\hat{M}$ will decrease with $x$ towards $\hat{M} = 1$;
\item In both cases, the second law $\hat{\Pi}_{\compr} > 0$ or $\hat{\dot{\sigma}} > 0$ implies ${d s}/{d x} > 0$, so the specific entropy $s$ will increase with $x$ towards $\hat{M}=1$.  Integrating \eqref{eq:Pi_compr}, this terminates at the maximum specific entropy $s_a$;
\item In the sonic limit $\hat{M} \to 1^\mp$, $\Theta \to 0$ and $d\hat{M}/dx \to \pm \infty$, but these limits combine to give $ \lim_{\hat{M} \to 1} \hat{\Pi}_{\compr} = \tfrac{1}{2} (\gamma -1)f > 0$ from either direction.
\end{enumerate}

The sonic point $x=L^*$ and $M(x)$ can be calculated numerically from the integrated friction equation \citep{Shapiro_1953, Pao_1961, White_1986, Greitzer_etal_2004, Douglas_etal_2011}:
\begin{align}
\begin{split}
\frac{f (L^*-x)}{\ell}=\dfrac{\gamma +1}{2 \gamma} \ln \dfrac{ (\gamma +1 )\hat{M}(x)^{2}}{ 2+(\gamma -1)\hat{M}(x)^{2}} 
+ \dfrac{1-\hat{M}(x)^{2}}{\gamma \hat{M}(x)^{2}} \YY
\end{split}
\label{eq:f_compr}
\end{align}
Flows in conduits longer than $L^*$ undergo {\it frictional choking}, producing a lower subsonic entry Mach number or 
supersonic flow with a normal shock, so that the flow exits at $\hat{M}=1$ \citep{Shapiro_1953, White_1986}. 
Clearly, such flows are controlled by their entropy production: since they are adiabatic, they cannot export heat, so each fluid element can only achieve a positive local entropy production $\hat{\dot{\sigma}}>0$ by increasing its specific entropy $s$ in \eqref{eq:Pi_compr}, via permissible changes in $p$ and $T$. When $s$ reaches its maximum, no solution to \eqref{eq:Pi_compr} with $ds/dx>0$ is physically realisable, to enable a positive entropy production. 
This triggers unsteady flow to create the choke. 
For isothermal flows, flows with heat fluxes, other non-fluid entropy fluxes or chemical reactions, extensions of \eqref{eq:Pi_compr}-\eqref{eq:f_compr} are required \citep{Shapiro_1953, Pao_1961, White_1986}.

For frictional compressible external flow, the entropy production  \eqref{eq:EP_ext_I_vect} 
due to inertial drag and lift \eqref{eq:CD_vect} is:
\begin{align}
\dot{{\sigma}}^{\text{compr}}_{\text{ext},I}
= \frac{\vec{F}_D \cdot \U}{T} 
= \frac{ \frac{1}{2} \rho A_s \, \vec{C}_D \cdot \U \, ||\U||^2 }{T} 
\label{eq:sigma_I_ext_compr}
\end{align}
where $\vec{U}$ is the velocity of the fluid relative to the solid [m s$^{-1}$]. For an airfoil, the area $A_s = \ell b$, where $\ell$ is the chord [m] and $b$ is the span [m] \citep{Shapiro_1953}. 
Eq.\ \eqref{eq:sigma_I_ext_compr} can be scaled by sonic conditions to give the entropic group:
\begin{align}
{\Pi}^{\text{compr}}_{\text{ext},I} 
=  \frac{ \dot{{\sigma}}^{\text{compr}}_{\text{ext},I} }{ \dot{\sigma}^\text{sonic}_{\text{ext},I}}
= \frac{  \rho_{\infty}   \, \vec{C}_D \cdot \vec{M} \, ||\vec{M} ||^2  / T_{\infty}}{  \rho_a  \, {C}_{Da} / T_a}
\sim  \vec{C}_D \cdot \vec{M} \, ||\vec{M} ||^2
\label{eq:Pi_I_compr}
\end{align}
where $\dot{\sigma}^\text{sonic}_{\text{ext},I}$ is the sonic inertial entropy production and $\vec{M}$ is a summary vector Mach number. 
Generally, the drag coefficient increases significantly beyond a critical Mach number $||\vec{M}_c|| \lesssim 1$, due to the local onset of supersonic flow and the formation of shock waves, and then falls to an asymptotic value with increasing $||\vec{M}|| > 1$ \citep{Pao_1961, Anderson_2001, Douglas_etal_2011}. In contrast, the lift coefficient of an airfoil exhibits a gradual rise and sudden fall over $||\vec{M}|| < ||\vec{M}_c|| < 1$, also increasing with the angle of attack \citep{Shapiro_1953}. 
From the cubic relation \eqref{eq:Pi_I_compr}, ${\Pi}^{\text{compr}}_{\text{ext},I}$ increases dramatically with $||\vec{M}||$.


\subsubsection{\it Blast waves} \label{sect:blast_waves}

For chemical combustion in a fluid or solid, the reaction is driven by a {\it combustion wave} or {\it blast wave} that moves relative to the reactants at the explosive or detonation velocity $U_{\explos}$ [m s$^{-1}$]. From \eqref{eq:Pi_infm2}, this can be scaled by the acoustic velocity within the reactants $a_R$, giving the information-theoretic group \citep{Shapiro_1953}:
\begin{align}
\Pi_{a_R} =  M_{\explos} = \frac{U_{\explos}}{a_R}
\label{eq:Mach_explos}
\end{align}
This defines an explosive Mach number, which discriminates between {\it detonation} of a high explosive for $M_{\explos}>1$ (typically $M_{\explos} \gg1)$ in a (compressive) supersonic shock front, or {\it deflagration} of a low explosive 
for $M_{\explos}<1$ in a (rarefaction) subsonic flame front \citep{Shapiro_1953, Sedov_1959, Churchill_1980}. 

Explosions in a compressible fluid can be modelled by the one-dimensional conservation equations used for a normal shock \eqref{eq:Pi_shock}, adding the reaction enthalpy and a minimum entropy assumption \citep{Shapiro_1953, Churchill_1980, Williams_2018}. This predicts alternative incoming velocities corresponding to detonation or deflagration; for the former, the outgoing combustion products are expelled at the acoustic velocity relative to the shock front. A kinetic model of detonation (ZND theory) extends this finding, with compression of the reactants at the shock front, causing ignition, heat release and acceleration of the combustion products to the choke point \citep{Zeldovich_1940, Neumann_1942, Neumann_1963, Doring_1943}. 


A large chemical, gas or nuclear explosion in the atmosphere will generate a {\it spherical shock wave}  expanding radially from the source. This provides a famous example of the use of dimensional scaling. Consider a point explosion with shock wave radius $R$ [m] governed only by the energy $E$ [J], initial density $\rho_0$ [kg m$^{-3}$] and time $t$ [s]. Dimensional reasoning gives the self-similar solution $R \propto (Et^2/\rho_0)^{1/5}$; with the conservation of mass, momentum and energy for inviscid flow,  
this yields power-law relations for $u$, $p$, $\rho$ and $T$ with time $t$ and radius $r$ \citep{Sedov_1946, Sedov_1959, Taylor_1950, Neumann_1963, Zeldovich_R_1968, Barenblatt_1996, Barenblatt_2003, Hornung_2006}. For short times, these reveal strong heating and the near-evacuation of air from the epicentre, and its accumulation behind the shock front. 

Surprisingly few authors have examined explosions from an entropic perspective \citep[e.g.,][]{Eyring_etal_1949, ByersBrown_Braithwaite_1996, Kuzmitskii_2012, Muller_2013, Muller_etal_2016, Zhang_etal_2016, Hafskjold_etal_2021}, despite its role as their driving force, and the use of minimum \citep{Scorah_1935, Duffey_1955} or maximum \citep{ByersBrown_1998} entropy closures. 
From \S\ref{sect:acoustic}, we suggest the local entropic group $\hat{\Pi}_{\text{explos}}(x)$ or $\hat{\Pi}_{\text{explos}}(r)$, extending \eqref{eq:Pi_compr}-\eqref{eq:f_compr} to include shock wave, heating and chemical reaction processes.

\subsubsection{\it Pressure waves} \label{sect:pressure}

Also related to acoustic waves is the phenomenon of {\it water hammer}, an overpressure (underpressure) wave 
in an internal flow of a liquid or gas, caused by rapid closure of a downstream (upstream) valve or pump \citep{Street_etal_1996, Douglas_etal_2011}. By reflection at the pipe ends, this causes the cyclic propagation of overpressure and underpressure waves along the pipe, commonly analysed by the method of characteristics. 
For flow of an elastic liquid in a thin-walled elastic pipe, the magnitude of the change in pressure and the acoustic velocity are, respectively \citep{Joukowsky_1900, Street_etal_1996, Douglas_etal_2011}:
\begin{align}
\Delta p = \rho a_H \Delta U, \YY
\hspace{20pt}
a_H =\sqrt{ \frac{K}{\rho} \biggl(1+\dfrac{Kd}{E\theta}(1-\nu_p^2) \biggr)^{-1} } \YY
\label{eq:hammer_p_c}
\end{align}
where $\Delta U$ is the magnitude of the change in mean velocity [m s$^{-1}$], 
$K$ is the bulk elastic modulus of the fluid [Pa], 
$E$ is the elastic modulus of the pipe [Pa],
$d$ is the pipe diameter [m],
$\theta$ is the pipe wall thickness [m] and
$\nu_p$ is Poisson's ratio for the pipe material [-].
An extended relation is available for gas flows \citep{Douglas_etal_2011}. In liquids, the underpressure wave can cause cavitation, the formation of vapour bubbles, leading to additional shock waves when these collapse at higher pressures \citep{Douglas_etal_2011}. 

Eqs.\ \eqref{eq:Pi_infm2} and \eqref{eq:hammer_p_c} suggest the information-theoretic dimensionless group:
\begin{align}
{{\Pi}}_{a_H} = \dfrac{\Delta U}{a_H} = \dfrac{\Delta p}{\rho a_H^2} 
\label{eq:Pi_hammer}
\end{align}
which can be recognised as an Euler number defined for water hammer. By frictional damping in accordance with the Darcy-Weisbach equation \eqref{eq:DW}, the pressure pulse $\Delta p$ -- hence also the wave speed $\Delta U$ and the group ${{\Pi}}_{a_H}$ -- will diminish with time.

\subsubsection{\label{sect:stress} \it Stress waves}  

Related to acoustic and pressure waves, a variety of waves can occur in solids, liquids and/or along phase boundaries due to the transport of compressive normal, transverse shear or torsional stresses generated by a sudden failure, expansion or impact. These can be divided into {\it elastic} or {\it inelastic waves}, involving reversible or irreversible solid deformation \citep{Kolsky_1963, Nowacki_1978, Hazell_2022}, and also classified into various types of {\it seismic} or {\it earthquake waves}. Stress waves can be analysed by information-theoretic constructs such as \eqref{eq:Pi_infm2} to identify the flow regime, but generally are not associated with the mean motion of a fluid, and so are not examined further here. Blast or shock waves can also be generated within a solid by an explosion or impact, as discussed in \S\ref{sect:blast_waves}. 

\subsubsection{\label{sect:grav} \it Surface gravity waves}  


On the surface of a liquid, energy can be carried by {\it gravity waves}, involving circular or elliptical rotational oscillations of the fluid in the plane normal to the surface, reducing in scale with depth. 
These can be classified as {\it standing waves}, which remain in place, or {\it progressive waves}, which move across the surface. 
Using Airy (linear) 
wave theory, the angular frequency and individual wave (phase) celerity of a two-dimensional progressive surface gravity wave are given by \citep{Henderson_1966, Dean_Dalrymple_1991, Chadwick_Morfett_1993, Kundu_Cohen_2002, Sutherland_2010}:
\begin{align}
\omega^2 = gk \tanh (ky), \YY
\hspace{10pt}
c_s =\frac{\omega}{k} 
=\sqrt{ \frac{g}{k} \tanh (k y)  } 
= \sqrt{ \frac{\lambda g}{2 \pi} \tanh \frac{2 \pi y}{\lambda}  } \YY
\label{eq:c_Airy_waves}
\end{align}
where $\omega$ is the angular frequency [s$^{-1}$], 
$k=2 \pi/\lambda$ is the angular wavenumber [m$^{-1}$], 
$\lambda$ is the wavelength [m] and $y$ is the liquid depth [m]. 
For an ambient flow with mean horizontal velocity $U$, applying information-theoretic similarity \eqref{eq:Pi_infm2} gives the generalised Froude number:
\begin{align}
\Pi_{c_s} = Fr_{s}
= \dfrac{U}{c_s}  
= \dfrac{U}{\sqrt{ \frac{g}{k} \tanh (k y)}} \Y
= \dfrac{U}{\sqrt{ \frac{\lambda g}{2 \pi} \tanh \frac{2 \pi y}{\lambda}  } } \Y
\label{eq:Pi_Froude}
\end{align}
However, due to wave interferences (beats), gravity waves generally travel in wave groups, with individual waves advancing faster than the group (termed {\it wave dispersion}). 
The group celerity, also equivalent to the speed of energy transmission \citep{Dean_Dalrymple_1991, Sutherland_2010}, and its corresponding Froude number are:
\begin{align}
c^{\group}_{s} = \frac{d\omega}{dk} \YY 
= \frac{c_s}{2} \biggl( 1 + \frac{ 2 k y }{\sinh (2 k y)} \biggr),\YY
\hspace{10pt} 
\Pi_{c_s}^{\group} = Fr^{\group}_{s}
= \dfrac{U}{c^{\group}_{s}} 
= 2 Fr_{s} \biggl( 1 + \frac{ 2 k y }{\sinh (2 k y)} \biggr)^{-1} \Y
\label{eq:Pi_surf_grav_Froude}
\end{align}
The Froude numbers in \eqref{eq:Pi_Froude}-\eqref{eq:Pi_surf_grav_Froude} are not in common use. Variants of the celerities in \eqref{eq:c_Airy_waves}-\eqref{eq:Pi_surf_grav_Froude} are available for gravity waves on the interface between two liquids \citep{Turner_1973, Kundu_Cohen_2002}.



Usually, three cases of surface gravity waves are distinguished: 
\begin{enumerate}

\item For {\it deepwater (deep liquid)} or {\it short waves}: $ky \gtrsim \pi$ or $\lambda/y \lesssim 2$ thus $\tanh (ky) \to 1$ in \eqref{eq:c_Airy_waves}, hence \citep{Henderson_1966, Dean_Dalrymple_1991, Chadwick_Morfett_1993, Kundu_Cohen_2002, Hornung_2006, Lemons_2017}:
\begin{align}
c_{\lambda} =\sqrt{ \frac{g}{k} } = \sqrt{ \frac{\lambda g}{2 \pi}  },\YY
\hspace{20pt} 
\Pi_{c_s} \to Fr_{\lambda}
= \dfrac{U}{c_{\lambda}} 
= U \sqrt{\frac{k}{g}}
= U \sqrt{\frac{2 \pi}{\lambda g}} \Y
\label{eq:deep_waves}
\end{align}
Such waves move freely by circular motions of the fluid, with little net horizontal transport. Deep waves travel in wave groups; taking the limit $\sinh (2 k y) \to 2 ky$ in \eqref{eq:Pi_surf_grav_Froude} gives the group celerity $c^{\group}_{\lambda} = \tfrac{1}{2} c_{\lambda} \YY$ and group Froude number $Fr^{\group}_{\lambda} = 2 Fr_{\lambda}  \Y$. Despite their simplicity, neither $Fr^{\group}_{\lambda}$ nor $Fr_{\lambda}$ are in common use. For wave drag on a ship, the Froude number $Fr_{\ship}=U/\sqrt{g L}$ is used, where $U$ is the ship velocity [m s$^{-1}$] and $L$ is the ship length [m] \citep{Pao_1961}.

\item For {\it transitional waves}: $\pi/10 \lesssim ky \lesssim \pi$ or $2 \lesssim \lambda/y \lesssim 20$, the wave motion is impeded by contact with the bottom, producing elliptical motions of the fluid. Such waves form in natural water bodies by the shoaling of deepwater waves as they approach the shoreline. The generalised group celerity and Froude number in \eqref{eq:Pi_surf_grav_Froude} apply. More complicated (nonlinear) wave descriptions can also be used, including {\it Stokesian waves} for $\lambda/y \lesssim 10$, a superposition of cosine wave forms, and {\it cnoidal waves} for $\lambda/y \gtrsim 10$, comprising horizontally asymmetric waveforms with pointed crests 
\citep{Henderson_1966}. 

\item For {\it shallow} or {\it long waves}: $ky \lesssim \pi/10$ or $\lambda/y \gtrsim 20$ thus $\tanh (ky) \to ky$ in \eqref{eq:c_Airy_waves}, giving \citep{Chow_1959, Pao_1961, Chadwick_Morfett_1993, Street_etal_1996,  Kundu_Cohen_2002, Hornung_2006, Sutherland_2010, Lemons_2017}:
\begin{align}
c_y =\sqrt{ g y },\YY
\hspace{20pt} 
\Pi_{c_s} \to Fr_y
= \frac{U}{\sqrt{gy}} \YY
\label{eq:shallow_waves}
\end{align}
In the same limit, $\sinh (2ky) \to 2ky$ and $Fr^{\group}_{y} \to Fr_y$ in \eqref{eq:Pi_surf_grav_Froude}, so there is no separate velocity for shallow wave groups (termed {\it non-dispersive waves}). Eq.\ \eqref{eq:shallow_waves} is applied to open channel flows with rectangular cross-section. For channels of low slope and arbitrary cross-section (of low aspect ratio), \eqref{eq:shallow_waves} is commonly generalised as \citep{Chow_1959, Pao_1961, French_1985, Chadwick_Morfett_1993}:
\begin{align}
c_{y_h} =\sqrt{ g y_h },\YY
\hspace{20pt} 
\Pi_{c_s} \to Fr_{y_h}
= \frac{U}{\sqrt{g y_h}} \YY
\label{eq:shallow_waves_gen}
\end{align}
where $y_h = A/B$ is the hydraulic mean depth [m], $A$ is the channel cross-sectional area [m$^2$] and $B$ is the channel top width [m]. For a rectangular channel, $y_h = y$. Traditionally, the Froude number in \eqref{eq:shallow_waves} or \eqref{eq:shallow_waves_gen}
is interpreted by dynamic similarity as the ratio of inertial to gravity forces \citep{Pao_1961, Chadwick_Morfett_1993, Street_etal_1996}. 
\end{enumerate}

Steady incompressible open channel flows commonly satisfy the condition for shallow waves. 
The Froude number in \eqref{eq:shallow_waves} or \eqref{eq:shallow_waves_gen}
then discriminates between two flow regimes \citep{Chow_1959, Pao_1961, Street_etal_1996}:
\begin{enumerate}
\item {\it Subcritical flow} ($Fr_y $ or $Fr_{y_h} < 1$), subject to the influence of downstream obstructions, of lower velocity $U$ and higher water height $y$; and  
\item {\it Supercritical flow} ($Fr_y $ or $Fr_{y_h} < 1$), which cannot be influenced by downstream obstructions, of higher velocity $U$ and lower water height $y$.
\end{enumerate}
Locally $Fr_y$ or $Fr_{y_h} = 1$ is termed {\it critical flow}, occurring at
the {\it critical depth} $y_c$ [m]. 

The above flow regimes can in principle be extended to flows with deepwater or transitional waves, discriminated by $Fr^{\group}_{s}$ in \eqref{eq:Pi_surf_grav_Froude}. Such flow systems may be influenced more strongly by internal gravity waves, examined in \S\ref{sect:grav_int}. 

In many open channel flows it is possible to effect a smooth transition between subcritical and supercritical flow (or vice versa) using a pinched channel (Venturi flume) or stepped bed, described as a {\it choke} \citep{Chow_1959, Henderson_1966, Chadwick_Morfett_1993}. However, the transition from supercritical to subcritical flow is often manifested as an {\it hydraulic jump}, with sharp changes in $U$ and $y$. 
For a macroscopic control volume extending across a jump in a rectangular channel, with inflow 1, outflow 2 and no non-fluid entropy fluxes, 
by the conservation of mass and momentum with energy loss \citep{Chow_1959, Henderson_1966, French_1985, White_1986, Chadwick_Morfett_1993}, the total entropy production is:
\begin{align}
\begin{split}
{\dot{\sigma}}_{\jump} &= \frac{\rho g Q \Delta E}{T} \Y=  \frac{\rho g Q}{T} \frac{(y_2-y_1)^3}{4 y_1 y_2},\Y 
\\ 
\text{with}  \hspace{10pt}
\frac{y_2}{y_1} &= \tfrac{1}{2} \Bigl(\sqrt{1 + 8 Fr_{y_1}^2}-1 \Bigr) =2 \Bigl(\sqrt{1 + 8 Fr_{y_2}^2}-1 \Bigr)^{-1} \YY
\end{split}
\label{eq:EP_jump} 
\end{align}
where 
$\Delta E$ is the loss in energy per unit weight [J N$^{-1}$ = m].
%
%
%
Scaling by the entropy flow rate gives the entropic group:
\begin{align}
\Pi_{\jump} 
= \frac{\dot{\sigma}_{\jump}}{\rho c_p Q} \Y 
= \frac{ g \Delta E}{c_p T} \Y
=  \frac{g}{c_p T} \frac{(y_2-y_1)^3}{4 y_1 y_2},\Y 
\label{eq:Pi_jump}
\end{align}
From \eqref{eq:EP_jump}-\eqref{eq:Pi_jump}, ${\Pi}_{\jump} > 0$ and ${{\dot{\sigma}}}_{\jump} > 0$ for $Fr_{y_1} > 1$ and $Fr_{y_2} < 1$, allowing the formation of an entropy-producing hydraulic jump in the transition from supercritical to subcritical flow. However, ${\Pi}_{\jump} < 0$ and ${{\dot{\sigma}}}_{\jump} < 0$ for $Fr_{y_1} < 1$ and $Fr_{y_2} > 1$, so the formation of a hydraulic jump in the transition from subcritical to supercritical flow (a reverse jump $y_2<y_1$) is prohibited by the second law\footnote{The astute reader will note the parallel wording in \S\ref{sect:acoustic} on shock waves in compressible flow.}. 


By inwards deflection of supercritical flow or by interaction with wall boundaries, it is also possible to form an {\it oblique hydraulic jump}, a sharp transition to a different supercritical or subcritical flow at the angle $\beta \in (0,\pi)$ to the flow centreline \citep{Henderson_1966}. 
This satisfies the same relations and entropy production \eqref{eq:EP_jump}-\eqref{eq:Pi_jump} as a normal jump, but written in terms of the velocities $U_1$ and $U_2$ normal to the jump, thus with normal Froude numbers $Fr_{y, ni}  = Fr_{yi} \sin \beta \Y$ for $i \in \{1,2\}$ \citep{Henderson_1966}.
From the second law $\Pi_{\jump} > 0$, an oblique jump is permissible for $Fr_{y1} > (\sin \beta)^{-1} > Fr_{y2} \Y$, 
so in general with a transitional Froude number $Fr_{yc} = (\sin \beta)^{-1} \ge 1$.

Frictional open channel flows at steady state can be classified as (i) {\it uniform flow}, of constant water elevation $y=y_0$ due to equilibrium between frictional and gravitational forces established in a long channel; (ii) {\it gradually-varied flow}, with a smooth flow profile $y(x)$, where $x$ is the flow direction [m]; or (iii) {\it rapidly-varied flow}, with a sharp change in $y(x)$ in response to a sudden constriction \citep{Chow_1959, Henderson_1966, French_1985}.
The resistance for arbitrary cross-sections is often represented by the Manning equation $U= R_H^{2/3} \sqrt{\slope}/n$, where 
$R_H=A/P_w$ is the hydraulic radius [m], 
$P_w$ is the wetted perimeter [m], 
$n$ is Manning's constant for the channel type [s m$^{-1/3}$] 
and 
$\slope=dH_L/dx>0$ is the hydraulic slope [-] 
\citep{Chow_1959, Henderson_1966, French_1985, Chadwick_Morfett_1993}.
The entropy production per unit channel length [J K$^{-1}$ m$^{-1}$ s$^{-1}$] $\Y$ by inertial dispersion \eqref{eq:EP_int} is:
\begin{align}
\widetilde{\dot{\sigma}}_{\open} (x)
= \frac {\rho g \slope Q }{T} 
= \frac {\rho g  Q }{T} \frac{dH_L}{dx}
= \frac{\rho g Q^3 n^2 P_w^{4/3}}{A^{10/3} T} \Y
\label{eq:EP_int_open_frict}
\end{align}
For a rectangular channel of bottom slope $\slope_0$ [-], constant width $B$ and constant flow rate $Q=qB \Y$, where $q=Uy$ is the flow rate per unit width [m$^2$ s$^{-1}$], substituting $A=By$, $P_w=B+2y$, $Fr_y = q/(y^{3/2} \sqrt{g}) \Y$ and the energy equation 
$\slope - \slope_0= (Fr_y^2 -1) \, dy/dx \Y$ 
into \eqref{eq:EP_int_open_frict} 
and rescaling gives the entropic group:
\begin{align}
\begin{split}
\widetilde{\Pi}_{\open}(x)
&= \frac{\widetilde{\dot{\sigma}}_{\open} (x) }{\rho c_p q} \Y
=\frac{ g B \slope(x)}{c_p T}   \Y
=\frac{ g B \slope_0(x)}{c_p T}  + \frac{ g B (\slope-\slope_0)}{c_p T}  \Y
\\&=\frac{ g B \slope_0(x)}{c_p T}  + \frac{ g B \bigl(Fr_y(x)^2 - 1 \bigr)}{c_p T} \frac{dy(x)}{dx} \Y
\\&= \frac{ g B \slope_0(x)}{c_p T} - \frac{ 2 B \bigl(Fr_y(x)^2-1 \bigr) (gq)^{2/3} }{ 3 c_p T \, Fr_y(x)^{5/3} } \frac{dFr_y(x)}{dx} \Y
= \frac{Fr_y(x)^2 \, g^2 n^2 \, (B+2y(x))^{4/3}}{c_p T \,  (B y(x))^{1/3}}  \Y
\end{split}
\label{eq:Pi_int_open_frict}
\raisetag{50pt}
\end{align}
Analysis of \eqref{eq:Pi_int_open_frict} reveals the following effects of friction:
\begin{enumerate}
\item The last term in \eqref{eq:Pi_int_open_frict} is positive for all $Fr_y>0$ and $y>0$, hence $\widetilde{\Pi}_{\open} > 0$ and $\widetilde{\dot{\sigma}}_{\open} > 0$; i.e., the entropy production cannot be zero for finite flow. 

\item In contrast to frictional compressible flows (\S\ref{sect:acoustic}), frictional open channel flows are subject to a larger set of upstream and downstream boundary conditions. These in combination with the channel steepness, flow rate and flow regime --  under the constraint of a positive entropy production -- determine the flow profile $y(x)$ that will be realised. 
Some profiles terminate or start at the critical depth $y=y_c$ or uniform depth $y=y_0$, some start from a (theoretical) zero depth $y=0$, and some terminate in a horizontal water surface \citep{Chow_1959, Henderson_1966, French_1985, Chadwick_Morfett_1993}. 


\item For subcritical flow $Fr_y<1$ and $y>y_c$, from the second law $\widetilde{\Pi}_{\open} > 0$ or $\widetilde{\dot{\sigma}}_{\open} > 0$ in \eqref{eq:Pi_int_open_frict}:
\newcounter{Lcount10}
\begin{list}{(\alph{Lcount10})}{\usecounter{Lcount10} \topsep 3pt \itemsep 0pt \parsep 2pt \leftmargin 40pt \rightmargin 0pt \listparindent 0pt \itemindent 20pt}
\item For uniform flow $\slope = \slope_0$, \eqref{eq:Pi_int_open_frict} implies constant $y=y_0$ and $Fr_y = Fr_{y0}$; 
\item For $\slope_0<\slope$, \eqref{eq:Pi_int_open_frict} implies ${d y}/{d x} < 0$ and ${dFr_y}/{dx} > 1$, so $Fr_y$ will increase with $x$, while $y(x)$ will decrease with $x$ (a drawdown curve); 
\item For $0<\slope<\slope_0$, \eqref{eq:Pi_int_open_frict} implies ${d y}/{d x} > 0$ and ${dFr_y}/{dx} < 1$, so $Fr_y$ will decrease with $x$, while $y(x)$ will increase with $x$ (a backwater curve); 
\end{list}

\item For supercritical flow $Fr_y>1$ and $y<y_c$, from the second law $\widetilde{\Pi}_{\open} > 0$ or $\widetilde{\dot{\sigma}}_{\open} > 0$ in \eqref{eq:Pi_int_open_frict}:
\newcounter{Lcount11}
\begin{list}{(\alph{Lcount11})}{\usecounter{Lcount11} \topsep 3pt \itemsep 0pt \parsep 2pt \leftmargin 40pt \rightmargin 0pt \listparindent 0pt \itemindent 20pt}
\item For uniform flow $\slope = \slope_0$, \eqref{eq:Pi_int_open_frict} implies constant $y=y_0$ and $Fr_y = Fr_{y0}$; 
\item For $\slope_0<\slope$, \eqref{eq:Pi_int_open_frict} implies ${d y}/{d x} > 0$ and ${dFr_y}/{dx} < 1$, hence $Fr_y$ will decrease with $x$, while $y(x)$ will increase with $x$ (a backwater curve); 
\item For $0<\slope<\slope_0$, \eqref{eq:Pi_int_open_frict} implies ${d y}/{d x} < 0$ and ${dFr_y}/{dx} > 1$, hence $Fr_y$ will increase with $x$, while $y(x)$ will decrease with $x$ (a drawdown curve); 
\end{list}

\item In the critical limit $Fr_y \to 1^\mp$ and $y \to y_c^\pm$,  $(Fr_y^2-1) \to 0 \Y$ and $dy/dx \to \mp \infty \Y$, but these limits combine to give 
$\lim_{Fr_y \to 1} \widetilde{\Pi}_{\open} \to {g^2 n^2 (B+2y_c)^{4/3}}/{c_p T \, (B y_c)^{1/3}} >0 \Y$ from either direction. 
A special case of critical uniform flow ($y=y_0=y_c$ and $Fr_y=Fr_{y0}=1$) can form, 
but otherwise critical flow will occur as a limiting case at $x=x_c$.

\end{enumerate}

For gradually-varied flows, the flow profile $y(x)$ and $Fr_y(x)$ can be calculated by numerical integration of the friction equation used in \eqref{eq:Pi_int_open_frict} \citep{Chow_1959, Henderson_1966, French_1985, Chadwick_Morfett_1993}. 
For profiles terminating at $y_c(x_c)$, if the channel is longer than the critical length $x_c$, the flow will undergo  a process similar to {\it frictional choking} (\S\ref{sect:acoustic}) to enable it to pass through $Fr_y=1$.  
Such flows are controlled by their entropy production: each cross-sectional fluid element can only achieve a positive entropy production $\widetilde{\dot{\sigma}}_{\open} > 0$ by altering its depth in accordance with an individual flow profile, as defined in (iii)-(v) above. When $y(x)$ reaches the critical depth $y_c$, no solution to \eqref{eq:Pi_int_open_frict} along that profile with $dy/dx \gtrless 0$ is physically realisable, to enable a positive entropy production.
This triggers the choke, manifested as a transition to a different flow profile (commonly, a smooth transition from subcritical to supercritical flow, or an hydraulic jump). 
For different flow sections, changes in slope, channels of variable width or spatially-varied flow rates, extensions of \eqref{eq:Pi_int_open_frict} are required \citep{Chow_1959, Henderson_1966, French_1985}.



\subsubsection{\it Surface capillary waves} \label{sect:tens}

Energy can also be carried by {\it surface capillary waves} held by surface tension on a gas-liquid surface. 
The angular frequency, phase celerity and group celerity of mixed transitional surface gravity-capillary waves are \citep{Henderson_1966,  Dean_Dalrymple_1991, Kundu_Cohen_2002}:
\begin{align}
\begin{split}
\omega^2 &= k \biggl( g + \dfrac{\varsigma k^2}{\rho} \biggr) \tanh (ky), \YY
\hspace{10pt}
c_{s} 
=\frac{\omega}{k}
= \sqrt{ \biggl( \dfrac{g}{k} + \dfrac{\varsigma k}{\rho} \biggr) \tanh (ky) }, \YY
\hspace{10pt}
\\
c_{s}^{\group} 
&= \dfrac{d\omega}{dk} 
= \dfrac{c_{s}}{2} \biggl( \dfrac{3 \varsigma  k^{2}+ \rho g}{ \varsigma  k^{2}+ \rho g}+\frac{2 k y}{\sinh (2 k y)} \biggr) \Y
\end{split}
\label{eq:Pi_Froude_grav_tens}
\end{align}
These can be expressed in terms of a wave E\"otv\"os or Bond number $Bo = \rho g/ \varsigma k^2 = \rho g \lambda^2/4 \pi \varsigma$ (compare \ref{eq:Pi_disp_tension}), to give the 
corresponding Froude numbers \eqref{eq:Pi_infm2}:
\begin{gather}
\begin{split}
\Pi_{c_{s}} &= Fr_{s} = \dfrac{U}{c_{s}}
= \dfrac{U} {\sqrt {\dfrac{\varsigma k }{\rho}  (Bo + 1) \tanh (ky)}} \Y
= \dfrac{U} {  \sqrt{  \dfrac{g}{k} ({Bo}^{-1}+1)  \tanh (ky)}} \Y
\\
\Pi_{c_{s}}^{\group}  
&= Fr_{s}^{\group}   
= \dfrac{U}{c_{s}^{\group}  } 
=  \dfrac{U}{ \frac{c_{s}}{2} \bigl( \frac{Bo+3}{Bo+1}+\frac{2 k y}{\sinh (2 k y)} \bigr) } \Y
\end{split}
\label{eq:Pi_Froude_grav_tens}
\end{gather}
These have two sets of limits: 
\begin{enumerate}
\item Deepwater waves for  $ky \gtrsim \pi$ hence $\tanh (ky) \to 1$, or shallow waves for 
$ky \lesssim \pi/10$ hence $\tanh (ky) \to ky$ (see \S\ref{sect:grav}); and 
\item Pure surface gravity waves \eqref{eq:c_Airy_waves}-\eqref{eq:Pi_surf_grav_Froude} for $Bo \to \infty$, or pure capillary waves for $Bo \to 0$. 
\end{enumerate}
For deepwater pure capillary waves, the phase celerity \citep{Kundu_Cohen_2002, Hornung_2006, Lemons_2017} and Froude number are obtained as:
\begin{align}
c_{\lambda(\varsigma)} 
= \sqrt{ \dfrac{\varsigma k}{\rho}} \Y
= \sqrt{ \dfrac{2 \pi \varsigma}{\rho \lambda}},  \Y
\hspace{10pt} 
\Pi_{c_{\lambda(\varsigma)}} = Fr_{\lambda(\varsigma)} 
= \dfrac{U}{c_{\lambda(\varsigma)}}  
= {U} {\sqrt{ \dfrac{\rho} {\varsigma k}}} \Y
= {U} {\sqrt{ \dfrac{\rho \lambda}{2 \pi \varsigma}}} \Y
\label{eq:Pi_Froude_tens_deep}
\end{align}
The group celerity and Froude number are $c^{\group}_{\lambda(\varsigma)} = \tfrac{3}{2} c_{\lambda(\varsigma)} \Y$ \citep{Hornung_2006} and $Fr^{\group}_{\lambda(\varsigma)} = \tfrac{2}{3} Fr_{\lambda(\varsigma)} \Y$, so these exhibit anomalous dispersion, with individual waves advancing more slowly than the group. 
In contrast, for shallow pure capillary waves:
\begin{align}
c_{y(\varsigma)} 
=\sqrt{\frac{y \varsigma  k^{2}}{\rho}} \Y
=2 \pi  \sqrt{\frac{y \varsigma}{\rho  \,\lambda^{2}}}, \Y
\hspace{10pt} 
\Pi_{c_{y(\varsigma)}} = Fr_{y(\varsigma)} 
= \dfrac{U}{c_{y(\varsigma)}}  
= U \sqrt{\dfrac{\rho}{y \varsigma  k^{2}}} \Y
= \dfrac{U}{2 \pi} \sqrt{\dfrac{\rho \lambda^2}{y \varsigma  }} \Y
\label{eq:Pi_Froude_tens_shal}
\end{align}
with the group celerity $c^{\group}_{y(\varsigma)} = 2 c_{y(\varsigma)} \Y$ and group Froude number $Fr^{\group}_{y(\varsigma)} = \tfrac{1}{2} Fr_{y(\varsigma)} \Y$.


\subsubsection{\label{sect:grav_int} \it Internal gravity waves}  

Related to surface waves are {\it internal (gravity) waves}, transverse waves within a density-stratified fluid, including the oceans due to temperature and/or salinity variations, or the atmosphere due to pressure and temperature gradients \citep{Miropolsky_2001, Sutherland_2010}. 
In the atmosphere, internal waves are often revealed by stationary (lenticular) clouds or 
repeating cloud patterns (herringbone or mackerel sky) \citep{Downing_2013}. 

For a vertically stratified fluid with horizontal and vertical coordinates $\x = [x,z]^\top$ in the plane of internal wave motion, under the Boussinesq approximation of small changes in density (which excludes sound waves), the celerities of individual waves and wave groups are \citep{Turner_1973, Cushman-Roisin_1994, Kundu_Cohen_2002, Houghton_2002, Nappo_2002, Pedlosky_2003, Holton_2004, Satoh_2004, Carmack_2007, Sutherland_2010}:
\begin{align}
\begin{aligned}
\vec{c}_i 
&=\frac{\omega \k}{||\k||^2} \YY
=\pm \frac{N_0 k_x \k}{||\k||^3},\YY 
&  \vec{c}_i^{\group} 
&= \vnabla_{\k} \omega \YY
= \pm \frac{N_0 k_z \mathcal{R} \k }{||\k||^3},\Y 
\\
\text{with} \hspace{5pt} 
\omega^2 &=   \frac{N_0^2 k_x^2}{||\k||^2},\YY 
&N_0^2 &
\simeq {- \dfrac{g}{\rho_0} \dfrac{d \rho}{d z}}  \YY
\simeq { \dfrac{g}{T_0}\biggl( \dfrac{dT}{d z} + \Gamma \biggr)} \YY
\end{aligned}
\label{eq:c_grav_wave_int}
\end{align}
where 
$\omega$ is the intrinsic angular frequency [s$^{-1}$], 
$\k=[k_x,k_z]^\top$ is the vector wavenumber [m$^{-1}$] in the plane of individual wave migration, 
$\mathcal{R}=\bigl( \begin{smallmatrix} 0&1\\ -1&0 \end{smallmatrix} \bigr)$ is a 90 degree clockwise rotation matrix, 
$N_0$ is the Brunt–V\"ais\"al\"a or buoyancy frequency [s$^{-1}$], a characteristic frequency of the stratified fluid, 
$\Gamma = g (\gamma-1)/R^* \gamma$ is the adiabatic lapse rate [K m$^{-1}$], and
subscript $0$ indicates a reference value. 
The first form of $N_0$ in \eqref{eq:c_grav_wave_int} applies to a density-stratified liquid such as a saline ocean, and the second to the atmosphere or a temperature-stratified ocean. For fluids with two gradients, a composite relation for $N_0$ may be needed \citep{Carmack_2007}. 
Other celerity relations can be derived for different assumptions, such as non-uniformly stratified fluids, non-Boussinesq fluids or nonlinear waves \citep{Satoh_2004, Sutherland_2010}. 
A curious feature of \eqref{eq:c_grav_wave_int} is that the two celerities are orthogonal, $\vec{c}_i \cdot \vec{c}_i^{\group} \propto \k^{\top} \mathcal{R} \k =0 \Y$. In consequence, internal wave groups (and the wave energy) propagate normal to the individual waves, aligned with the wave crests and troughs (travelling in opposing directions) \citep{Turner_1973, Kundu_Cohen_2002, Nappo_2002, Pedlosky_2003, Holton_2004, Satoh_2004, Sutherland_2010}. 

For flow with mean velocity $\bar{\u}$, applying information-theoretic similarity \eqref{eq:Pi_infm2} to \eqref{eq:c_grav_wave_int} in the direction of a unit normal $\n$ gives the directional phase and group Froude numbers:
\begin{align}
\begin{split}
{\Pi}_{\vec{c}_i} (\n) 
&= {Fr}_{\vec{c}_i} (\n)
= \dfrac{\bar{\u} \cdot \n}{\vec{c}_i \cdot \n} 
= \pm \dfrac{||\k||^3 \, \bar{\u} \cdot \n }{N_0 k_x \k \cdot \n}, \Y
\\
{\Pi}_{\vec{c}_i}^{\group} (\n)
&= {Fr}_{\vec{c}_i }^{\group} (\n)
= \dfrac{\bar{\u} \cdot \n}{\vec{c}_i^{\group} \cdot \n}  
= \pm \dfrac{||\k||^3 \, \bar{\u} \cdot \n }{N_0 k_z (\mathcal{R} \k) \cdot \n}  \Y
\end{split}
\label{eq:Pi_Froude_grav_int}
\end{align}
These do not appear to have been defined previously, but suggest the existence of complicated directional subcritical and supercritical internal wave flow regimes, extending those observed for surface gravity waves (\S\ref{sect:grav}). 
Common summary definitions include
{$Fr_y = U/N_0 y$}, 
$Fr_{\ell}=U/N_0 \ell$, 
$Fr_{k}=Uk/N_0$
or the Long or Russell number $Lo= N_0 h/U$, 
where 
$U$ is a summary velocity [m s$^{-1}$], 
$\ell$ is a horizontal length scale [m]
and $h$ is vertical step length scale [m] \citep{Cushman-Roisin_1994, Sutherland_2010, Mayer_Fringer_2017}. 
These have been used to distinguish simple subcritical ($Fr<1$) and supercritical ($Fr>1$) flow regimes (\S\ref{sect:grav}). 
Furthermore, the gradient Richardson number 
$Ri = {N_0^2}/{(\partial u/\partial z)^2}$ 
discriminates between buoyancy-dominated ($Ri \gg 1$) and shear-dominated ($Ri \ll 1$) flows \citep{Turner_1973}. 

\subsubsection{\label{sect:inertial_waves} \it Inertial waves}  

For a fluid on a rotating body such as the Earth, the apparent force (Coriolis effect) created by the non-inertial frame of reference can act as a restoring force for wave motion. There are three main categories: (i) {\it pure inertial waves} with transverse oscillations, dominated by the rotation rate; (ii) {\it inertia-gravity waves} with elliptical oscillations, due to the action of both rotation and buoyancy, and (iii) large-scale {\it Rossby} or {\it planetary waves}, caused by variation of the Coriolis effect with latitude \citep{Houghton_2002, Holton_2004}. 

On a planetary surface, the local strength of rotation can be represented by the Coriolis parameter $f=2 \Omega \sin \varphi$ [s$^{-1}$], where $\Omega$ is the angular frequency of rotation [rad s$^{-1}$] and $\varphi$ is the latitude \citep{Cushman-Roisin_1994, Holton_2004}. 
%
%
For two-dimensional inertia-gravity waves subject to the Boussinesq approximation, the phase celerity and frequency can be obtained as 
\citep{Nappo_2002, Pedlosky_2003, Satoh_2004, Sutherland_2010, Buhler_2014}: 
\begin{align}
\vec{c}_i 
&=\frac{\omega \k}{||\k||^2} \YY
=\pm \frac{ \sqrt{N_0^2 k_x^2 + f^2 k_z^2 }\; \k}{||\k||^3},\Y 
\hspace{10pt}
\text{with}
\hspace{10pt}
\omega^2 =   \frac{N_0^2 k_x^2 + f^2 k_z^2 }{||\k||^2} \YY
\label{eq:c_inertial_wave}
\end{align}
with limits $f \to 0$ for pure gravity waves \eqref{eq:c_grav_wave_int} and $N_0 \to 0$ for pure inertial waves. 
Applying \eqref{eq:c_grav_wave_int} to obtain the group celerity $\vec{c}_{i}^{\group}$, we again find $\vec{c}_{i} \cdot \vec{c}_{i}^{\group} = 0 \Y$ \citep{Satoh_2004}. 
In contrast, for three-dimensional Rossby waves on Earth with coordinates $\x=[x,y,z]^\top$ and wavenumber $\k=[k_x,k_y,k_z]^\top$ oriented east, north and upwards respectively, by the conservation of absolute vorticity subject to the Boussinesq approximation and $f=f_0 + \beta y$ with 
parameter 
$\beta >0$ [m$^{-1}$ s$^{-1}$], 
the celerity relative to the mean flow is \citep{Cushman-Roisin_1994, Houghton_2002, Pedlosky_2003, Satoh_2004, Buhler_2014}: 
\begin{align}
\vec{c}_{Ro} = {\omega} [k_x^{-1}, k_y^{-1}, k_z^{-1}]^\top,  \YY
\hspace{10pt}
\text{with}
\hspace{10pt}
\omega = - \frac{\beta k_x}{k_x^2 + k_y^2 + k_z^2 f_0^2/N_0^2} \YY
\label{eq:c_Rossby_wave}
\end{align}
Since $c_{Ro,x}<0$, Rossby waves on Earth migrate westwards relative to the mean flow, while since $f_0^2/N_0^2 \ll 1$, the vertical component is small \citep{Houghton_2002, Satoh_2004}. Applying \eqref{eq:c_grav_wave_int}, 
we obtain $\vec{c}_{Ro} \cdot \vec{c}_{Ro}^{\group} \ne 0 \Y$; i.e., individual Rossby waves and wave groups are neither parallel nor orthogonal, but travel at an oblique angle \citep{Satoh_2004, Buhler_2014}.  

The directional phase and group Froude numbers for inertia-gravity and Rossby waves can be defined by \eqref{eq:Pi_Froude_grav_int} from each celerity above, 
again suggesting the existence of directional subcritical and supercritical flow regimes. 
More simplistically, applying information-theoretic similarity \eqref{eq:Pi_infm2} based on distinct stratification and rotational celerity scales $c_{\strat}$ and $c_{\rot}$ respectively gives the Froude and Rossby numbers \citep[c.f.,][]{Cushman-Roisin_1994}:
\begin{align}
\Pi_{c_{\strat}} &= \frac{U}{c_{\strat}} \sim Fr_{N_0} = \frac{U}{N_0 h} \YY &\to \hspace{10pt} Fr_{h} = \frac{U}{\sqrt{g h}}, \\
\Pi_{c_{\rot}} &= \frac{U}{c_{\rot}} \sim Ro_f = \frac{U}{f \ell} \Y &\to \hspace{10pt} Ro_{\Omega}= \frac{U}{\Omega \ell} \YY
\label{eq:Rossby_no}
\end{align}
where $h$ and $\ell$ are vertical and horizontal length scales [m]. These have been used to map the flow regimes of inertia-gravity systems, as discriminators respectively of inertial transport relative to wave motion governed by stratification or rotation \citep{Cushman-Roisin_1994}. The Burger number $Bu=Ro^2/Fr^2 = (N_0 h/f \ell)^2 \YY$  directly ranks the last two effects.
%

Many other inertial wave formulations are also available, including for non-Boussinesq fluids, acoustic and inertia-gravity waves in combination, bounded waves, baroclinic waves, thermocline effects, Kelvin waves, equatorial waves, lee and solitary waves, solar and lunar tidal waves, and seasonal and climatic oscillations \citep[e.g.,][]{Eckert_1960, Cushman-Roisin_1994, Houghton_2002, Nappo_2002, Pedlosky_2003, Satoh_2004, Sutherland_2010, Buhler_2014}. 

\subsubsection{\it Electromagnetic waves} \label{sect:electromag}

Electromagnetic waves consist of synchronised transverse oscillations of electric and magnetic fields, quantised in the form of photons.  By information-theoretic reasoning \eqref{eq:Pi_infm2}, the speed of a material object $U$ emitting electromagnetic waves can be ranked against the speed of light $c$ [m s$^{-1}$], both relative to an external inertial frame of reference:
\begin{align}
\Pi_{c} =  \frac{U}{c} 
\end{align}
From special relativity, no phenomenon with $\Pi_{c} > 1$ is permissible in the universe. Instead, a material object with $U \to c$ will undergo time dilation and length contraction in accordance with the Lorentz transformations, while the simultaneity of events for different observers is broken \citep{Collier_2014}. A new technology to achieve $\Pi_{c}>1$ remains a dream of science fiction writers and movie producers. 

Electromagnetic waves transmitted through a material also exhibit wave dispersion, creating {wave packets} with the group celerity $c^{\group} = d\freq/dk$, where $\freq$ is the wave frequency [s$^{-1}$] 
\citep{Brillouin_1960}.
This gives the information-theoretic group $\Pi_{c}^{\group} =  {U}/{c^{\group}}$.
By normal wave dispersion $c^{\group}<c$ and $\Pi_{c}^{\group}>\Pi_{c}$, so it is possible to achieve $\Pi_{c}^{\group}>1$ with respect to the speed of energy transmission. Indeed, the interference of two opposing waves can produce stationary wave packets with $c^{\group}=0$ and $\Pi_{c}^{\group}=\infty$. 

The transport of entropy by electromagnetic radiation (or subatomic particles) and its resulting entropy production are important phenomena, often omitted from standard analyses of radiative processes. 
The {\it radiative energy flux} or {\it energy irradiance} $\vec{j}_{E,\rad}$ [W m$^{-2}$] and the {\it radiative entropy flux} or {\it entropy irradiance} $\vec{j}_{S,\rad}$ [W K$^{-1}$ m$^{-2}$] of electromagnetic radiation striking an infinitesimal area with unit normal $\vec{n}$ are given respectively by  \citep{Planck_1914, Essex_1984a, Essex_1984b, Callies_Herbert_1988, Pelkowski_1994, Incropera_DeWitt_2002, Niven_Noack_2014}: 
\begin{align}
\vec{j}_{E,\rad} &= \vec{n}  \; \int\limits_{0}^{\infty}  \iint\limits_{\Omega}   I_{\rad} (\vec{m})  \; \vec{m} \cdot \vec{n}  \;  d \Omega (\vec{m})  \, d\freq  \YY
\label{eq:rad_energy_flux}
\\
\vec{j}_{S,\rad} &= \vec{n}  \; \int\limits_{0}^{\infty}  \iint\limits_{\Omega}   L_{\rad} (\vec{m})  \; \vec{m} \cdot \vec{n}  \;   d \Omega (\vec{m}) \, d\freq \YY
\label{eq:rad_entropy_flux}
\end{align}
where $I_{\rad}(\vec{m})$ is the {specific energy intensity} or {energy radiance} [W m$^{-2}$ s sr$^{-1}$], the radiation energy per unit frequency travelling through an infinitesimal area of unit normal $\vec{m}$ and infinitesimal solid angle per unit time,
$L_{\rad}(\vec{m})$ is the analogous {specific entropy intensity} or {entropy radiance}  [W K$^{-1}$ m$^{-2}$ s sr$^{-1}$] and
$\Omega$ is the solid angle [sr].
As evident from \eqref{eq:rad_entropy_flux}, the entropy irradiance is a property of the radiation itself, independent of the entropy produced by its conversion to heat.
Thus in the presence of electromagnetic radiation, in addition to the non-radiative (material) local entropy production \eqref{eq:EPdef_local}, there is a separate radiative component \citep{Essex_1984a, Essex_1984b, Essex_1987, Callies_Herbert_1988, Pelkowski_1994}:
\begin{align}
\hat{\dot{\sigma}}_{\rad} 
&=   \frac{\partial }{\partial t} \hat{S}_{\rad} + \nabla \cdot \vec{j}_{S,\rad} \YY
\label{eq:EP_rad}
\end{align}
where $\hat{S}_{\rad}$ is the radiation entropy concentration [J K$^{-1}$ m$^{-3}$].
The Clausius heating term $\vec{j}_{E,\rad}/T$ due to the radiative energy flux must also be added to the thermodynamic entropy flux $\J_S$ in  \eqref{eq:EPdef_local} \citep{Essex_1984a, Essex_1984b, Callies_Herbert_1988}. 

For unpolarised bosons such as photons, the specific entropy intensity is given by 
\citep{Essex_1984a, Essex_1984b, Callies_Herbert_1988, Pelkowski_1994, Goody_Abdou_1996}:
\begin{align}
L_{\rad} (\vec{m}) &= \frac{2 k_B \freq^2}{{c}^2} \biggl[ \biggl(\frac{{c}^2 I_{\rad} (\vec{m})}{2 h \freq^3} +1 \biggr) \ln  \biggl(\frac{{c}^2 I_{\rad} (\vec{m})}{2 h \freq^3} +1 \biggr) - \frac{{c}^2 I_{\rad} (\vec{m})}{2 h \freq^3} \ln  \frac{{c}^2 I_{\rad} (\vec{m})}{2 h \freq^3}  \biggr] \YY
\label{eq:entropy_radiance}
\end{align}
where $k_B$ is Boltzmann's constant [J K$^{-1}$] 
and $h$ is the Planck constant [J s].  Other relations are available for polarised radiation \citep{Planck_1901, Planck_1914} 
or fermions \citep{Essex_Kennedy_1999}. 
For unpolarised bosons emitted by a black-body of radiative temperature $T_{\rad}$, $I_{\rad}$ is given by Planck's law \citep{Planck_1901, Planck_1914, Eckert_Drake_1972, Callies_Herbert_1988, Goody_Abdou_1996}: 
\begin{align}
I_{\rad}(\vec{m}) &
=  \frac{2 h \freq^3}{{c}^2} \frac{1}{\exp (h \freq /k_B T_{\rad}) -1}  \YY
\label{eq:Planck_law}
\end{align}
Substituting \eqref{eq:Planck_law} and \eqref{eq:entropy_radiance} into \eqref{eq:rad_energy_flux}-\eqref{eq:rad_entropy_flux} and integrating over the frequency and a hemisphere $\Omega \in [0, 2 \pi]$, using $\m \cdot \n = \cos \vartheta$ and $d\Omega = \sin \vartheta d\vartheta d\varphi$, where $\vartheta$ and $\varphi$ are respectively the colatitude and longitude in spherical coordinates, gives: 
\begin{align}
\j_{E,\rad} =  \dfrac{2 \pi^{5} k_B^{4} T_{\rad}^{4}}{15 c^{2} h^{3}} \n \Y = \Stef T_{\rad}^{4} \, \n, \Y
\hspace{10pt}
\j_{S,\rad} = \dfrac{8  \pi^{5}  k_B^{4} T_{\rad}^{3}}{45 c^{2} h^{3}} \Y = \frac{4}{3} \Stef T_{\rad}^{3} \, \n \Y
\label{eq:fluxes_rad}
\end{align}
where $\Stef ={2 \pi^{5} k_B^{4}}/{15 c^{2} h^{3}}$ is the Stefan-Boltzmann constant \citep{Essex_1984a, Eckert_Drake_1972}. 
For emissions by or interactions with a surface, \eqref{eq:fluxes_rad} are multiplied by a fractional emissivity $\emiss$, absorptivity $\abs$, reflectivity $\refl$ or transmissivity $\transm$ [-], with $\emiss=\abs$ and $\abs+\refl+\transm=1$ \citep{Incropera_DeWitt_2002, Cengel_etal_2012}, 

We can now construct entropic dimensionless groups from the entropy fluxes for radiation \eqref{eq:fluxes_rad}, the fluid-borne entropy flux $\rho s \u$ or individual diffusion processes \eqref{eq:fluxS}-\eqref{eq:fluxS2}:
\begin{align*}
\begin{aligned}
\hatPi_{\j_{S,\rad}/\j_{E,\rad}} 
&= \dfrac{ ||\j_{S,\rad} || }{ || \j_{E,\rad} /T ||}
= \dfrac{4 T}{3 T_{\rad}},\Y
%
&\hatPi_{\j_{S,\rad}/\j_{S,f}} 
&= \dfrac{ ||\j_{S,\rad}  ||}{|| \j_{S,f} ||} 
\sim \frac{\Stef T_{\rad}^{3}}{\rho  s || \u ||},\Y
\\
\hatPi_{\j_{S,\rad}/\j_{S,\alpha}} 
&= \dfrac{ ||\j_{S,\rad} || }{|| \j_{S,\alpha}|| } 
\sim \frac{\Stef T \, T_{\rad}^{3}}{|| \j_Q ||} \Y
\sim \frac{\Stef T_{\rad}^{3}}{\alpha  \rho  c_{p} T \, || \vnabla T^{-1} ||} \Y
&\to 
Sk  &= \dfrac{\Stef \ell \, T_{\rad}^3}{\alpha \rho c_p} = \dfrac{\Stef \ell \, T_{\rad}^3}{k},\YY
\end{aligned}
\end{align*}
\begin{align}
\begin{aligned}
\hatPi_{\j_{S,\rad}/\j_{S,\Dc}} 
&= \dfrac{ ||\j_{S,\rad}  || }{ || \j_{S,\Dc} ||}
\sim \frac{\Stef T \, T_{\rad}^{3}}{\mu_{c} \, || \j_c ||} \Y
\sim \frac{\Stef R T \, T_{\rad}^{3}}{\Dc \rho  m_{c} \mu_{c} \, || \vnabla  \frac{\mu_c}{T} ||},\Y
\\
\hatPi_{\j_{S,\rad}/\j_{S,\Dk}} 
&= \dfrac{ ||\j_{S,\rad}  || }{ || \j_{S,\Dk} ||}
\sim \frac{ \Stef T \, T_{\rad}^{3} }{ \Phi \, || \i_k ||} \Y 
\sim \frac{ \Stef R T^2 \, T_{\rad}^{3} }{ \Dk z_{k}^{2} F^{2} C_{k} \Phi \, || \vnabla \Phi ||} \Y
\end{aligned}
\label{eq:Pi_radiation_diffusion}
\end{align}
where $\ell$ is a length scale [m]. 
The first group shows that the energy and entropy radiances, for the former converted to heat at $T=T_{\rad}$, respectively carry $\frac{3}{7}$ and $\frac{4}{7}$ of the radiation entropy \citep{Essex_1984a}. 
The second group, comparing radiative transfer and fluid flow, is related to the reciprocal of the Boltzmann or Thring number $Th = \rho c_p U/\emiss \Stef T_{\rad}^3$ with fluid velocity $U$ [m s$^{-1}$] 
\citep{Catchpole_Fulford_1966}. 
The third group reduces to the Stefan or Stark number $Sk$ for heat transfer by radiation relative to conduction \citep{Catchpole_Fulford_1966}.
The final two groups relate radiative transfer to chemical or charge diffusion, with application to photochemical and photovoltaic processes; these are less readily interpreted by dynamic similarity. 

 Additional entropic groups for radiation can be defined using the entropy fluxes for thermodynamic cross-phenomena \eqref{eq:Onsager}, or directly from the entropy production of radiation \eqref{eq:EP_rad}, diffusion \eqref{eq:hatsigma_Q2}-\eqref{eq:hatsigma_k2} and chemical reaction \eqref{eq:hatsigma_chem} processes. 
The group $h_Q/\Stef T_{\rad}^3$ to compare heat convection \eqref{eq:rels_convection} and radiation was given by \cite{Shati_etal_2012}. 


\subsection{\label{sect:univ}Universal diffusion processes}

Consider the common interpretation of the universe based on five dimensions, represented by the set of fundamental SI units $\{$kg, m, s, K,  A$\}$. By dimensional considerations, these must be expressed in terms of five universal constants, commonly taken as:
\begin{enumerate}
\item $c = 299 \, 792 \, 458$ m s$^{-1}$, the speed of light in a vacuum;
\item $\hbar = 1.054\,571\,817 \times 10^{-34}$ J s (or kg m$^2$ s$^{-1}$), the reduced Planck constant;
\item $G = 6.67430 \times 10^{-11}$ m$^3$ kg$^{-1}$ s$^{-2}$, the gravitational constant; 
\item $e = 1.602\, 176\, 634  \times 10^{-19}$ C (or A s), the elementary positive charge, and
\item $k_B = 1.380\,649 \times 10^{-23}$ J K$^{-1}$ (or kg m$^2$ s$^{-2}$ K$^{-1}$) , the Boltzmann constant,
\end{enumerate}
written in the 2019 redefinition of SI units \citep{BIPM_2019}. 
These can be used to define natural or Planck units of mass, length, time, temperature, force and other quantities by dimensional reasoning \citep{Planck_1899}. 
They also define a universal diffusion coefficient:
\begin{align}
\D_{\univ} \sim \sqrt {\frac{\hbar G}{c}} = 4.845\,410\,655 \times 10^{-27} \text{ m$^2$ s$^{-1}$} \Y
\label{eq:diff_universal}
\end{align}
This expresses a minimum bound for diffusion processes in the universe. {Eq.\ \eqref{eq:diff_universal} also suggests an additional conjugate pair for Heisenberg's uncertainty principle $\delta m \, \delta \D \ge \tfrac{1}{2}\hslash$, based on uncertainties in the mass $\delta m$ and diffusion coefficient $\delta \D$ of a physical particle.}

Applying entropic similarity,  we can construct universal dimensionless groups to compare molecular diffusion processes with universal diffusion. For the diffusion of heat, momentum, species $c$ or charged particle $k$ under constant gradients, this gives:
\begin{align}
\begin{aligned}
\hat{\Pi}_{\alpha/\D_{\univ}} 
&= \frac{\hat{\dot{\sigma}}_{\alpha} }{{\hat{\dot{\sigma}}_{\D_{\univ}}}}
\sim \frac{\alpha}{\D_{\univ}}, 
 &
\hat{\Pi}_{\nu/\D_{\univ}} 
&= \frac{\hat{\dot{\sigma}}_{\nu} }{{\hat{\dot{\sigma}}_{\D_{\univ}}}}
\sim \frac{\nu}{\D_{\univ}}, 
\\
\hat{\Pi}_{\Dc/\D_{\univ}} 
&= \frac{\hat{\dot{\sigma}}_{\Dc} }{{\hat{\dot{\sigma}}_{\D_{\univ}}}}
\sim \frac{\Dc}{\D_{\univ}}, 
&
\hat{\Pi}_{\Dk/\D_{\univ}} 
&= \frac{\hat{\dot{\sigma}}_{\Dk} }{{\hat{\dot{\sigma}}_{\D_{\univ}}}}
\sim \frac{\Dk}{\D_{\univ}}
\end{aligned}
\end{align}
In a typical fluid, these will be strongly dominated by microscopic diffusion, but in small or quantum systems the universal dispersion may become important.


\section{\label{sect:concl}Conclusions}

This study proposes an additional category of dimensionless groups based on the principle of {\it entropic similarity}, involving ratios of entropic terms. Since {all} processes involving work against friction, dissipation, diffusion, dispersion, mixing, separation, chemical reaction, gain of information or other irreversible changes are driven by (or must overcome) the second law of thermodynamics, it is appropriate to analyse these processes directly in terms of competing entropy-producing and transporting phenomena and the dominant entropic regime, rather than indirectly in terms of their associated forces.  
The theoretical foundations of entropy are examined in \S\ref{sect:theory}, following which the principle of entropic similarity is established in \S\ref{sect:entrop_sim}, to give three definitions of an entropic dimensionless group: (i) a ratio of entropy production terms; (ii) a ratio of entropy flow rates or fluxes; or (iii) an information-theoretic definition based on a ratio of information fluxes. 
In \S\ref{sect:EP}, these definitions are used to derive entropic groups for a number of entropy-producing and transporting phenomena, including diffusion and chemical reaction processes (\S\ref{sect:diffusion}), an assortment of dispersion mechanisms (\S\ref{sect:dispersion}) and a variety of wave phenomena (\S\ref{sect:wave}). 

Comparing the derived entropic dimensionless groups to those obtained by kinematic or dynamic similarity, or by other means, we can draw several conclusions:

\begin{enumerate}

\item For many groups defined by dynamic similarity as ratios of forces $\Pi = F_1/F_2$ \eqref{eq:dyn_sim}, their reformulation in terms of macroscopic entropy production terms $\Pi = {\dot{\sigma}_1}/{\dot{\sigma}_2}$ \eqref{eq:entropic_sim_EP} recovers the same or a similar dimensionless group. Examples in this category include the Reynolds numbers for internal, external or rotational flows (\S\ref{sect:inertial}), the P\'eclet numbers for diffusion (\S\ref{sect:inertial}), the Grashof and Rayleigh numbers in convection (\S\ref{sect:conv}), and the Weber and E\"otv\"os or Bond numbers for bubbles and droplets (\S\ref{sect:dispersed}). 
However, in all these cases the entropic formulation recovers the product of a friction factor, drag or torque coefficient (or its square) and the dimensionless group, rather than the standalone group given by dynamic similarity. The Archimedes number (\S\ref{sect:conv}) is also obtained as a composite group. 
In addition, the entropic perspective provides an alternative interpretation of these groups: thus the Reynolds number \eqref{eq:Re} expresses the ratio of inertial dispersion and viscous diffusion coefficients, while the Grashof and Archimedes numbers \eqref{eq:Pi_I_nu_conv} and \eqref{eq:Pi_Frdisp_Ar} are the square of this ratio. Furthermore, the various P\'eclet numbers \eqref{eq:Pe} can be interpreted as ratios of the inertial dispersion coefficient to the heat, mass or charge diffusion coefficients, while the Rayleigh number \eqref{eq:Pi_I_nu_alpha_conv} is a mixed ratio of inertial dispersion, viscous and heat diffusion coefficients. The Weber and E\"otv\"os numbers are obtained directly from the ratios of entropy production terms for inertial dispersion and surface or interfacial tension. 

\item In contrast, many entropic groups formed from ratios of local entropy production terms $\hat{\Pi} = {\hat{\dot{\sigma}}_1}/{\hat{\dot{\sigma}}_2}$ \eqref{eq:entropic_sim_EP} or entropy fluxes $\hat{\Pi} = {|| \j_{S,1} ||}/{|| \j_{S,2} ||}$ \eqref{eq:entropic_sim_flux} do not appear to have simple interpretations by dynamic similarity. These include ratios of diffusion terms leading to the Prandtl, Schmidt and Lewis numbers (\S\ref{sect:diffusion_phenom}), most chemical reaction groups including the Damk\"ohler number (\S\ref{sect:chem_rn}), cross-phenomenological groups (\S\ref{sect:cross-diffusion}), ratios of turbulent dispersion coefficients (\S\ref{sect:turb}), hydrodynamic dispersion groups leading to P\'eclet numbers (\S\ref{sect:hydrodynamic}), shear-flow dispersion groups (\S\ref{sect:shear_disp}) and radiative heat groups (\S\ref{sect:electromag}). 
Historically, such groups have usually been identified directly by dimensional analysis or by non-dimensionalisation of the governing equations; the principle of entropic similarity provides a more natural basis for their interpretation. Furthermore, the entropic perspective yields extended formulations of these groups, essential for some problems, as well as many entirely new groups. Examples of the latter include 
the primary definitions of the diffusion groups \eqref{eq:Pi_diffusion} and \eqref{eq:Pi_diffusion_fluxes}, 
hybrid diffusion groups containing a flux and a gradient such as \eqref{eq:Pi_diffusion_mixed}, 
groups for diffusion relative to fluid entropy transport \eqref{eq:Pi_diffusion_fluxes}, 
the primary chemical reaction groups \eqref{eq:Pi_reaction_EP} and \eqref{eq:Pi_reaction_fluxes},
the hybrid group for settling and dispersion of a sediment \eqref{eq:Pi_hydrid_sediment}, 
and groups for the competition between radiation and diffusion \eqref{eq:Pi_radiation_diffusion}.

\item Many groups commonly defined as ratios of the global or local fluid and wave velocities $\Pi = U/c$ or $\hat{\Pi}=\u \cdot \n/ \vec{c} \cdot \n \to ||\u||/||\vec{c}||$ can be reinterpreted from an information-theoretic perspective as ratios of the fluid velocity to the prevalent signal velocity \eqref{eq:Pi_infm_gen} or \eqref{eq:Pi_infm2}. For $\Pi<1$ or $\hat{\Pi}<1$, it is possible for a signal (manifested by a wave) to be transported upstream, thereby influencing the flow, while for $\Pi>1$ or $\hat{\Pi}>1$ this is not possible, leading to two distinct downstream- and upstream-controlled information-theoretic flow regimes. Example groups in this category include Mach numbers for acoustic waves \eqref{eq:Pi_Mach} and blast waves \eqref{eq:Mach_explos}, 
defining the transition from subsonic to supersonic flow, 
and various Froude numbers for 
surface gravity waves \eqref{eq:Pi_Froude}, \eqref{eq:deep_waves} or \eqref{eq:shallow_waves_gen}, 
surface gravity-capillary waves \eqref{eq:Pi_Froude_grav_tens},
internal gravity waves \eqref{eq:Pi_Froude_grav_int} and
inertial waves (\S\ref{sect:inertial_waves}), 
defining the transition from subcritical to supercritical flow. 
Traditionally, by dynamic similarity,  
the Mach number is interpreted as the ratio of inertial and elastic forces, 
while the Froude number is interpreted as the ratio of inertial and gravity forces. 
Similarly the Rossby number \eqref{eq:Rossby_no} is interpreted as the ratio of inertial and rotational forces in planetary flows. 
The entropic perspective also provides a more natural interpretation of a sharp transition between the two flow regimes and their frictional behaviour, including the occurrence of shock waves \eqref{eq:Pi_shock} and frictional choking \eqref{eq:Pi_compr} in compressible flows, and the occurrence of hydraulic jumps \eqref{eq:Pi_jump} and different flow profiles \eqref{eq:Pi_int_open_frict} in open channel flows.
Due to wave dispersion, many flows also have distinct individual and group wave celerities, with corresponding Froude numbers and flow regimes; these can become quite complicated, such as direction-dependent flow regimes for fluids with internal gravity or inertial waves (\S\ref{sect:grav_int}-\ref{sect:inertial_waves}). 

\item A number of dimensionless groups admit multiple interpretations by entropic and other forms of similarity, leading to some fascinating new insights. For example, the Nusselt and Sherwood numbers for convection processes \eqref{eq:Pi_convection} are here derived by entropic similarity based on entropy fluxes, but are usually obtained by kinematic similarity as ratios of the convective and molecular fluxes of heat, species or charge. They can also be interpreted by geometric similarity as dimensionless temperature, concentration or electrical potential gradients
\citep{Schlichting_1968, Incropera_DeWitt_1990, Incropera_DeWitt_2002, Streeter_etal_1998, White_2006}. 
Similarly, the densimetric particle Froude number \eqref{eq:Pi_Frdisp_Ar} is here derived as the ratio of entropy production terms for inertial dispersion by external flow and the dispersed phase, but is traditionally interpreted by dynamic similarity as the ratio of inertial to buoyancy forces. It can also be identified as an inverse Richardson number, which distinguishes free and forced convection (\S\ref{sect:conv}). An equivalence between the Grashof and Archimedes numbers, as square ratios of inertial dispersion to viscous diffusion terms respectively for convection processes or a dispersed phase, is also established. The information-theoretic group for pressure waves (water hammer) \eqref{eq:Pi_hammer} can be identified as an Euler number, traditionally identified as a ratio of pressure to inertial forces. Several other groups commonly used for heat or mass transfer (appendix \ref{sect:apx_plethora}), including the Biot, Fourier, Stefan, Eckert, Brinkman and Stanton numbers, can also be variously interpreted by kinematic, dynamic and entropic similarity. 

\end{enumerate}

To conclude, it is shown that the principle of entropic similarity enables the derivation of new dimensionless groups, beyond those accessible by geometric, kinematic and dynamic similarity, as well as the reinterpretation of many known dimensionless groups. These significantly expand the scope of dimensional analysis and similarity arguments for the resolution of new and existing problems across all branches of science and engineering.

Throughout this study, a concerted effort has been made to examine the transfer of electrical charge, to place this on an equal footing with the better-known relations for mass, momentum and energy transfer processes and chemical reactions. Charge transfer phenomena have important applications to electrolytic, electrochemical and photovoltaic processes -- especially in the presence of fluid turbulence and convection -- needed for the world energy transition from fossil fuels. 
Emphasis is also placed on the vector or tensor basis of the underlying physical phenomena, including velocities, forces, fluxes and gradients, which demands the use of a modern vector-tensor mathematical framework for the construction of dimensionless groups.

Finally, while this work examines a number of important entropic phenomena in mass, momentum, energy and charge transfer processes, chemical reactions, dispersion processes and wave propagation relevant to fluid flow systems, it is not claimed to be complete. Many important phenomena and systems are not examined, including 
mixing and separation unit operations in chemical and environmental engineering 
\citep[e.g.,][]{Seader_Henley_1998}, 
radioactive decay and nuclear processes \citep[e.g.,][]{Sitenko_T_1975}, 
gravitation \citep[e.g.,][]{Misner_etal_2017}, 
hydraulic and hydrological systems \citep[e.g.,][]{Singh_2014, Singh_2015}, 
biological growth, evolutionary and planetary processes \citep[e.g.,][]{Prigogine_1967, Kleidon_2004, Harte_2011}, 
transport systems \citep[e.g.,][]{Ortuzar_Willumsen_2011, Niven_etal_2019}
and economic systems and industrial ecology \citep[e.g.,][]{Ayres_1994}. 
Further research is required on the derivation of entropic dimensionless groups to represent these and many other natural, engineered and human phenomena.

\begin{acknowledgments}
The author thanks all students in his UNSW Canberra undergraduate courses in fluid mechanics, hydraulics, environmental engineering, contaminant hydrogeology, soil mechanics and material science over the past two decades, for lively discussions on dimensional analysis and the phenomena examined herein. 

This research was supported by UNSW and by French sources including Institute Pprime, R\'egion Poitou-Charentes and l'Agence Nationale de la Recherche Chair of Excellence (TUCOROM), Poitiers, France. 

\end{acknowledgments}

\appendix

\section{\label{sect:apx_elec} Electrochemical and Charge Carrier Relations}

To formulate the charge density form of Ohm's law \eqref{eq:Ohm_v2}, it is necessary to adopt a consistent notation for the diffusion of charged species (``drift'') in an electric field, which accounts for both positive and negative ions. For the molar formulation, we first identify the charge flux for the $k$th species as $\i_k=z_k F \j_k \YY = z_k F \v_k C_k \YY$, where $\j_k$ is the molar flux [(mol species) m$^{-2}$ s$^{-1}$] and $\v_k$ is the velocity [m s$^{-1}$] of the $k$th species \citep{Levine_1978,  Newman_1991}. For negative ions $z_k<0$ we have $\sign(\j_k)<0$ and $\sign(\v_k)<0$ in a positive field, whence $\sign(\i_k)>0$; the sign reversal indicates that a flux of negative charge in the negative direction is equivalent to a flux of positive charge in the positive direction. The velocity can further be represented by $\v_k = -z_k F \mu_k \vnabla \Phi \YY= - u_k \vnabla \Phi \YY$, where $\mu_k$ is the charged species mobility [(mol species) m$^{2}$ J$^{-1}$ s$^{-1}$] and $u_k$ is the electric mobility [m$^{2}$ V$^{-1}$ s$^{-1}$], connected by $u_k = z_k F \mu_k \YY$ \citep[c.f.,][]{Levine_1978,  Newman_1991}. (Note the unfortunate overlap between the symbols commonly used for mobilities with other quantities defined in this study.) This yields $\i_k = - z_k^2 F^2 \mu_k C_k \vnabla \Phi \YY = - z_k F u_k C_k \vnabla \Phi \YY$ \citep[c.f.,][]{Newman_1991}. For negative ions $z_k<0$   we have $\mu_k>0$ and $u_k<0$, in a positive electric field $\sign(\vec{E})=-\sign(\vnabla \Phi)>0$, again giving $\sign(\v_k)<0$ and $\sign(\i_k)>0$. Finally, adopting the Sutherland-Einstein relation for ion diffusion $\mu_k=\Dk/RT$ \citep{Sutherland_1905, Einstein_1905} gives $\i_k  = - z_k^2 F^2 \Dk C_k  \vnabla \Phi/R T \YY$ \citep[c.f.,][]{Miomandre_etal_2011}. Moving all terms except $\Dk$ inside the gradient -- thus allowing for spatial inhomogeneity in the concentration and temperature -- gives the first charge density relation in \eqref{eq:Ohm_v2}. 

For the individual carrier formulation, we start with $\i_k=n_k q_k \v_k \YY$ \citep{Halliday_etal_2007}, hence $z_k F c_k = n_k  q_k \YY$ and $\i_k=- q_k^2 n_k^2 \mu_k \vnabla \Phi / C_k \YY$. For negative ions $q_k<0$, we again obtain $\sign(\v_k)<0$ and $\sign(\i_k)>0$ in a positive field. Substituting for the mobility $\mu_k$ and recognising $C_k R=n_k k_B \YY$ then gives $\i_k=- q_k^2 n_k D_k \vnabla \Phi / k_B T \YY$. Moving all terms except $\Dk$ inside the gradient gives the second relation in \eqref{eq:Ohm_v2}. 

\section{\label{sect:apx_thermod_diffusion} Thermodynamic Diffusion Relations}
 
To reduce the thermodynamic diffusion parameter $\D'_c$ in \eqref{eq:Fick1_th}, consider the relationship between the chemical potential $\mu_c$ [J (mol species)$^{-1}$], (relative) chemical activity $\alpha_c$ [--] and specific molar concentration $m_c$ [(mol species) kg$^{-1}$] for the $c$th species in solution \citep{Lewis_Randall_1961, Guggenheim_1967, Levine_1978, Atkins_1982}:
\begin{align}
\alpha_c = \exp \biggl( {\frac{ \mu_c - \mu_c^\minuso}{RT}} \biggr) \YY
= \gamma_c \frac{ m_c}{m_c^\minuso }  \YY
\label{eq:activity}
\end{align}
where $\gamma_c$ is the activity coefficient [--]  and $\minuso$ refers to a quantity at a defined reference state. Commonly $m_c^\minuso=$1 [(mol species) kg$^{-1}$] is chosen to enable the cancellation of units \citep{Atkins_1982}. 
Alternatively, for gaseous species:
\begin{align}
\alpha_c = \exp \biggl( {\frac{\mu_c - \mu_c^\minuso}{RT}} \biggr) \YY
= \frac{f_c}{p_c^\minuso} \YY
= \hat{\gamma}_c \frac{p_c}{p_c^\minuso } \YY
\label{eq:fugacity}
\end{align}
where $f_c$ is the fugacity [Pa], $\hat{\gamma}_c$ is the fugacity coefficient [--] and $p_c=RT C_c$ is the partial pressure [Pa]. 
Rearranging and differentiating \eqref{eq:activity}-\eqref{eq:fugacity} for a spatially invariant reference state gives:
\begin{align}
\vnabla m_c &= m_c \biggl[  \vnabla {\frac{(\mu_c - \mu_c^\minuso)}{RT}}  - \vnabla \ln \gamma_c \biggr]  \YY
\label{eq:dm_c}
\\
\vnabla p_c &= p_c \biggl[  \vnabla  {\frac{(\mu_c - \mu_c^\minuso)}{RT}}  - \vnabla \ln \hat{\gamma}_c \biggr] \YY
\label{eq:dp_c}
\end{align}
Substitution into \eqref{eq:Fick1} gives, respectively:
\begin{align}
\j_c  &= - \rho m_c \Dc \biggl[   \vnabla  \frac{(\mu_c -\mu_c^\minuso)}{R T}  -  \vnabla \ln \gamma_c + \vnabla \ln \rho \biggr]   \YY 	
\label{eq:Fick1_v3_mc}
\\
\j_c  &= - \frac{ p_c \Dc}{R T} \biggl[  \vnabla  {\frac{(\mu_c - \mu_c^\minuso) }{RT}} 
-   \vnabla \ln \hat{\gamma}_c + \vnabla \ln  \frac{1}{T} \biggr]	\YY
\label{eq:Fick1_v3_pc}
\end{align}
We see that the practical relation \eqref{eq:Fick1} conforms to the thermodynamic \eqref{eq:Fick1_th} for a spatially-invariant (or small logarithmic gradients in the) activity or fugacity coefficient, fluid density and temperature, giving $\D'_c \simeq \rho m_c \Dc /R \YY=  p_c \Dc /R^2 T \YY$ respectively for solutes or gaseous species. Neglecting the spatial variation of activity or fugacity coefficients and temperature, the entropy production \eqref{eq:hatsigma_c2} becomes:
\begin{align}
\begin{split}
\hat{\dot{\sigma}}_{\Dc}  
& \simeq \dfrac{ \rho R \Dc }{m_c} \, \bigl | \bigl| \vnabla  {m_c} \bigr| \bigr| ^2 \Y
= \dfrac{ \Dc}{p_c T} \, \bigl | \bigl| \vnabla  {p_c} \bigr| \bigr| ^2 \Y
\end{split}
\label{eq:hatsigma_c2_approx}
\end{align}
for a gradient-controlled system, respectively for diffusion in a liquid or a gas.

\section{\label{sect:apx_plethora} Additional Entropic Groups in Heat and Mass Transfer}

The analysis of convection in heat and mass transfer in \S\ref{sect:conv} can be used to derive a number of related entropic dimensionless groups. 
For example, for ``lumped system analysis'' of the heating of a solid with surface area $A_s$ [m$^2$], volume $V_s$ [m$^3$] and length scale $\ell_s \sim V_s/A_s$ [m], we obtain \citep[c.f.,][]{Bejan_1993, Cengel_etal_2012}:
\begin{align}
\begin{aligned}
\Pi_{h_Q/k_s}
&=\dfrac{||\j_{S,h_Q}||}{||\j_{S,\alpha_s}||}
=\dfrac{||{\widetilde{\j_Q}}|| \frac{1}{T}}{||\j_{Q,s}|| \frac{1}{T} }
= \dfrac{h_Q \Delta T}{k_s ||(\vnabla T)_s || } 
&&\to 
Bi = \dfrac{h_Q \ell_s}{k_s}, \YY 
\\
\Pi_{k_s/(\rho c_p)_s}
&= \dfrac{\F_{S,\cond}^{\net}}{\F_{S,\stor}^{\net}}
= \dfrac{\F_{Q,\cond}^{\net} \frac{1}{T}}{\F_{Q,\stor}^{\net} \frac{1}{T}}
= \dfrac{k_s ||(\vnabla T)_s || A_s} {(\rho c_p)_s (\Delta T)_s V_s /t}
&&\to 
Fo = \dfrac{k_s t}{\rho_s c_{p,s} \ell_s^2} = \dfrac{\alpha_s t}{\ell_s^2} \YY
\end{aligned}
\raisetag{50pt}
\end{align}
where $Bi$ and $Fo$ are the Biot and Fourier numbers, 
$\F_{S}^{\net}$ is a net outwards entropy flow rate {[J K$^{-1}$ m$^{-3}$]} \eqref{eq:EPdef} and 
$\F_{Q}^{\net}$ is a net outwards heat flow rate [J m$^{-3}$], 
with subscripts $cond$ denoting conduction, $stor$ storage and $s$ a solid. $Bi$ represents competition between heat convection at the surface and conduction into the solid, while $Fo$ represents competition between heat conduction into and storage of internal energy within the solid \citep{Eckert_Drake_1972, Cengel_etal_2012}. 
Similarly, applying \eqref{eq:hatsigma_Q2} and \eqref{eq:hatsigma_chem} to a phase change yields:
\begin{align}
\begin{aligned}
\hatPi_{d/\alpha,i} 
&= \dfrac{ \hat{\dot{\sigma}}_{d} }{ \hat{\dot{\sigma}}_{\alpha,i}}
= \dfrac{{\hat{\dot{\xi}}_d} {|\Delta G_{d}|}/{T}} {{\alpha_i \rho_i c_{p,i} T^2} \, || \vnabla {T}^{-1} ||^2}  \Y
&&\to 
{Ste}_i = \dfrac {|\Delta \widetilde{H}_{\lat,d}| / M_{d,i}}{ c_{p,i} \Delta T } \YY
\end{aligned}
\end{align}
where $d$ is the reaction index, $i$ is the index of the final phase, $\Delta \widetilde{H}_{\lat,d}$ is the molar enthalpy (latent heat) of the phase change [J mol$^{-1}$], $M_{d,i}$ is the molar mass [kg mol$^{-1}$] and ${Ste}_i$ is the phase Stefan or Jakob number (or its reciprocal), which assesses the competition between latent and sensible heat during a fluid-solid or gas-liquid phase change \citep{Lock_1969, Bejan_1993}. 
Note that $Ste_i$ is different for each direction of the phase change. 
%
For high-speed flow past a solid surface:
\begin{align}
\begin{aligned}
\Pi_{\text{int}, \ad}
&=\dfrac{|| {\j_{S,\ad}}  ||}{|| {\j_S} ||}
=\dfrac{|| {\j_{Q,\ad}}  || \frac{1}{T} }{|| {\j_Q}  ||  \frac{1}{T} }
\sim \dfrac{ (\Delta T)_{\ad}}{\Delta T}
&&\to
Ec = \frac{(\Delta T)_{\ad}}{\Delta T} = \frac{U_{\infty}^2}{c_p \Delta T}, \YY
\\
\Pi_{\text{int},\nu/\alpha}
&= \dfrac{\dot{\sigma}_{\text{int},\nu}}{\dot{\sigma}_{\text{int},\alpha}}
\sim \dfrac{\rho \nu \ell U_{\infty}^2 / T}{k  \ell^3 || \vnabla T ||^2  /T^2} 
\sim \dfrac{\rho \nu  U_{\infty}^2 T}{k  (\Delta T )^2 } 
&&\to
Br = \frac{\rho \nu U_{\infty}^2}{k \Delta T} 
= \frac{\nu U_{\infty}^2}{ \alpha c_p \Delta T} \YY
= Pr Ec
\end{aligned}
\raisetag{40pt}
\end{align}
where $Ec$ and $Br$ are the Eckert and Brinkman numbers, $U_{\infty}$ is the free-steam velocity [m s$^{-1}$] and subscript $ad$ indicates adiabatic. $Ec$ compares the entropy flux or temperature difference to that of an adiabatic process, controlling the direction of heat transfer \citep{Schlichting_1968, Eckert_Drake_1972}, while $Br$ represents the ratio of macroscopic entropy production by viscous dissipation \eqref{eq:EP_int_lam} to heat conduction \eqref{eq:hatsigma_Q2} \citep[c.f.,][]{Bird_etal_2006}. 
For wall heat transfer, the ratio of convection to inertial dispersion from \eqref{eq:Pe} and \eqref{eq:Pi_convection} is:
\begin{align}
\begin{split}
\Pi_{h_Q/I}
&=\dfrac{\dot{\sigma}_{\text{int},h_Q}}{\dot{\sigma}_{\text{int},I} }
=\dfrac{\dot{\sigma}_{\text{int},h_Q} / \dot{\sigma}_{\text{int},\alpha}  }{  \dot{\sigma}_{\text{int},I} / \dot{\sigma}_{\text{int},\alpha} } 
\sim 
\dfrac{ h_Q \Delta T / k ||\vnabla T|| } { f Pe_\alpha}
\hspace{10pt}
\to 
\frac{1}{f} \, St
\\ 
\text{with } \hspace{5pt} 
&St 
= \dfrac{Nu}{Pe_{\alpha}} 
= \dfrac{Nu}{Re Pr} 
= \dfrac{h_Q}{\rho c_p U} \YY
= \dfrac{h_Q \alpha}{k U} 
\end{split}
\label{eq:heat_mass_groups}
\end{align}
where $St$ is the Stanton number, often used instead of $Nu$. Applying the ``Reynolds analogy'' between momentum and heat transfer, this is correlated with $f$, $Re$ and $Pr$ \citep{Bosworth_1956, Eckert_1963, Schlichting_1968, Eckert_Drake_1972, Incropera_DeWitt_1990, Holman_1990, Bejan_1993, White_2006}. 

Mass-transfer analogues of $Bi$, $Fo$ and $St$, and a momentum-transfer analogue of $Fo$, have also been defined \citep{Eckert_1963, Eckert_Drake_1972, Bear_Bachmat_1991, Incropera_DeWitt_2002}. 


\end{document}